\documentclass[english]{JHEP3}
\usepackage[T1]{fontenc}
\usepackage[latin9]{inputenc}
\usepackage{float}
\usepackage{bm}
\usepackage{amsmath}
\usepackage{graphicx}
\usepackage{amssymb}
\usepackage{esint}

\makeatletter

\newcommand{\noun}[1]{\textsc{#1}}
\newcommand{\lyxdot}{.}

\usepackage{slashed}
\usepackage{cite}
\usepackage[mathscr]{euscript}

\preprint{
DESY 11-133\\
IPPP/11/43\\
DCPT/11/86\\
MCnet/11/17\\
LPN11-43}

\title{Practical improvements and merging of P\Large{OWHEG} \LARGE simulations for vector boson production.}

\author{Simone Alioli \\ Deutsches Elektronen-Synchrotron DESY \\ Platanenallee 6, D-15738 Zeuthen, Germany \\ Email: \email{simone.alioli@desy.de}}

\author{Keith Hamilton \\ INFN, Sezione di Milano Bicocca, Piazza della Scienza 3, 20126 Milan, Italy. \\ Email: \email{keith.hamilton@mib.infn.it}}

\author{Emanuele Re \\ Institute for Particle Physics Phenomenology, Department of Physics \\ University of Durham, Durham, DH1 3LE, UK \\ Email: \email{emanuele.re@durham.ac.uk}}

\abstract{
In this article we generalise \textsc{Powheg} next-to-leading order parton shower (\textsc{Nlops}) simulations of vector boson production and vector boson production in association with a single jet, to give matrix element corrected \textsc{Menlops} simulations. In so doing we extend and provide, for the first time, an exact and faithful implementation of the \textsc{Menlops} formalism in hadronic collisions. We also consider merging the resulting event samples according to a phase space partition defined in terms of an effective jet clustering scale. The merging scale is restricted such that the component generated by the associated production simulation does not impact on the \textsc{NLO} accuracy of inclusive vector boson production observables. The dependence of the predictions on the unphysical merging scale is demonstrated. Comparisons with Tevatron and LHC data are presented.}


\newcommand{\splusminus}{{\mathchoice%
{\vplusminus\displaystyle}%
{\vplusminus\scriptstyle}%
{\vplusminus\scriptscriptstyle}%
{\vplusminus\scriptscriptstyle}%
}}

\newdimen\hbigcirc
\newdimen\wbigcirc
\newcommand{\vplusminus}[1]{%
\settoheight{\hbigcirc}{$#1\bigcirc$}%
\settowidth{\wbigcirc}{$#1\bigcirc$}%
\makebox[\wbigcirc]{%
\makebox[0pt]{\rule[0.4\hbigcirc]{0.5\wbigcirc}{0.05\hbigcirc}}%
\makebox[0pt]{\rule[0.1\hbigcirc]{0.5\wbigcirc}{0.05\hbigcirc}}%
\makebox[0pt]{\rule[0.1\hbigcirc]{0.05\wbigcirc}{0.6\hbigcirc}}%
\makebox[0pt]{$#1\bigcirc$}}%
}

\makeatother

\usepackage{babel}

\begin{document}

\section{Introduction\label{sec:Introduction}}

In order to discover and distinguish the possible production of new
particles at the Tevatron and LHC, a thorough understanding of the
backgrounds is fundamental. In this respect vector boson production
is an all too common obstacle in searches for physics beyond the standard
model. Furthermore, given the high energies afforded by these colliders,
in particular the LHC, and the high-$p_{{\scriptscriptstyle \mathrm{T}}}$
regions of phase space to which new physics searches are sensitive,
the emission of associated QCD radiation, manifest as hard and soft
jets, is ubiquitous. 

The role played by W and Z boson production at hadron colliders is
of course not simply negative. Production of these particles in such
large quantities facilitates precision measurements of \emph{e.g.}
the W boson mass, compensating their obscuring of direct new physics
signals by probing indirect ones. W and Z production are also an important
tool in furthering our knowledge of other fundamentals such as parton
distributions, their uncertainties and also detector / jet energy
calibration. From a purely practical point of view there is therefore
a clear need for accurate and detailed simulations of these processes. 

Of course vector boson production is not the only standard model process
for which precision event generators are well motivated and so it
comes as little surprise, that in recent years the nearing of LHC
data-taking has fueled a resurgence in the research and development
of Monte Carlo simulations. Ground-breaking work carried out by various
groups in 2000-2004 demolished significant long standing problems,
radically improving the fully exclusive, hadron-level, description
of parton shower simulations to consistently include multi-leg tree-order
matrix elements \cite{Catani:2001cc,Mangano:2001xp,Lonnblad:2001iq,Krauss:2002up,Mrenna:2003if}
and, separately, exact next-to-leading order corrections \cite{Frixione:2002ik,Nason:2004rx,Frixione:2007vw};
these are referred to as\noun{ Meps }and \noun{Nlops }simulations
respectively.

In the following years many simulation programs have been constructed,
implementing these new methods and techniques, for a wide variety
of collider physics processes, and their worth proven in various experimental
applications. Progress in this direction has rapidly evolved to the
point where there presently exist three public computer packages automating,
or partially automating, the construction of\noun{ Nlops} programs,
such that little or no expert knowledge is required on the part of
the user to produce a simulation for a given process \cite{Alioli:2010xd,Hoche:2010pf,Frederix:2011zi,Frederix:2011qg}.
Moreover, these packages and the methods underlying them can now be
considered to have withstood the test of complexity, with the recent
arrival of public \noun{Nlops }programs for processes involving rich
colour structures and infrared divergences at leading order \cite{Alioli:2010xa,Alioli:2010qp,Kardos:2011qa},
as well as high multiplicity processes \cite{Melia:2011gk,Oleari:2011ey,Frederix:2011zi,Frederix:2011qg}. 

Having reached this degree of maturity, with systematic differences
of \emph{\noun{Nlops}} schemes now well understood and code production
reaching industrial scales, it becomes relevant to discuss whether
further improvements are possible and how the output of the new automated\emph{
}\emph{\noun{Nlops }}packages may be best used to best effect. In
Refs.\,\cite{Nagy:2005aa,Giele:2007di,Lavesson:2008ah,Bauer:2008qh,Bauer:2008qj,Giele:2011cb}
a number of novel and sophisticated methods for matching parton shower\noun{
}simulations to \emph{multiple} higher order one-loop and tree level
matrix elements have already been put forward. Applications of these
techniques to hadron collider processes are eagerly anticipated.

More recently, two somewhat less ambitious \noun{Menlops }methods
have been introduced, in Refs.\,\cite{Hamilton:2010wh} and \cite{Hoche:2010kg},
for simple processes, combining a \emph{single} \emph{\noun{Nlops}}
event generator with a \noun{Meps} simulation, enhancing the description
of multi-jet events in the former. What these methods lack in accuracy
with respect to their forerunners appears to be offset by their relative
simplicity, making use of, and requiring little modification to, existing
\emph{\noun{Nlops}} and \noun{Meps} programs, with a number of applications
to hadron collider processes having been performed in the referenced
articles. 

We take the view that there is still more to be gained by a continued
bottom-up approach to improving on the established crop of \noun{Nlops
}and \noun{Meps} simulations, as found in the case of \noun{Menlops}
simulations, building on existing programs rather than starting from
scratch.

Our first aim in this work has been to see if we can improve on the
\noun{Menlops }implementations in Refs.~\cite{Hamilton:2010wh} and
\cite{Hoche:2010kg}, to realize the method exactly, simply and without
approximation. In other words, we seek to implement the \noun{Menlops}
method for vector boson production without the introduction of an
unphysical merging scale, in a way which manifestly respects the unitarity
of the \noun{Powheg }method and thus NLO accuracy.

Our second objective has been to extend the \noun{Menlops }method
to the case of processes involving a final-state jet at the leading
order. Since the leading order matrix elements in such cases typically
contain infrared divergences, these simulations require unphysical
transverse momentum cut-offs to make event generation possible. This
has two important practical consequences. Firstly, the description
of the region in which the vector boson has low transverse momentum,
where the great bulk of vector bosons are produced, is also unphysical,
being highly sensitive to the cut and lacking all-orders resummation
of large logarithms in that region. Secondly, when studying jet-associated
production as a signal process, the users of these programs must always
check that the predictions from the program do not depend on the cut.
Typically this mandates using a generation cut far below those defining
all observables, resulting in a very substantial fraction of events
that never make it into any part of the analysis. Thus, here, by extending
the \noun{Menlops} method, we mean effectively including the relevant
lower order, sub-leading matrix element, and the all-orders resummation
required for a physical description of the low $p_{{\scriptscriptstyle \mathrm{T}}}$
region. Note that the methodology here borrows much from the Sudakov
reweighting / vetoing procedures in \noun{Meps} merging schemes. 

Lastly, we have considered merging event samples from the two \noun{Menlops}
simulations, to give one NLO accurate in the description of both fully
inclusive and inclusive vector boson plus jet production observables.
To this end we apply similar reasoning to Ref.\,\cite{Hamilton:2010wh},
populating the region in which the vector boson has a low transverse
momentum with events generated by the single vector boson \noun{Menlops}
simulation and elsewhere with events from the vector boson plus jet
\noun{Menlops }program. While requiring that \emph{both} classes of
observables be described with NLO accuracy puts constraints on the
values that the unphysical merging scale can take, in all other respects
the merging scale dependence is never worse than in the CKKW(-L) /
MLM approaches. However, discounting the fact that renormalization
and factorization scales are perhaps not always chosen optimally in
fixed order calculations, concerns about the scale dependence may
be more a matter of theoretical correctness, rather than practical
importance. To minimize the dependence on the merging scale beyond
that here, in particular, to retain NLO accuracy for both classes
of observable independently of the merging scale, is to our understanding,
tantamount to achieving genuine NNLO-parton shower matching; a goal
far in advance of this simple study, beyond our humbler aims of systematically
getting the most from existing and forthcoming simulations by minimal
and modular interventions.

In section~\ref{sec:Method} we describe the extension of the \noun{Powheg}
\noun{Nlops} simulations to \noun{Menlops }ones, as well as the business
of merging the resulting event samples. Here, when possible, we shall
aim to be conceptual, proving that all relevant technical details
have been sufficiently understood and controlled by reference to Monte
Carlo validation plots. We kindly ask the reader to bear in mind that
the \noun{Powheg }formalism has been derived in considerable detail
\cite{Nason:2004rx,Nason:2006hfa,Frixione:2007vw,Alioli:2010xd},
so a fully self-contained presentation here is not feasible. In section~\ref{sec:Results}
we demonstrate the improvements to be gained through comparisons to
relevant Tevatron and LHC measurements. Our findings and conclusions
are summarized in section~\ref{sec:Conclusions}.

\section{Method\label{sec:Method}}

In this section we elaborate on the steps involved in formulating
event samples NLO accurate in their description of fully inclusive
and single-jet inclusive vector boson production observables. As we
have described in the introduction, this proceeds by merging the output
of two \noun{Menlops} simulations. Thus, our presentation is organized
here as follows. First, in section~\ref{sub:Preliminaries}, we introduce
some basic ideas and notation for \noun{Powheg Nlops }matching, we
then go on to describe how these simulations of W and Z production
may be enhanced to produce genuine \noun{Menlops} simulations in section~\ref{sub:v_menlops},
this is followed by the case of the associated production channel
in section~\ref{sub:vj_menlops}. Finally, the combination of the
\noun{Menlops} samples is described in section~\ref{sub:v-vj_menlops}. 

Since we will already show in this section results from the programs
we have used and created, for illustrative and validation purposes,
we first declare some of the main technical parameters used to run
them. All \noun{Powheg }and \noun{Menlops} events in this paper have
been generated using the CTEQ6M \cite{Pumplin:2002vw} parton distribution
functions, interfaced via \noun{Lhapdf} \cite{Whalley:2005nh}, with
the corresponding value of $\Lambda_{{\scriptscriptstyle \mathrm{QCD}}}$.
In the case of vector boson production, the renormalization and factorization
scales assumed in evaluating the $\bar{B}$ functions, distributing
the \emph{underlying Born }kinematics (Sect.~\ref{sub:Preliminaries}),\emph{
}are given by the vector boson invariant mass, while for associated
production the transverse momentum of the initial $hh\rightarrow Vj$
state is used. The \noun{Powheg~Box} defaults for the W and Z boson
masses and widths are: $m_{{\scriptscriptstyle \mathrm{W}}}=80.398\,\mathrm{GeV}$,
$\Gamma_{\mathrm{W}}=2.141\,\mathrm{GeV}$, $m_{{\scriptscriptstyle \mathrm{Z}}}=91.188\,\mathrm{GeV}$,
$\Gamma_{\mathrm{Z}}=2.486\,\mathrm{GeV}$. The value of the QED coupling
used is given by $\alpha\left(m_{{\scriptscriptstyle \mathrm{Z}}}\right)=1/128.93$.
Throughout the article we have used the \noun{Rivet }analysis framework
\cite{Buckley:2010ar}, including the \noun{FastJet} package \cite{Cacciari:2005hq},
to study our event samples. The Monte Carlo validation analyses in
this section then include a set of minimal cuts on the pseudorapidities
and transverse momenta of the charged leptons emanating from the boson
decays, $\left|\eta\right|<3.5,\, p_{{\scriptscriptstyle \mathrm{T}}}>25\,\mathrm{GeV}$
and, in the case of W production, a missing transverse energy cut,
$\slashed E_{{\scriptscriptstyle \mathrm{T}}}\ge25\,\mathrm{GeV}$.
Lastly, in order to develop the bare \emph{\noun{Powheg }}and \noun{Menlops
}events to include effects of parton showering, hadronization and
multiple interactions we have used the \noun{Pythia~8.150 }\cite{Corke:2010zj,Corke:2010yf,Corke:2011yy}
program with its default tune and PDF set.

\subsection{Preliminaries\noun{\label{sub:Preliminaries}}}

At the heart of the \noun{Powheg~Box} simulation of vector boson
production, as with all \noun{Powheg} simulations, is the so-called
hardest emission cross section \cite{Nason:2004rx} \emph{i.e.} the
differential cross section governing the distribution of the hardest
(highest $p_{T}$) parton branching in each event:\begin{eqnarray}
d\sigma_{_{V}} & = & \bar{B}\left(\bm{\Phi}_{{\scriptscriptstyle V}}\right)\, d\bm{\Phi}_{{\scriptscriptstyle V}}\,\left[\Delta\left(\bm{\Phi}_{{\scriptscriptstyle V}},\, p_{_{\mathrm{T}}}^{\mathrm{min}}\right)+\frac{R\left(\bm{\Phi}_{{\scriptscriptstyle Vj}}\right)}{B\left(\bm{\Phi}_{{\scriptscriptstyle V}}\right)}\,\delta\left(k_{_{\mathrm{T}}}\left(\bm{\Phi}_{{\scriptscriptstyle Vj}}\right)-p_{{\scriptscriptstyle \mathrm{T}}}\right)\,\Delta\left(\bm{\Phi}_{{\scriptscriptstyle V}},\, p_{{\scriptscriptstyle \mathrm{T}}}\right)\, d\bm{\Phi}_{{\scriptscriptstyle j_{1}}}\, dp_{{\scriptscriptstyle \mathrm{T}}}\right]\nonumber \\
 &  & \phantom{\mathrm{phantom\, stuff}}\label{eq:V_hardest_emission_xsec}\end{eqnarray}
In Eq.\,\ref{eq:V_hardest_emission_xsec} $B(\bm{\Phi}_{{\scriptscriptstyle V}})$
stands for the leading order cross section, dependent on the underlying
Born kinematics $\bm{\Phi}_{{\scriptscriptstyle V}}$. Analogously,
$R(\bm{\Phi}_{{\scriptscriptstyle Vj}})$ is the real emission cross
section depending on the real kinematics, $\bm{\Phi}_{{\scriptscriptstyle Vj}}$.
The real kinematics are defined by a mapping which takes as arguments
the leading order kinematics, $\bm{\Phi}_{{\scriptscriptstyle V}}$,
and the radiative phase space variables, $\bm{\Phi}_{{\scriptscriptstyle j_{1}}}$,
parametrizing the hardest parton branching with respect to the configuration
$\bm{\Phi}_{{\scriptscriptstyle V}}$,\begin{equation}
\bm{\Phi}_{{\scriptscriptstyle Vj}}\equiv\bm{\Phi}_{{\scriptscriptstyle Vj}}\left(\bm{\Phi}_{{\scriptscriptstyle V}},\bm{\Phi}_{{\scriptscriptstyle j_{1}}}\right)\,;\label{eq:Phi_Vj}\end{equation}
the $R(\bm{\Phi}_{{\scriptscriptstyle Vj}})$ term can be understood
to have absorbed the Jacobian associated to the factorisation $d\bm{\Phi}_{{\scriptscriptstyle Vj}}=d\bm{\Phi}_{{\scriptscriptstyle V}}\, d\bm{\Phi}_{{\scriptscriptstyle j_{1}}}$.
For the \noun{Powheg~Box }programs used in this study such mappings
are explicitly given in Ref.~\cite{Frixione:2007vw}.%
\footnote{As we will discuss in the next paragraphs, more generally the real
emission cross section needs to be separated according to the number
of counterterms required to make it finite on numerical integration,
each one having its own unique phase space parametrization. %
} $\bar{B}(\bm{\Phi}_{{\scriptscriptstyle V}})$ is the next-to-leading
order distribution of the same (infrared-safe) underlying Born kinematics\begin{equation}
\bar{B}\left(\bm{\Phi}_{{\scriptscriptstyle V}}\right)=B\left(\bm{\Phi}_{{\scriptscriptstyle V}}\right)+\left[V\left(\bm{\Phi}_{{\scriptscriptstyle V}}\right)+\int d\bm{\Phi}_{{\scriptscriptstyle j_{1}}}\, R\left(\bm{\Phi}_{{\scriptscriptstyle Vj}}\right)\right],\label{eq:V_Bbar}\end{equation}
where the virtual contribution $V(\bm{\Phi}_{{\scriptscriptstyle V}})$
here implicitly includes soft and collinear divergences, canceling
those present in the real term; to avoid digressing we simply state
that in general some regularization / subtraction scheme is adopted
rendering the square bracket term in Eq.\,\ref{eq:V_Bbar} finite
and regular. Lastly, the \noun{Powheg} Sudakov form factor is defined
as \begin{equation}
\Delta\left(\bm{\Phi}_{{\scriptscriptstyle V}},\, p_{_{\mathrm{T}}}\right)=\exp\left[-\int d\bm{\Phi}_{{\scriptscriptstyle j_{1}}}\,\frac{R\left(\bm{\Phi}_{{\scriptscriptstyle Vj}}\right)}{B\left(\bm{\Phi}_{{\scriptscriptstyle V}}\right)}\,\theta\left(k_{_{\mathrm{T}}}\left(\bm{\Phi}_{{\scriptscriptstyle Vj}}\right)-p_{_{\mathrm{T}}}\right)\right],\label{eq:V_Sudakov}\end{equation}
where $k_{_{\mathrm{T}}}(\bm{\Phi}_{{\scriptscriptstyle Vj}})$ tends
to the transverse momentum of the radiated parton in the collinear
and soft limits. As in conventional parton shower simulations $k_{_{\mathrm{T}}}$
always has an implicit lower cut-off, $p_{_{\mathrm{T}}}^{\mathrm{min}}$,
of order $\Lambda_{_{\mathrm{QCD}}}$. 

Note that from the point of view of the next-to-leading order corrections,
the details concerning the internal parametrization of the Born kinematics
are completely irrelevant, the key point there being the mappings
defining the relationship between them and the real kinematics. As
an heuristic aid, neglecting, for a moment, the leptonic decay products,
$\bm{\Phi}_{{\scriptscriptstyle V}}$, could be taken to comprise
of the vector boson mass and rapidity, while the radiation phase space
in the \noun{Powheg~Box} is mapped by the polar and azimuthal angles
of the emitted parton with respect to the singular direction (here
the beam axis) together with its energy.

We now quickly remind the reader of the two main properties of the
hardest emission cross section, Eq.\,\ref{eq:V_hardest_emission_xsec},
through which it gives rise to \textsc{NLO} predictions. Firstly,
as the transverse momentum of the hardest branching increases beyond
the Sudakov peak region, the form factor, $\Delta$, tends to one
and, neglecting higher order terms of relative order $\alpha_{{\scriptscriptstyle \mathrm{S}}}^{2}$,
the kinematics are distributed according to the real emission cross
section $R(\bm{\Phi}_{{\scriptscriptstyle Vj}})$: \begin{eqnarray}
\lim_{p_{_{\mathrm{T}}}\rightarrow\mathcal{O}\left(m_{{\scriptscriptstyle V}}\right)}\, d\sigma_{{\scriptscriptstyle V}} & = & \frac{\bar{B}\left(\bm{\Phi}_{{\scriptscriptstyle V}}\right)}{B\left(\bm{\Phi}_{{\scriptscriptstyle V}}\right)}\, R\left(\bm{\Phi}_{{\scriptscriptstyle Vj}}\right)\, d\bm{\Phi}_{{\scriptscriptstyle V}}\, d\bm{\Phi}_{{\scriptscriptstyle j_{1}}}\label{eq:V_NLO_accuracy_1}\\
 & = & R\left(\bm{\Phi}_{{\scriptscriptstyle Vj}}\right)\, d\bm{\Phi}_{{\scriptscriptstyle Vj}}+\mathcal{O}\left(\alpha_{{\scriptscriptstyle \mathrm{S}}}^{2}\right)\,.\nonumber \end{eqnarray}
More subtly and, from the point of view of this study, more significantly,
the second key feature of Eq.\,\ref{eq:V_hardest_emission_xsec}
through which \textsc{NLO} accuracy attained is its \emph{unitarity}
with respect to the Born kinematics, specifically, in the present
context, that the term in square brackets integrates to one for any
$\bm{\Phi}_{{\scriptscriptstyle V}}$. This can be readily deduced
by making use of the identity \begin{equation}
\frac{d\Delta\left(p_{_{\mathrm{T}}}\right)}{dp_{_{\mathrm{T}}}}=\Delta\left(p_{_{\mathrm{T}}}\right)\int d\bm{\Phi}_{{\scriptscriptstyle j_{1}}}\,\frac{R\left(\bm{\Phi}_{{\scriptscriptstyle Vj}}\right)}{B\left(\bm{\Phi}_{{\scriptscriptstyle V}}\right)}\,\delta\left(k_{_{\mathrm{T}}}\left(\bm{\Phi}_{{\scriptscriptstyle Vj}}\right)-p_{_{\mathrm{T}}}\right),\label{eq:V_NLO_accuracy_2}\end{equation}
where the full kinematic dependence of $\Delta$ has been suppressed
for ease of notation. 

For the case of jet-associated vector boson production the basic form
of the hardest emission cross section is unchanged. The principal
difference is in the treatment of the real phase space and real cross
section. In general the real cross section is separated into pieces,
$R=\sum_{\alpha}R^{\alpha}$, one for each infrared counterterm required
to render the cross section regular and numerically integrable, moreover,
for each contribution the real phase space is parametrized differently,
such that the counterterms lend themselves to being integrated analytically
over the radiative phase space \emph{i.e. }\begin{equation}
\bm{\Phi}_{{\scriptscriptstyle Vjj}}\rightarrow\bm{\Phi}_{{\scriptscriptstyle Vjj}}^{\alpha}\equiv\bm{\Phi}_{{\scriptscriptstyle Vjj}}\left(\bm{\Phi}_{{\scriptscriptstyle Vj}},\bm{\Phi}_{{\scriptscriptstyle j_{2}}};\,\alpha\right)\,.\label{eq:Phi_Vjj_alpha}\end{equation}
Thus, the \noun{Powheg }hardest emission cross section in the case
of vector boson plus jet production has the form \begin{eqnarray}
d\sigma_{_{Vj}} & = & \bar{B}\left(\bm{\Phi}_{{\scriptscriptstyle Vj}}\right)\, d\bm{\Phi}_{{\scriptscriptstyle Vj}}\,\left[\,\,\,\,\Delta\left(\bm{\Phi}_{{\scriptscriptstyle Vj}},\, p_{_{\mathrm{T}}}^{\mathrm{min}}\right)\phantom{+\frac{R\left(\bm{\Phi}_{{\scriptscriptstyle Vj}}\right)}{B\left(\bm{\Phi}_{{\scriptscriptstyle V}}\right)}\,\delta\left(k_{_{\mathrm{T}}}\left(\bm{\Phi}_{{\scriptscriptstyle Vj}}\right)-p_{{\scriptscriptstyle \mathrm{T}}}\right)\,\Delta\left(\bm{\Phi}_{{\scriptscriptstyle V}},\, p_{{\scriptscriptstyle \mathrm{T}}}\right)\, d\bm{\Phi}_{{\scriptscriptstyle j_{1}}}\, dp_{{\scriptscriptstyle \mathrm{T}}}}\right.\nonumber \\
 &  & \phantom{\bar{B}\left(\bm{\Phi}_{{\scriptscriptstyle Vj}}\right)\, d\bm{\Phi}_{{\scriptscriptstyle Vj}}\,}+\left.\Delta\left(\bm{\Phi}_{{\scriptscriptstyle V}j},\, p_{{\scriptscriptstyle \mathrm{T}}}\right)\,\sum_{\alpha}\,\frac{R^{\alpha}\left(\bm{\Phi}_{{\scriptscriptstyle Vjj}}^{\alpha}\right)}{B\left(\bm{\Phi}_{{\scriptscriptstyle Vj}}\right)}\,\delta\left(k_{_{\mathrm{T}}}^{\alpha}\left(\bm{\Phi}_{{\scriptscriptstyle Vjj}}^{\alpha}\right)-p_{{\scriptscriptstyle \mathrm{T}}}\right)\, d\bm{\Phi}_{{\scriptscriptstyle j_{2}}}\, dp_{{\scriptscriptstyle \mathrm{T}}}\right]\label{eq:Vj_hardest_emission_xsec}\end{eqnarray}
where\begin{equation}
\bar{B}\left(\bm{\Phi}_{{\scriptscriptstyle Vj}}\right)=B\left(\bm{\Phi}_{{\scriptscriptstyle Vj}}\right)+\left[V\left(\bm{\Phi}_{{\scriptscriptstyle Vj}}\right)+\sum_{\alpha}\,\int d\bm{\Phi}_{{\scriptscriptstyle j_{2}}}\, R^{\alpha}\left(\bm{\Phi}_{{\scriptscriptstyle Vjj}}^{\alpha}\right)\right],\label{eq:Vj_Bbar}\end{equation}
and\begin{equation}
\Delta\left(\bm{\Phi}_{{\scriptscriptstyle Vj}},\, p_{{\scriptscriptstyle \mathrm{T}}}\right)=\exp\left[-\sum_{\alpha}\,\int d\bm{\Phi}_{{\scriptscriptstyle j_{2}}}\,\frac{R^{\alpha}\left(\bm{\Phi}_{{\scriptscriptstyle Vjj}}^{\alpha}\right)}{B\left(\bm{\Phi}_{{\scriptscriptstyle Vj}}\right)}\,\theta\left(k_{_{\mathrm{T}}}^{\alpha}\left(\bm{\Phi}_{{\scriptscriptstyle Vjj}}^{\alpha}\right)-p_{_{\mathrm{T}}}\right)\right]\,.\label{eq:Vj_Sudakov}\end{equation}
Note that in order to ease the proliferation of indices, the functions
$B$, $V$, $R$, $\Delta$ and $\bar{B}$, for the direct and jet-associated
vector boson production processes, are implicitly distinguished by
their phase space arguments. Also, although such details are somewhat
tangential to our course, the $\alpha$-dependent Jacobians associated
with the phase space factorisation $d\bm{\Phi}_{{\scriptscriptstyle Vjj}}^{\alpha}\rightarrow d\bm{\Phi}_{{\scriptscriptstyle Vj}}d\bm{\Phi}_{{\scriptscriptstyle j_{2}}}$
can be understood to have been absorbed in the $R^{\alpha}$ terms,
unless otherwise stated. 

What is important to realize is that, fundamentally, the arguments
concerning how NLO accuracy manifests in this more general case are
unchanged with respect to that of vector boson production. In the
high $p_{{\scriptscriptstyle \mathrm{T}}}$ regime the Sudakov form
factor tends to one and combines with the the $\bar{B}$ prefactor
to adjust the real cross section by genuinely NNLO terms only. Likewise,
as with Eq.\,\ref{eq:V_hardest_emission_xsec}, by virtue of the
fact that the exponent of the Sudakov form factor, Eq.\,\ref{eq:Vj_Sudakov},
is precisely equal to the coefficient multiplying it in Eq.\,\ref{eq:Vj_hardest_emission_xsec},
integration over the radiative phase space reduces the square bracket
term to one, leaving a next-to-leading order distribution of the leading
order vector boson plus jet kinematics. Of course, the fact that the
integral over the radiative phase space returns just $\bar{B}$, does
not prove that NLO accuracy is obtained for general inclusive and
semi-inclusive observables but hopefully it is sufficiently conducive
that we can proceed here and refer the interested reader to section.~4.3
of Ref.~\cite{Frixione:2007vw}, where a rigorous proof based on
this property can be found. For the purposes of our discussion it
is sufficient to appreciate that this unitarity is a necessary condition
for the \noun{Powheg }method to yield NLO predictions, deviations
from unitarity amount to deviations from NLO accuracy in equal measure.

With preliminary concepts and notation in hand we now go on to discuss
how we augment the \noun{Powheg }simulations of vector boson production
and vector boson plus jet production.

\subsection{\noun{Menlops} improved simulation of vector boson production\label{sub:v_menlops}}

In Ref.~\cite{Hamilton:2010wh} it was described how one could instill
NLO accuracy in \noun{Meps}-type simulations for simple processes.
In that case the key consideration was that the effective hardest
emission cross section of \noun{Meps} simulations are not in general
unitary. The square bracket terms in the formulae analogous to Eq.\,\ref{eq:V_hardest_emission_xsec}
do not contain precisely the real emission cross section $R(\bm{\Phi}_{{\scriptscriptstyle Vj}})$
but something equivalent to it up to relative corrections $\mathcal{O}\left(\alpha_{{\scriptscriptstyle \mathrm{S}}}\right)$;
the modifications being due to several factors, the presence of higher
order tree-level matrix elements in the simulation being one of the
more obvious ones. Furthermore, the Sudakov form factor multiplying
the approximate real cross section aims to be that of the parton shower
simulation rather than one based on exponentiating $R(\bm{\Phi}_{{\scriptscriptstyle Vj}})$
or its \noun{Meps }approximate version. In the \noun{Meps}, therefore,
the integral over the radiative phase space then rather gives a function
$N(\bm{\Phi}_{{\scriptscriptstyle V}})\approx1+\mathcal{O}(\alpha_{{\scriptscriptstyle \mathrm{S}}})$.
Based on these assertions it was noted in Ref.~\cite{Hamilton:2010wh}
that \noun{Meps }event samples could be made to respect unitarity
and reproduce NLO predictions in the same way as \noun{Powheg }by
reweighting their events according to $\bar{B}(\bm{\Phi}_{{\scriptscriptstyle V}})/N(\bm{\Phi}_{{\scriptscriptstyle V}})$. 

Here we take a simpler different approach, albeit one adhering to
the same principles, viewing the \noun{Powheg }program for the jet-associated
vector boson production channel as a `perfect' (unitary)\emph{ }\emph{\noun{Meps
}}apparatus, merging vector boson plus jet and vector boson plus dijet
final states. 

To explain how we formulate our \noun{Menlops }simulation in this
section first note that, in practice, the hardest emission cross section
in a \noun{Powheg }simulation is implemented by initially generating
an unweighted set of leading order kinematics according to the next-to-leading
order $\bar{B}$ function, then, with this configuration in hand,
the radiative variables are generated with respect to it, according
to the distribution in square brackets in Eq.\,\ref{eq:V_hardest_emission_xsec}
/ \ref{eq:Vj_hardest_emission_xsec}. This latter radiation generation
mechanism is implemented in essentially the same way as the long established
\emph{matrix} \emph{element} \emph{corrections} technique used in
parton shower simulations \cite{Seymour:1994we,Seymour:1994df,Miu:1998ju,Corcella:1999gs},
with the relevant distributions, $\sim\Delta\, R/B$, being sampled
according to the same \emph{veto} \emph{algorithm}. Thus, here we
proceed by first generating a parton level configuration according
to the hardest emission cross section in the\noun{ }vector boson production
\noun{Powheg }code, Eq.\,\ref{eq:V_hardest_emission_xsec}, \emph{with}
\emph{no} \emph{alterations}, saving the events in the Les Houches
format. These events are then passed as input to a slightly modified
version of the vector boson plus jet simulation, which overrides the
generation of leading order configurations according to $\bar{B}(\bm{\Phi}_{{\scriptscriptstyle Vj}})$,
taking the kinematics and flavour structure instead from the Les Houches
event file input. By propagating the \emph{bare }\emph{\noun{Powheg}}
events output from the vector boson production program to the radiation
generating apparatus of the associated production simulation we obtain
doubly radiative events distributed according to \begin{eqnarray}
d\sigma_{_{\infty}} & = & \bar{B}\left(\bm{\Phi}_{{\scriptscriptstyle V}}\right)\, d\bm{\Phi}_{{\scriptscriptstyle V}}\,\left[\Delta\left(\bm{\Phi}_{{\scriptscriptstyle V}},\, p_{_{\mathrm{T}}}^{\mathrm{min}}\right)+\frac{R\left(\bm{\Phi}_{{\scriptscriptstyle Vj}}\right)}{B\left(\bm{\Phi}_{{\scriptscriptstyle V}}\right)}\,\delta\left(k_{_{\mathrm{T}}}\left(\bm{\Phi}_{{\scriptscriptstyle Vj}}\right)-p_{{\scriptscriptstyle \mathrm{T},1}}\right)\,\Delta\left(\bm{\Phi}_{{\scriptscriptstyle V}},\, p_{{\scriptscriptstyle \mathrm{T,1}}}\right)\, d\bm{\Phi}_{{\scriptscriptstyle j_{1}}}\, dp_{{\scriptscriptstyle \mathrm{T,1}}}\right.\nonumber \\
 &  & \phantom{BLANKBL}\times\left\{ \,\Delta\left(\bm{\Phi}_{{\scriptscriptstyle Vj}},\, p_{_{\mathrm{T}}}^{\mathrm{min}}\right)\phantom{+\frac{R\left(\bm{\Phi}_{{\scriptscriptstyle Vj}}\right)}{B\left(\bm{\Phi}_{{\scriptscriptstyle V}}\right)}\,\delta\left(k_{_{\mathrm{T}}}\left(\bm{\Phi}_{{\scriptscriptstyle Vj}}\right)-p_{{\scriptscriptstyle \mathrm{T}}}\right)\,\Delta\left(\bm{\Phi}_{{\scriptscriptstyle V}},\, p_{{\scriptscriptstyle \mathrm{T}_{1}}}\right)\, d\bm{\Phi}_{{\scriptscriptstyle j_{1}}}\, dp_{{\scriptscriptstyle \mathrm{T}}_{1}}}\right.\nonumber \\
 &  & \phantom{BLANKBL}\phantom{\times}\left.\left.\,+\,\Delta\left(\bm{\Phi}_{{\scriptscriptstyle V}j},\, p_{{\scriptscriptstyle \mathrm{T,2}}}\right)\,\sum_{\alpha}\,\frac{R^{\alpha}\left(\bm{\Phi}_{{\scriptscriptstyle Vjj}}^{\alpha}\right)}{B\left(\bm{\Phi}_{{\scriptscriptstyle Vj}}\right)}\,\delta\left(k_{_{\mathrm{T}}}^{\alpha}\left(\bm{\Phi}_{{\scriptscriptstyle Vjj}}^{\alpha}\right)-p_{{\scriptscriptstyle \mathrm{T,2}}}\right)\, d\bm{\Phi}_{{\scriptscriptstyle j_{2}}}\, dp_{{\scriptscriptstyle \mathrm{T,2}}}\right\} \right]\,;\nonumber \\
\label{eq:menlops_inf_xsec}\end{eqnarray}
this follows straightforwardly from the modular form of the hardest
emission cross sections and Monte Carlo implementation. 

In this way we exploit the unitarity of the radiation machinery in
the vector boson plus jet program to best effect. As described in
section.~\ref{sub:Preliminaries}, the final term in braces is an
exact differential with the integral over the radiative phase space
$\bm{\Phi}_{{\scriptscriptstyle j_{2}}}$ being identically equal
to one for all $\bm{\Phi}_{{\scriptscriptstyle Vj}}$, thus the distributions
of Born and real kinematics of the vector boson production \noun{Nlops}
should be completely untouched in this way. At a technical level,
that this is true for the real kinematics is trivial, since those
events are fed directly to the radiation generation routines of the
vector boson plus jet simulations, on the other hand a much more subtle
and, in fact, obligatory, verification of this procedure is to check
that the predictions for fully inclusive quantities are completely
unchanged by the generation of the secondary radiation in this way. 

Before proceeding to the validation let us also discuss the resummation
properties of Eq.~\ref{eq:menlops_inf_xsec}. Since the \noun{Powheg~Box
}programs implement the hardest emission cross sections using the
\noun{FKS }subtraction formalism, the real cross section $R$ is separated
into pieces singular in only one collinear direction. Moreover, the
damping factors giving rise to the separation are such that each $R^{\alpha}$
is relatively suppressed by a factor $1/p_{{\scriptscriptstyle \mathrm{T}}}^{2}$
as the direction of the emitted parton deviates from the collinear
one associated to $\alpha$, \emph{regardless} of the positioning
of other partons in the event. In fact, as other collinear singular
regions are approached $R^{\alpha}$ is defined to vanish \cite{Frixione:2007vw}.

At least from the point of view of FKS subtraction, the hardest emission
generation in the \noun{Powheg }programs and Eq.~\ref{eq:menlops_inf_xsec}
is not greatly different from a conventional parton shower simulation:
each leg $\left(\alpha\right)$ effectively having associated to it
its own Sudakov form factor, with $R^{\alpha}/B$ in the exponent
(equal to an Altarelli-Parisi splitting function in the collinear
limit), and its own local definition of $p_{{\scriptscriptstyle \mathrm{T}}}$.%
\footnote{See \emph{e.g.} Sect.~4.4.2 of Ref.~\cite{Frixione:2007vw} for
more details.%
} For each emission the radiating leg is, with high probability, the
one with the smallest relative transverse momentum separation and
the distribution carries essentially the same Sudakov suppression
as in a parton shower. Furthermore, the separation of the real cross
section into unique collinear singular regions, $R^{\alpha}$, and
the associated damping factors, effectively produce a strong $p_{{\scriptscriptstyle \mathrm{T}}}$
ordering $p_{{\scriptscriptstyle \mathrm{T},1}}>p_{{\scriptscriptstyle \mathrm{T},2}}$
in Eq.~\ref{eq:menlops_inf_xsec}.%
\footnote{This issue has also been addressed in the technical formulation of
the \noun{Powheg} method in Ref.~\cite{Frixione:2007vw} and again
in the case of Dijet production in Ref.~\cite{Alioli:2010xa}.%
} We stress, however, that while $R^{\alpha}$ is heavily damped away
from the collinear region associated to $\alpha$, and vanishes when
other collinear configurations are approached, there is no hard cut-off,
$p_{{\scriptscriptstyle \mathrm{T},1}}>p_{{\scriptscriptstyle \mathrm{T},2}}$,
as in a parton shower, so disordered configurations can occur, albeit
at a reduced rate --- indeed they must occur if one is to reproduce
the radiation pattern of the double emission matrix element. 

While in our opinion these points, and the formulae they are based
on, are very compelling evidence that the logarithmic accuracy of
the \noun{Menlops }simulation, defined by Eq.~\ref{eq:menlops_inf_xsec},
should be no worse than a conventional parton shower simulation ---
in particular, here, we mean that the Sudakov form factor is NLL accurate
--- we make no allusions about the fact that we have not presented
a rigorous proof of this. Hence, in what follows, we must check that
this is the case and that the quality of the vector boson production
Sudakov form factor is not degraded.

\begin{figure}[H]
\begin{centering}
\includegraphics[width=0.4\textwidth]{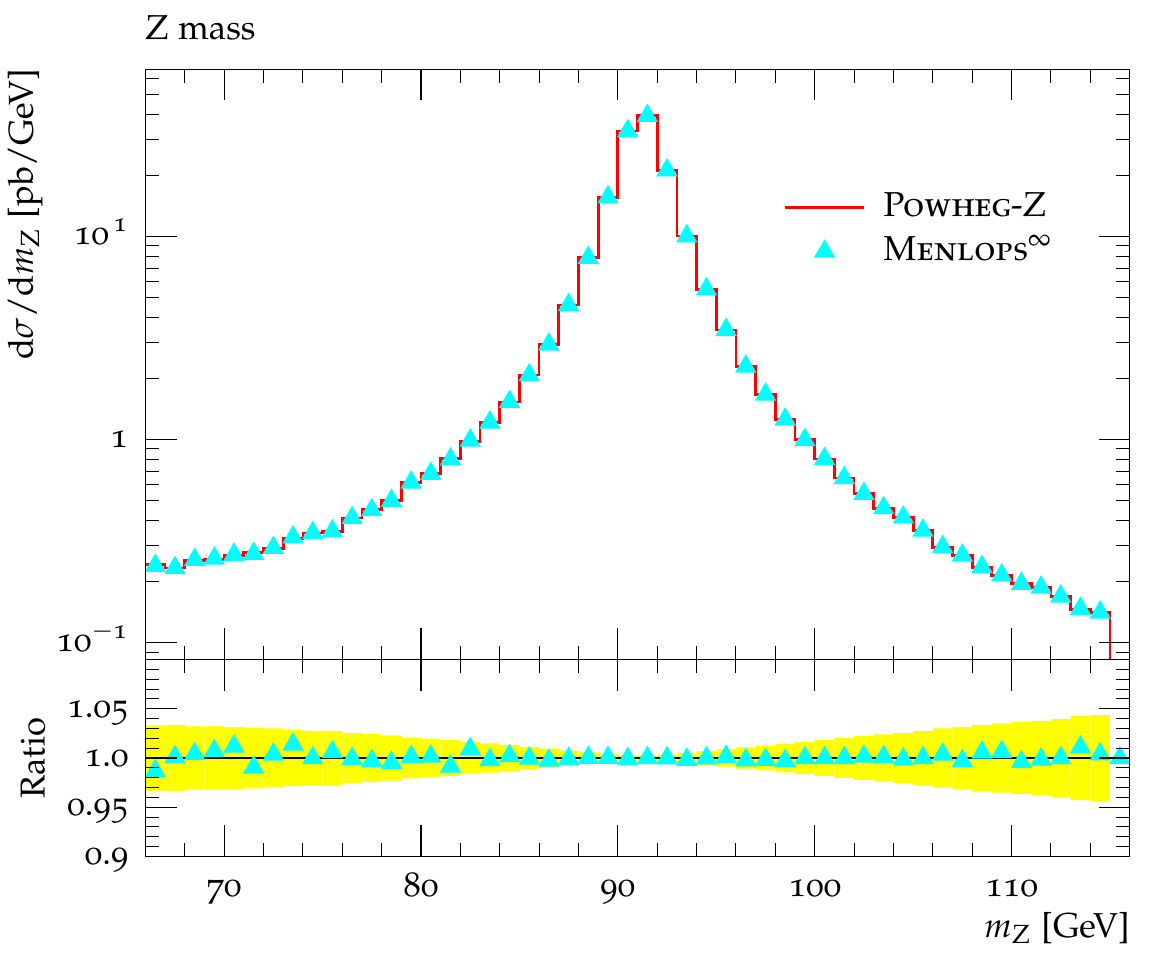}\hfill{}\includegraphics[width=0.4\textwidth]{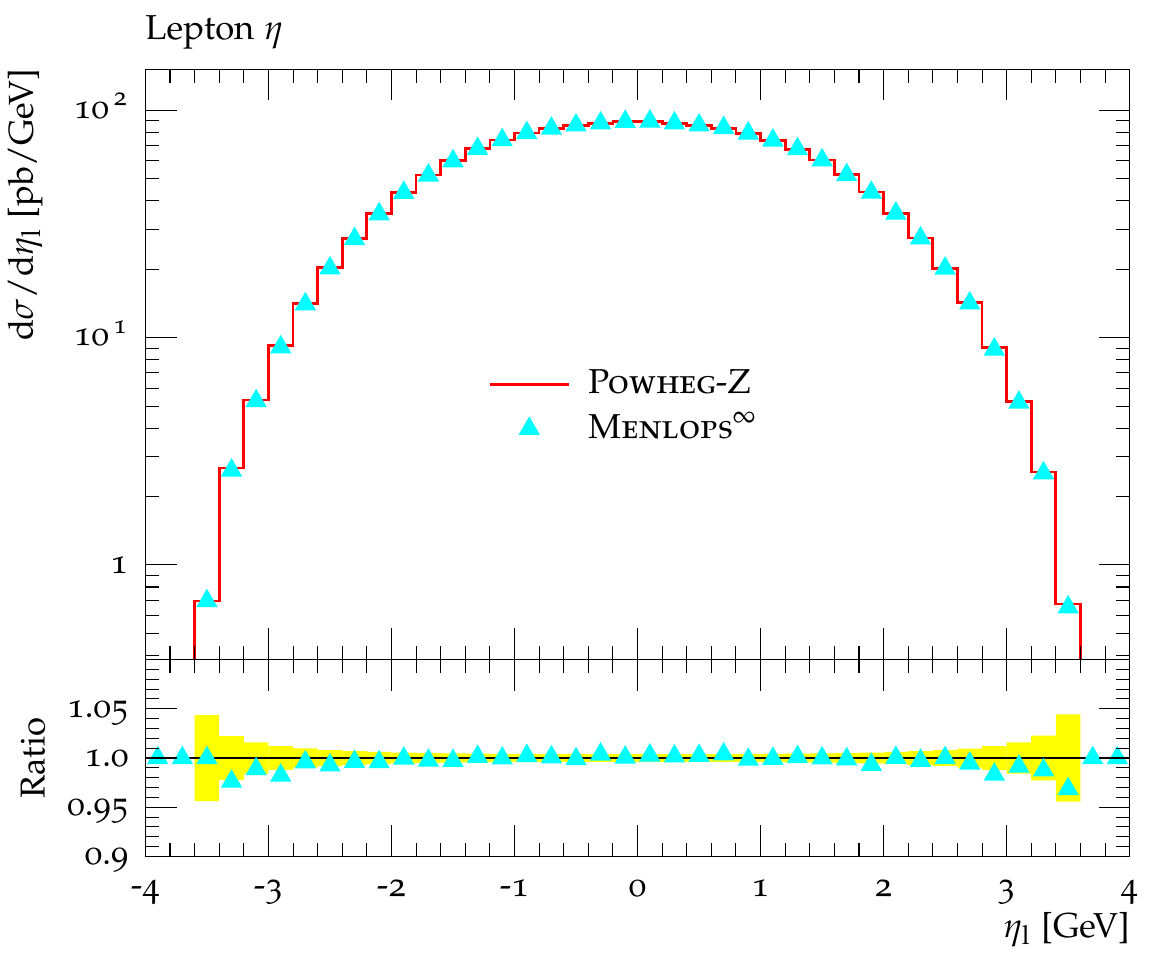}
\par\end{centering}

\begin{centering}
\includegraphics[width=0.4\textwidth]{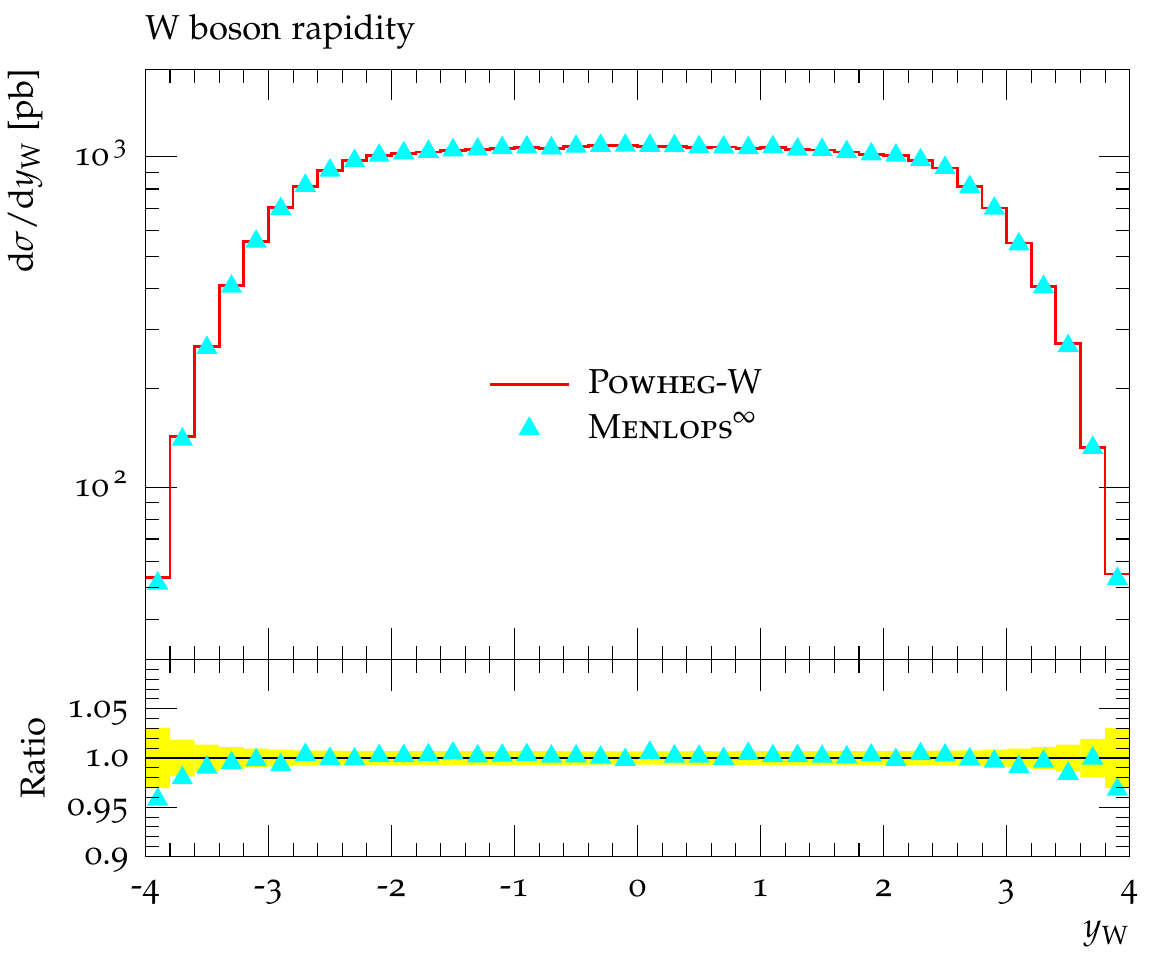}\hfill{}\includegraphics[width=0.4\textwidth]{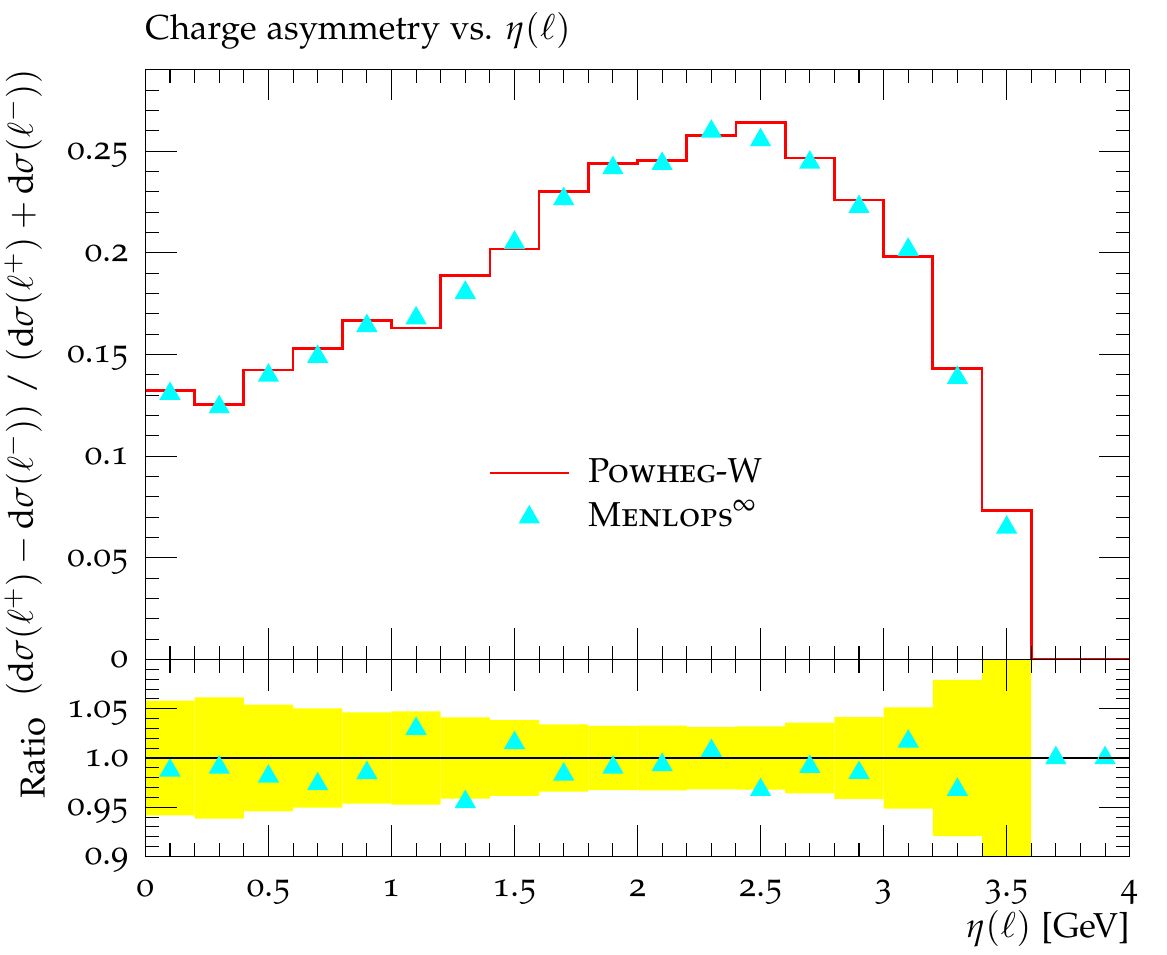}
\par\end{centering}

\caption{Here we display \noun{Powheg }(red line) and \noun{Menlops$^{\text{\ensuremath{\infty}}}$}
(cyan triangles) predictions for inclusive observables. In the upper
pair of plots we show the Z boson mass spectrum and the resulting
lepton pseudorapidity spectrum assuming a Tevatron collider configuration
with a centre-of-mass energy $\sqrt{s}=1.96\,\mathrm{TeV}$, on the
left and on the right respectively. The lower plots show the W boson
rapidity distribution (left) and the charge asymmetry (right), given
$pp$ beams at $\sqrt{s}=7\,\mathrm{TeV}$. The yellow band corresponds
to the projection of the Monte Carlo statistical errors on the reference
data (the first entry in the legend) onto the ratio plot. }

\label{fig:menlops_infty_validation_fully_inclusive} 
\end{figure}

Since we shall introduce a second, different, \noun{Menlops }simulation
in the next section, and in Sect.~\ref{sub:v-vj_menlops} a merging
of the two, we distinguish the one in this section with a suffix $\infty$,
referring to it throughout as \noun{Menlops$^{\text{\ensuremath{\infty}}}$,}
with the double-hard emission\emph{ }cross section $d\sigma_{{\scriptscriptstyle \infty}}$.
The $\infty$ labeling is indicative of the fact that the merged samples,
to be discussed in Sect.~\ref{sub:v-vj_menlops}, are exclusively
comprised of events from this simulation in the limit that the $p_{{\scriptscriptstyle \mathrm{T}}}$
scale associated with the phase space partitioning there tends to
infinity. 

In Fig.~\ref{fig:menlops_infty_validation_fully_inclusive} we present
a comparison of predictions for some inclusive observables obtained
with the idealized \noun{Menlops$^{\text{\ensuremath{\infty}}}$}
simulation described above and also the standard single emission \noun{Powheg~Box}
vector boson production simulation, at the level of the Les Houches
file events \emph{i.e.} the level at which events are input to shower
Monte Carlo programs. All of the distributions show a remarkable level
of agreement between the \noun{Menlops$^{\text{\ensuremath{\infty}}}$
}predictions (cyan triangles) and those obtained with the regular
\noun{Powheg~Box} vector production programs underlying them (red
lines). This is strong evidence that the double-hard emission cross
section, Eq.~\ref{eq:menlops_inf_xsec}, has been realized in practice,
substantiating our earlier claim that, by unitarity, the fully inclusive
\noun{Menlops$^{\text{\ensuremath{\infty}}}$} predictions, Eq.~\ref{eq:menlops_inf_xsec},
should be basically identical to those of the underlying \noun{Powheg
}program.

\begin{figure}[H]
\begin{centering}
\includegraphics[width=0.4\textwidth]{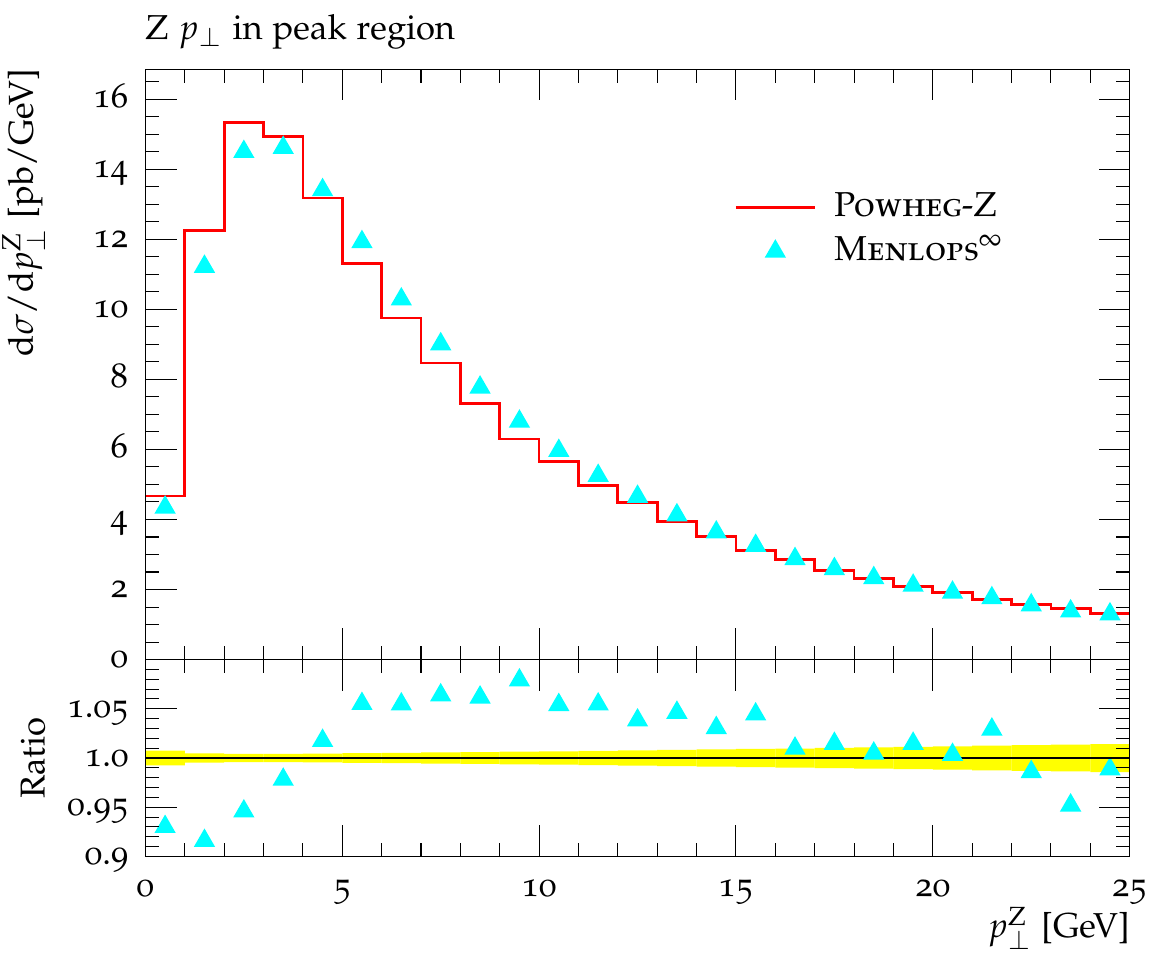}\hfill{}\includegraphics[width=0.4\textwidth]{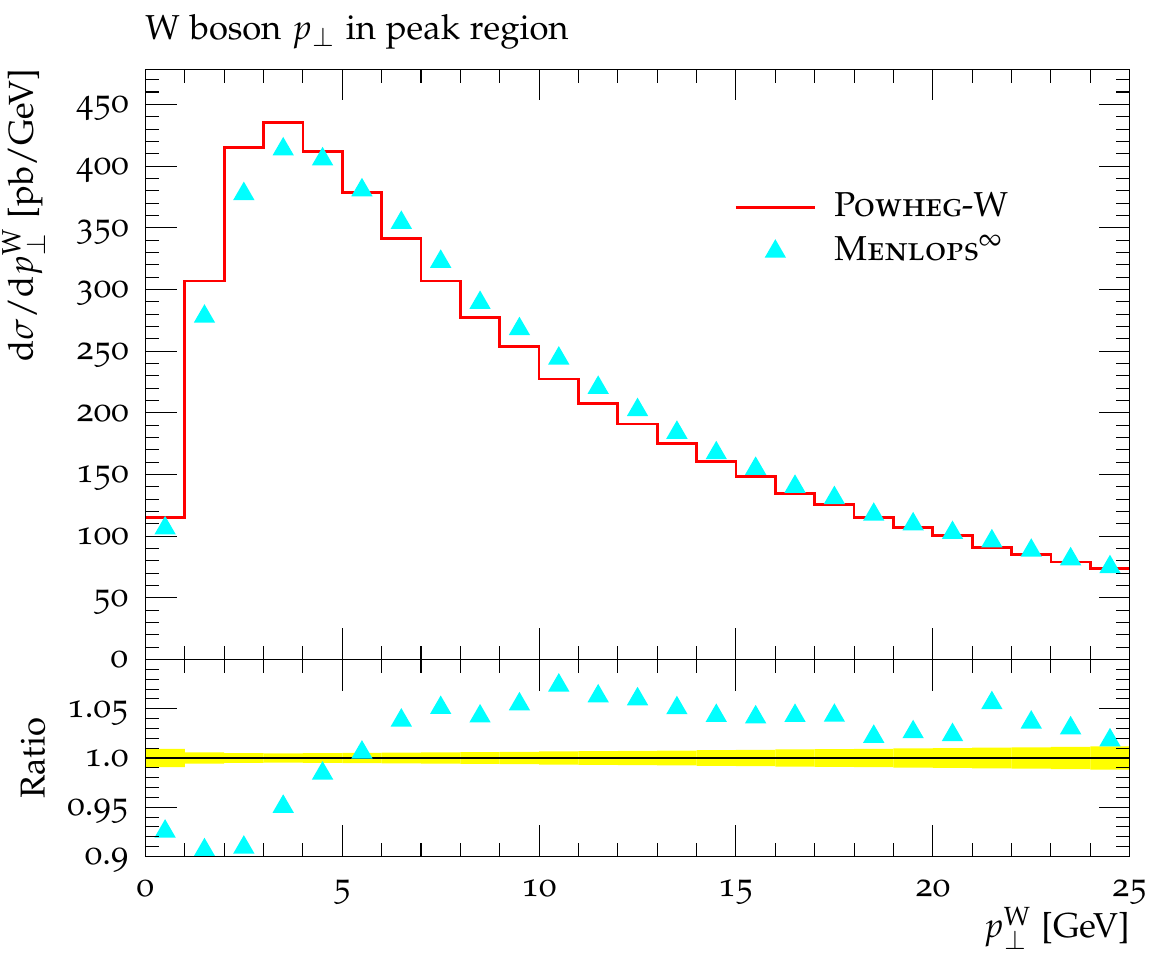}
\par\end{centering}

\caption{\noun{Powheg }(red line) and \noun{Menlops$^{\text{\ensuremath{\infty}}}$}
(cyan triangles) predictions for the low $p_{{\scriptscriptstyle \mathrm{T}}}$,
Sudakov peak region of the vector boson transverse momentum spectrum.
On the left the distributions have been obtained for the case of Z
boson production in $\sqrt{s}=1.96\,\mathrm{TeV}$ $p\bar{p}$ collisions,
while on the right they correspond to W production in $\sqrt{s}=7\,\mathrm{TeV}$
$pp$ reactions.}

\label{fig:menlops_infty_validation_sudakov_peak} 
\end{figure}

Figure~\ref{fig:menlops_infty_validation_sudakov_peak} shows distributions
of the Z and W boson $p_{{\scriptscriptstyle \mathrm{T}}}$ spectra
at Tevatron and LHC energies respectively. In both cases we have focused
on the low $p_{{\scriptscriptstyle \mathrm{T}}}$ Sudakov region,
with a mind to testing our earlier assertions regarding the quality
of the resummation in the \noun{Menlops$^{\text{\ensuremath{\infty}}}$}
prescription (Eq.~\ref{eq:menlops_inf_xsec}). Here, since the distribution
is sensitive to multiple emission effects, and because the \noun{Powheg}
vector boson production cross section contains just one emission,
while Eq.~\ref{eq:menlops_inf_xsec} contains two, to facilitate
a meaningful comparison we have evolved all events with \noun{Pythia}
8 to include parton shower effects (but not hadronization or multiple
interactions). Again, the predictions here vindicate our claims in
regards to how the \noun{Menlops$^{\text{\ensuremath{\infty}}}$}
procedure should preserve the quality of the NLL resummation in the
underlying \noun{Powheg }vector boson simulation, with the differences
in the peak region being $\lesssim10\,\%$. 

From a purely technical point of view, considering how the \noun{Powheg~Box
}program distributes radiation and constructs events \cite{Frixione:2007vw,Alioli:2010xd},
one expects some shuffling of the transverse momentum spectrum associated
with the single emission \noun{Powheg-V }events by the secondary emission
which this \noun{Menlops$^{\text{\ensuremath{\infty}}}$ }procedure
dresses on them.%
\footnote{We will hence use \noun{Powheg-V }and \noun{Powheg-Vj }to generically
refer to \noun{Powheg} vector boson and vector boson plus jet simulations
respectively. %
} More to the point it is natural that this agitation acts to smear
out the peak rather than sharpen it, as is seen to be the case here. 

Note that the same trend is seen to basically the same extent (a depression
of the peak region of $\lesssim10\,\%$) when the basic cuts on the
leptons in the final state in the \noun{Rivet }analysis\noun{ }are
removed. We also note a very similar sized downward shift in the Sudakov
peak is seen in the matched NNLL+NLO $p_{{\scriptscriptstyle \mathrm{T}}}$
spectra of Ref.~\cite{Bozzi:2010xn} with respect to the NLL+LO one
(again, with no cuts applied). Although a great deal more study would
be required before drawing firm conclusions in this regard, and although
we in no way claim to have approached a similar level of accuracy,
the changes seen in that case are intriguingly similar to those shown
here in going from the pure \noun{Powheg }description to this \noun{Menlops}
one. In fact, we believe the explanation for the effects given there
applies here too. In Ref.~\cite{Bozzi:2010xn} it is stated that
additive higher order corrections in the intermediate- and high-$p_{{\scriptscriptstyle \mathrm{T}}}$
regions, coupled with the fact that the NNLO total cross section is
only 3\% greater than the NLO one, serves to redistribute radiation
such that the Sudakov peak decreases.%
\footnote{Note that the resummed computation of Ref.~\cite{Bozzi:2010xn} employs,
in some sense, a sort of unitarity condition like that in \noun{Powheg}.%
} In our case, by unitarity, the total cross section is in fact completely
unchanged by the \noun{Menlops$^{\text{\ensuremath{\infty}}}$ }addition,
while the same higher order real corrections enter as in Ref.~\cite{Bozzi:2010xn},
thus one finds a similar redistribution of radiation.

From our point of view, however, the most significant point to take
away from Ref.~\cite{Bozzi:2010xn} is that the theoretical uncertainties
due to scale variations alone, for the NNLL+NLO calculation, already
bracket the $10\,\%$ effect we see here, while for the NLL+LO calculation
(which we would expect to have equivalent accuracy to \cite{Nason:2006hfa,Frixione:2007vw})
they can change the peak height by $\pm25\,\%$. We also remark that
when the full range of non-perturbative effects are activated in the
simulation the Sudakov peak is eroded further, with the differences
between the \noun{Powheg} and \noun{Menlops$^{\text{\ensuremath{\infty}}}$
}predictions becoming negligible. 

\begin{figure}[H]
\begin{centering}
\includegraphics[width=0.4\textwidth]{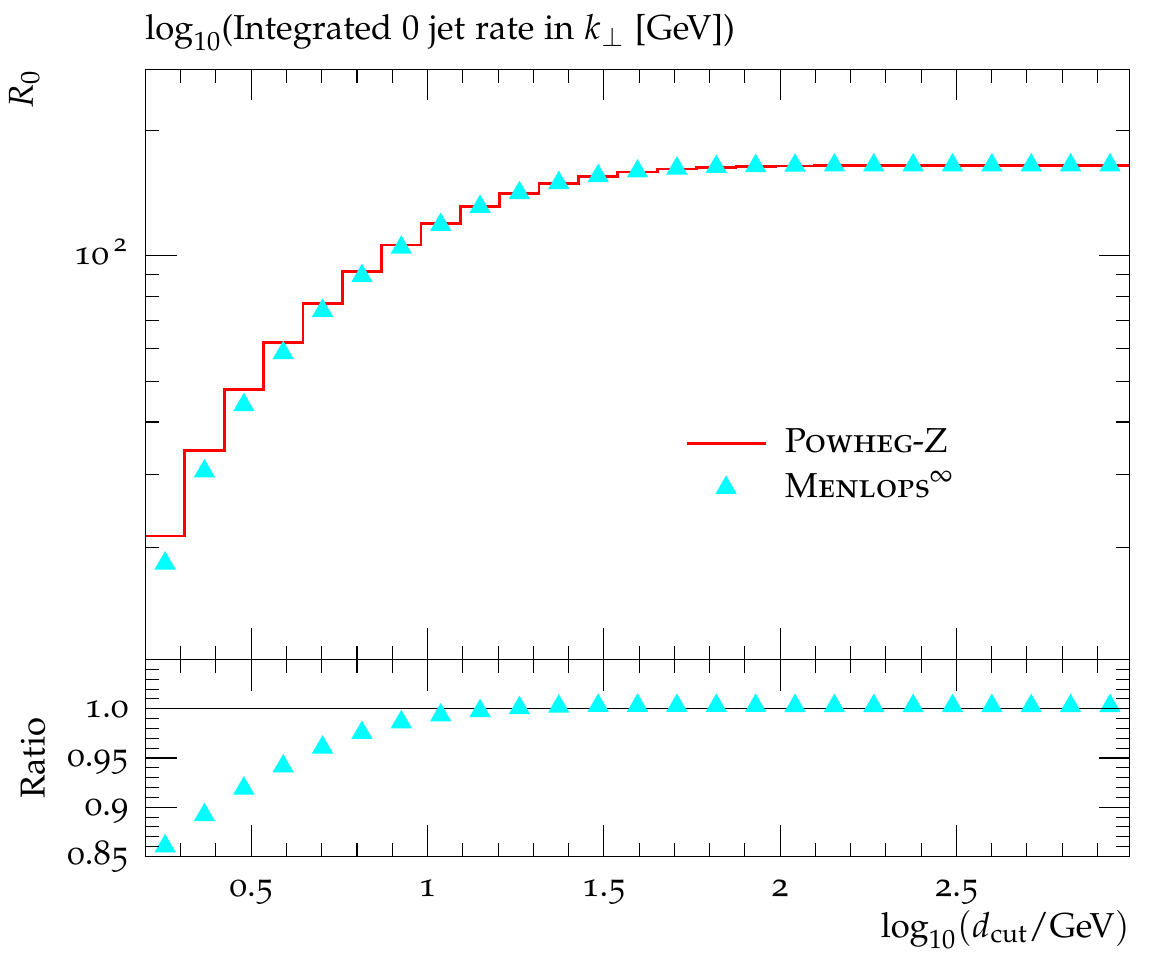}\hfill{}\includegraphics[width=0.4\textwidth]{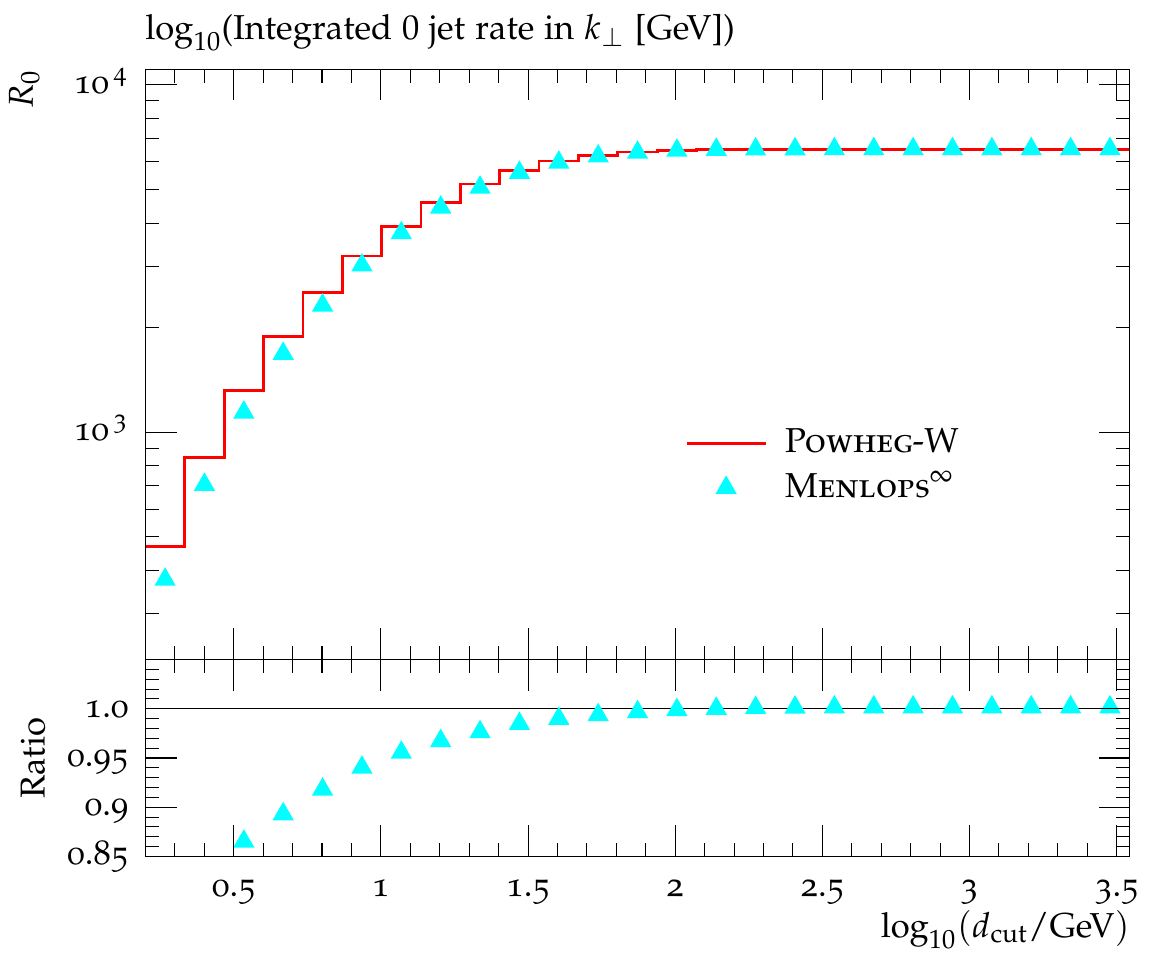}
\par\end{centering}

\caption{\noun{Powheg }(red line) and \noun{Menlops$^{\text{\ensuremath{\infty}}}$}
(cyan triangles) predictions for the integrated $0$-jet cross section
as a function of the $k_{{\scriptscriptstyle \mathrm{T}}}$-jet clustering
scale. As in Fig.~\ref{fig:menlops_infty_validation_sudakov_peak}
the left plot depicts results for Z boson production in $\sqrt{s}=1.96\,\mathrm{TeV}$
$p\bar{p}$ collisions, while those in the right-hand plot correspond
to W production in $\sqrt{s}=7\,\mathrm{TeV}$ $pp$ reactions.}

\label{fig:menlops_infty_validation_integrated_jet_rate} 
\end{figure}

Lastly, to end the discussion on the $p_{{\scriptscriptstyle \mathrm{T}}}$
spectrum, we draw attention to Fig.~\ref{fig:menlops_infty_validation_integrated_jet_rate},
displaying a related but more inclusive observable with which to expect
agreement, namely, the \emph{integrated }0-jet rate, as a function
of the $k_{{\scriptscriptstyle \mathrm{T}}}$-jet clustering scale.
This quantity is closely related to the integral of the vector boson
$p_{{\scriptscriptstyle \mathrm{T}}}$ spectrum. Here one can see
very small differences at the low end of the distribution, in keeping
with those seen in the $p_{{\scriptscriptstyle \mathrm{T}}}$ spectrum,
however, unlike that case, the more inclusive nature of the observable
(the fact that it is cumulative) allows for an averaging effect, and
sure enough, the two distributions rapidly become indistinguishable
like those in Fig.~\ref{fig:menlops_infty_validation_fully_inclusive}.
Again, this is a strong check of the unitarity and NLO accuracy of
the double-hard emission cross section and its implementation. 

\begin{figure}[H]
\begin{centering}
\includegraphics[width=0.4\textwidth]{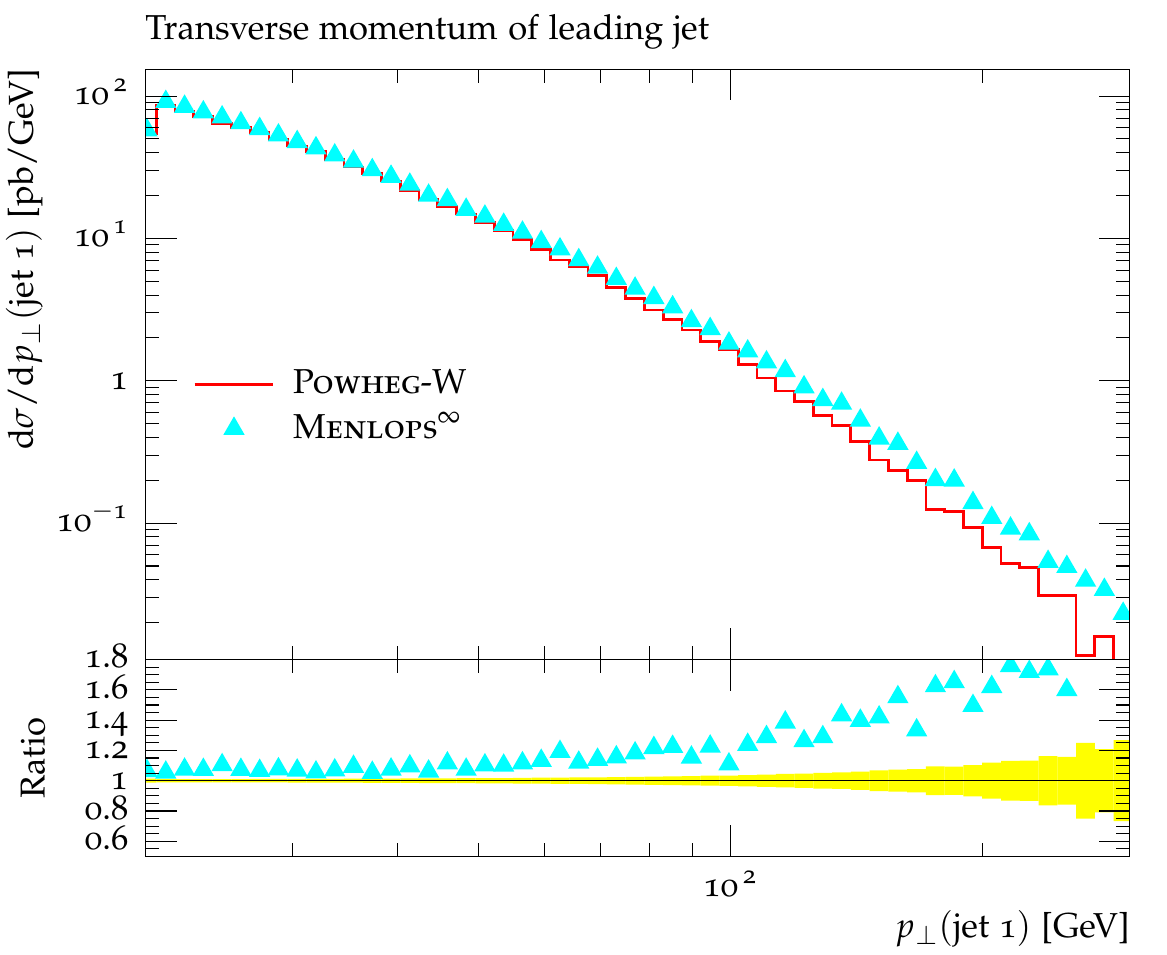}\hfill{}\includegraphics[width=0.4\textwidth]{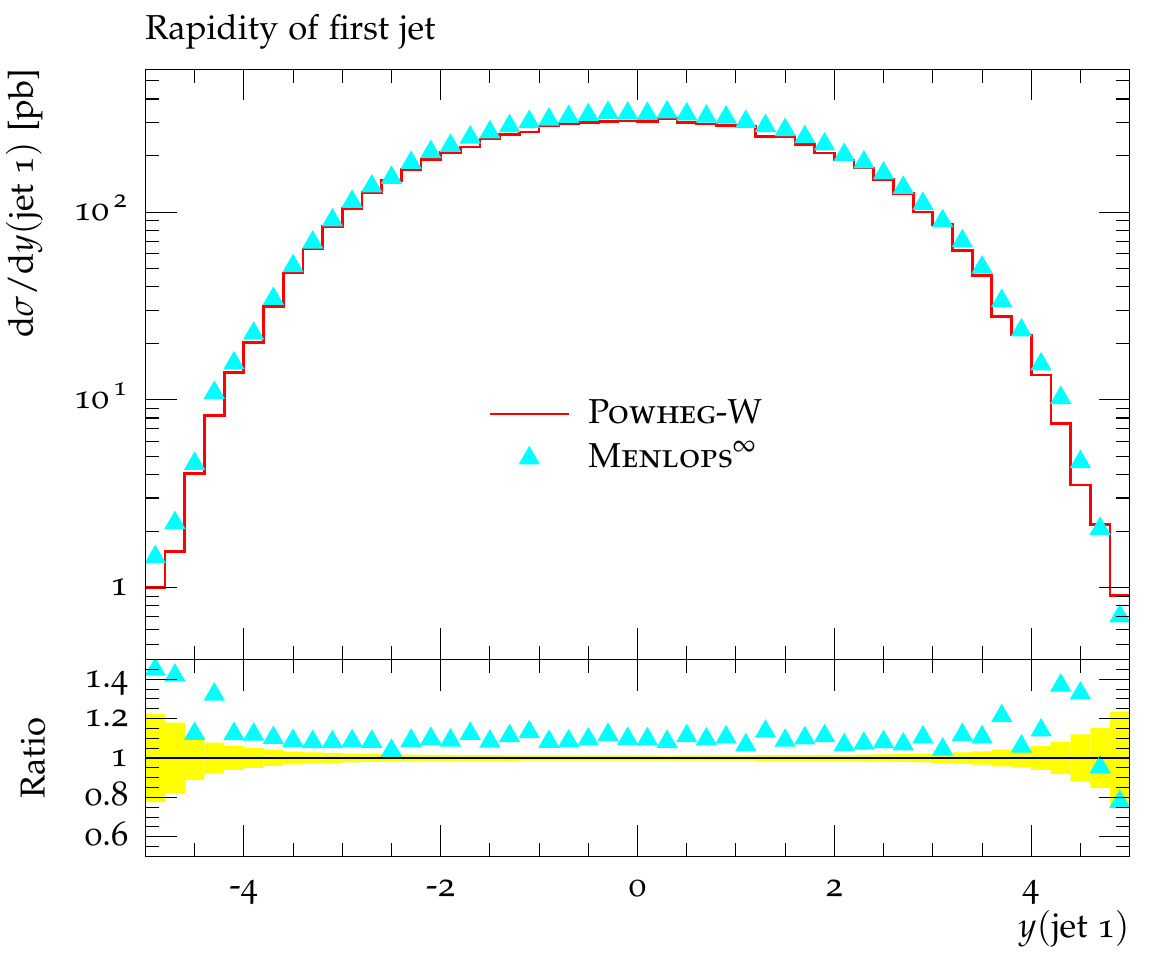}
\par\end{centering}

\caption{\noun{Powheg }(red line) and \noun{Menlops$^{\text{\ensuremath{\infty}}}$}
(cyan triangles) predictions for the leading jet transverse momentum
(left) and rapidity spectrum (right) in W boson production events,
assuming current LHC $pp$ beam energies.}

\label{fig:menlops_infty_validation_ge_1j} 
\end{figure}

Turning our attention to slightly less inclusive observables in Fig.~\ref{fig:menlops_infty_validation_ge_1j}
we show the leading jet transverse momentum and rapidity spectra.
Somewhat reassuringly the rapidity spectrum exhibits no discernible
differences between the two different simulations; since it is still
a rather inclusive observable and the distributions governing its
generation are essentially the same in either case, the good agreement
seen in that case is to be expected. The predictions for the leading
jet $p_{{\scriptscriptstyle \mathrm{T}}}$ spectrum also agree very
well in the region where one expects them to \emph{i.e.} the low-
/ intermediate-$p_{{\scriptscriptstyle \mathrm{T}}}$ region. On the
contrary, the distributions come apart at high transverse momentum,
the \noun{Menlops} simulation proving to be harder there, due to the
inclusion of the doubly radiative events there \cite{Rubin:2010xp}.

\begin{figure}[H]
\begin{centering}
\includegraphics[clip,width=0.4\textwidth]{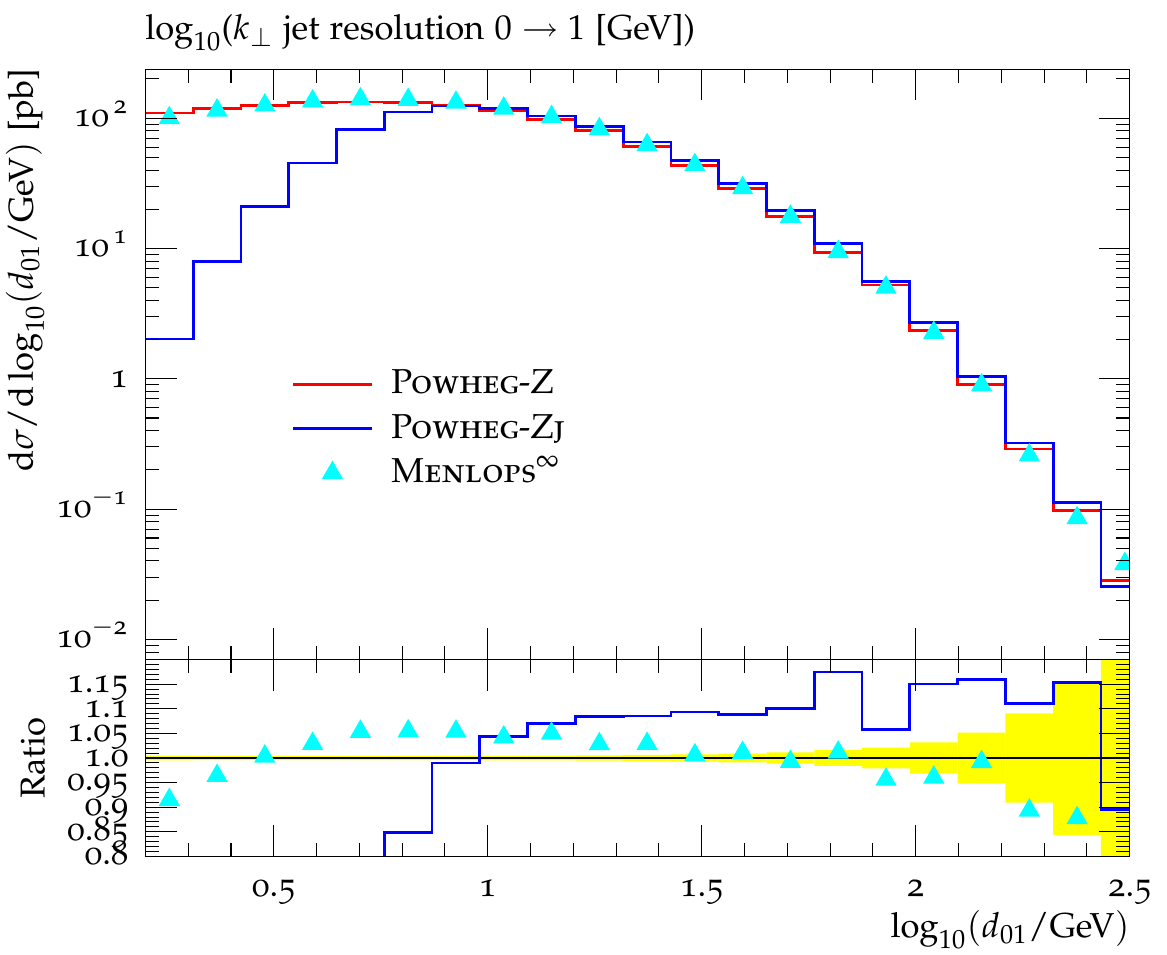}\hfill{}\includegraphics[clip,width=0.4\textwidth]{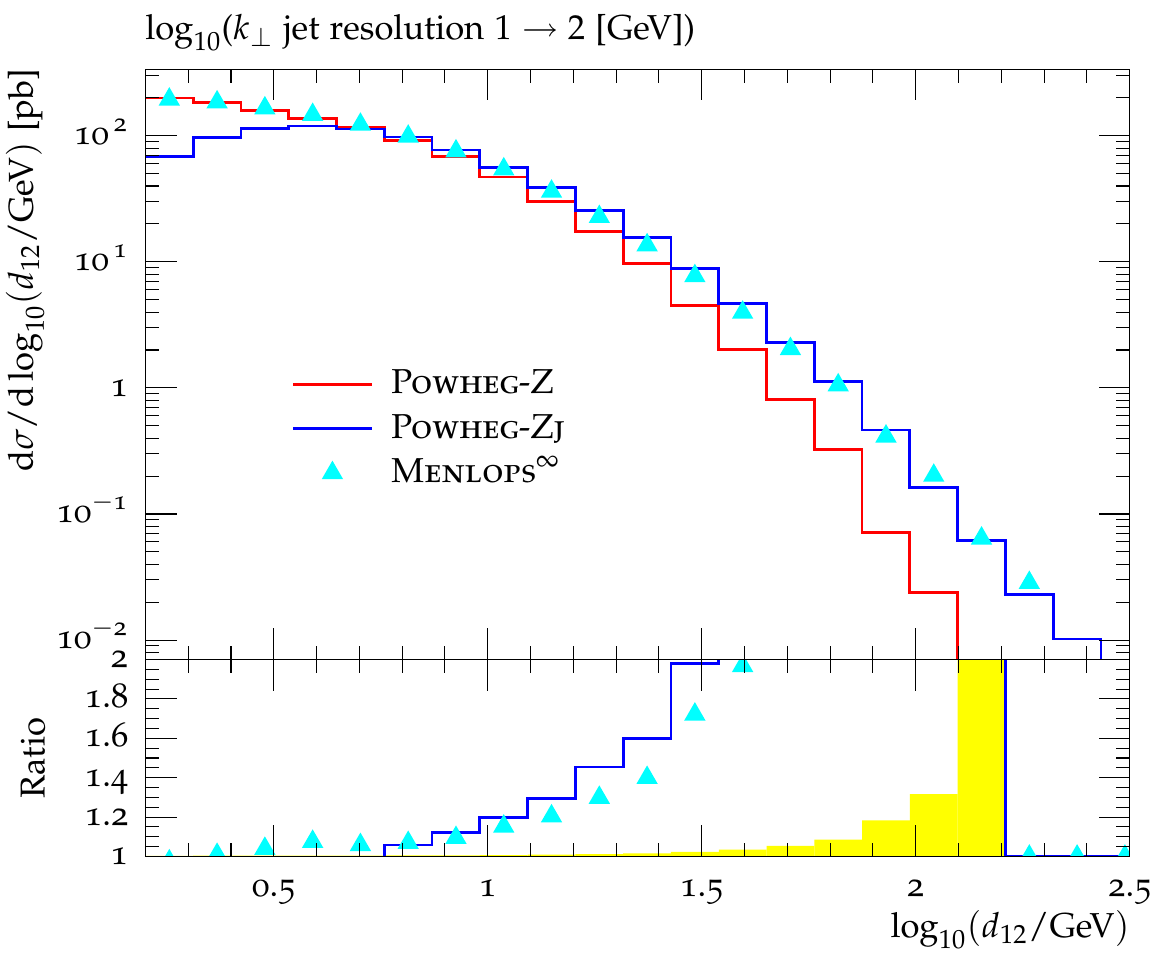}
\par\end{centering}

\begin{centering}
\includegraphics[clip,width=0.4\textwidth]{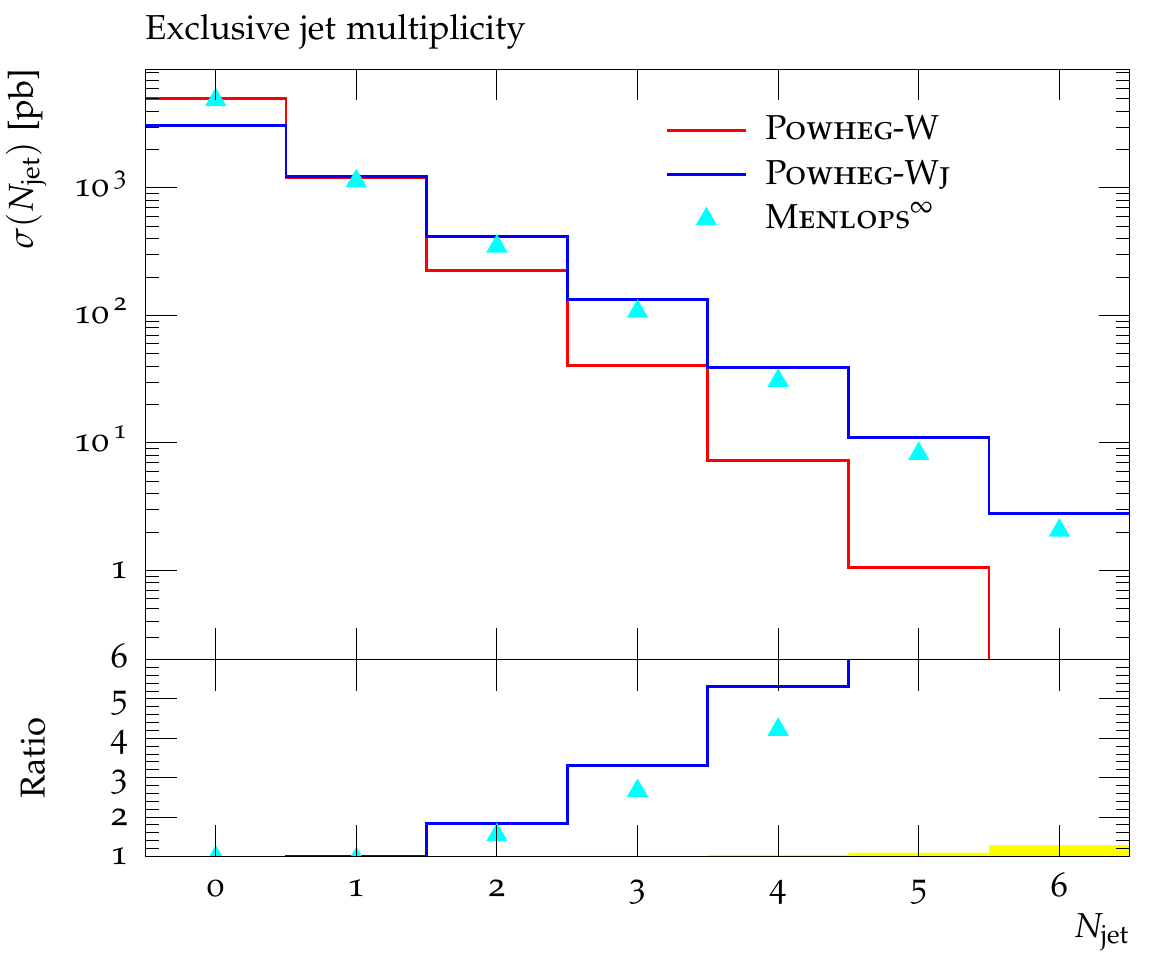}\hfill{}\includegraphics[clip,width=0.4\textwidth]{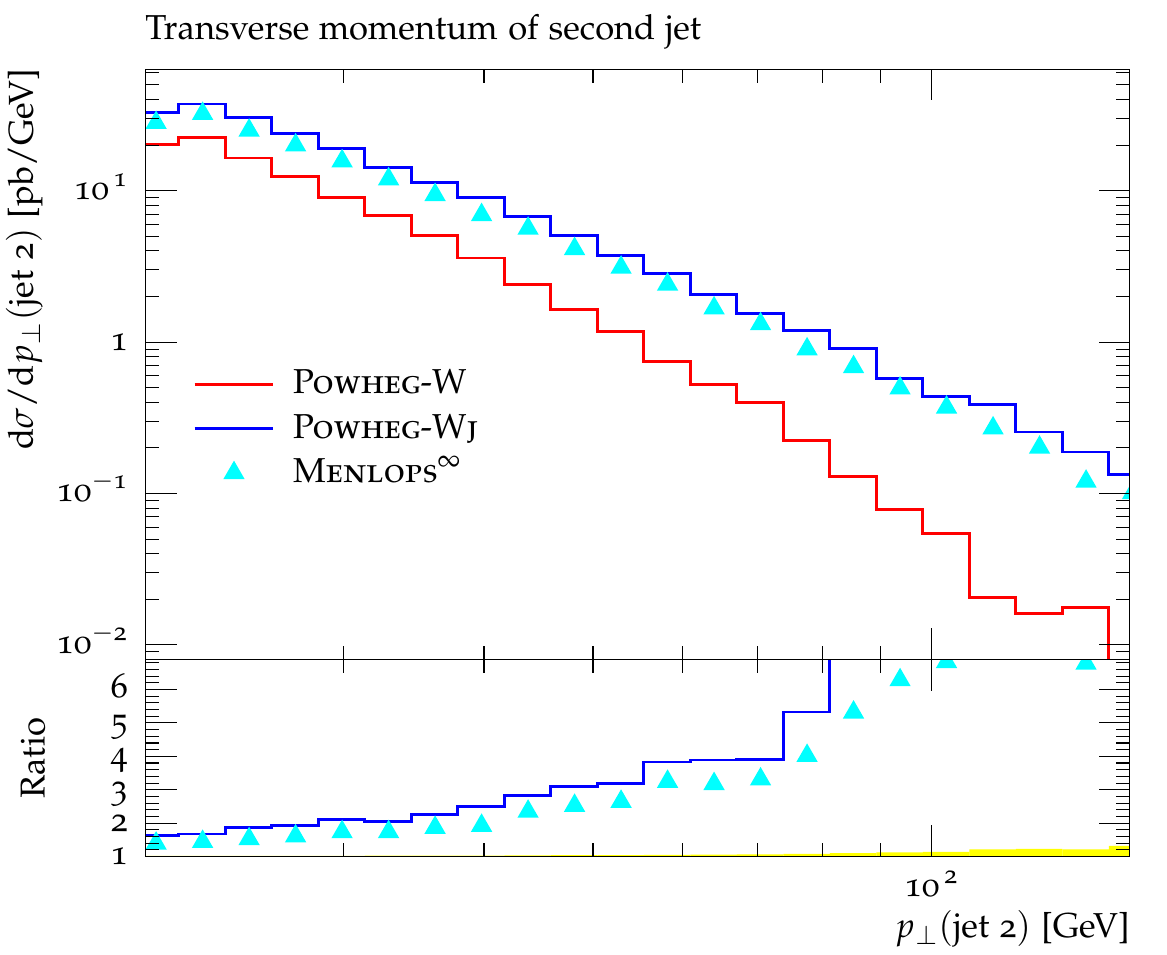}
\par\end{centering}

\vspace{5mm}

\caption{In this figure we show predictions for observables sensitive to the
emission of at least one hard jet in Z production at the Tevatron
and W production at the LHC. The colouring is as in the preceding
plots in this section, up to the addition of the blue lines, which
correspond to predictions made with the vector boson plus jet \noun{Powheg
}simulations.}

\label{fig:menlops_infty_validation_ge_2j} 
\end{figure}

In figure~\ref{fig:menlops_infty_validation_ge_2j} we shift focus
to quantities directly sensitive to the emission of two or more jets.
Since the \noun{Powheg }vector boson program only generates single
radiation events while the \noun{Menlops$^{\text{\ensuremath{\infty}}}$
}enhancement generates a further emission, in order to have a comparison
here we have evolved the events with \noun{Pythia} to include parton
shower effects (only). As with Fig.~\ref{fig:menlops_infty_validation_ge_1j}
the plots shown here lend themselves to a fairly intuitive understanding.
In particular, we see that since both \noun{Powheg-V} and\noun{ Menlops}
simulations include the first order real emission corrections to vector
boson production, the distributions of the 0-jet to 1-jet transition
rate (top left plot) given by the two approaches are in good agreement,
as are their predictions for the 0- and 1-jet cross sections (bottom
left plot). On the other hand, once the presence of a second hard
jet is probed, as in the 1-jet to 2-jet transition rate, in the upper-right
corner of the figure, and the next-to-leading jet $p_{{\scriptscriptstyle \mathrm{T}}}$
spectrum, in the lower-right corner, the softer spectrum of the \noun{Powheg-V
}simulation becomes apparent, with the \noun{Menlops }result tending
to that of the \noun{Powheg-Vj} programs at high $p_{{\scriptscriptstyle \mathrm{T}}}$.
The deficit of hard radiation from the \noun{Powheg-Z }and \noun{Powheg-W
}simulations in these distributions is simply due to the fact that
they are reliant on the soft-collinear parton shower approximation
to generate the second jet, whereas the \noun{Menlops }simulation
defers this dependency to the next-to-next-to-leading jet. The \noun{Menlops}
generator distributes the second jet according to the vector boson
plus dijet tree-level matrix element, in much the same way as the
\noun{Powheg-Vj} programs. As we shall now go on to discuss, and as
is clear from these plots, while the \noun{Powheg-Vj }simulations
describe multi-jet events well, they lack the resummation of Sudakov-logs
required to describe the low-$p_{{\scriptscriptstyle \mathrm{T}}}$
region and hence inclusive observables.%
\footnote{One can also see the \emph{full} set of NLO corrections to vector
boson plus jet production manifesting as a small enhancement over
the \noun{Menlops }prediction.%
} 

Note that the \noun{Menlops }simulation arrived at here is \emph{ideal
}in the sense that no unphysical scales have been introduced in its
formulation, in keeping with the theoretical proposal of Ref.~\cite{Hamilton:2010wh},
improving on the approximate implementations carried out there and
in Ref.~\cite{Hoche:2010kg}. 

Finally, before leaving this section, we wish to comment on the assignment
of the colour structure and interfacing to the parton shower. While
the flavour structure and $\bm{\Phi}_{{\scriptscriptstyle Vj}}$ are
read in from the Les Houches event file, the colour flow is assigned
to the event in the same way as in the unmodified \noun{Powheg-Vj}
program. This is a two stage process. Firstly, the large-$N_{{\scriptscriptstyle \mathrm{C}}}$
planar cross sections associated to the $hh\rightarrow Vj$ state
are evaluated at $\bm{\Phi}_{{\scriptscriptstyle Vj}}$ and a flow
assigned with a probability proportional to its cross section. Secondly,
the colour structure for the subsequently emitted radiation, parametrized
by $\bm{\Phi}_{j_{2}}$ and $\alpha$, is added as if it were a $1\rightarrow2$
parton branching from the leg of the $hh\rightarrow Vj$ state associated
to $\alpha$. This is also the method of colour assignment used in
Sect.~\ref{sub:vj_menlops}, the only technical difference there
being that the flavour structure and $\bm{\Phi}_{{\scriptscriptstyle Vj}}$
kinematics are generated by the program itself, as opposed to being
read in from a Les Houches file. 

Note that a more careful treatment of the colour structure would include
truncated showering effects, the absence of which has been suggested
as a potential cause of spurious high-mass clusters in hadronization
\cite{Mrenna:2003if}. While we have not investigated such effects
here, we have checked that at the hadron level we do not observe any
of the distortions in differential jet rates seen in Ref.~\cite{Mrenna:2003if}.
We suggest that this may be due to the fact that, as well as being
a very much suppressed effect, the matrix element-parton shower interface
occurs at a fixed scale in \noun{Meps} merging schemes like that studied
in Ref.~\cite{Mrenna:2003if}, whereas in \noun{Powheg }the analogous
scale is variable, being set by the transverse momentum of the emitted
radiation. We stress that the issue of truncated showers does not
arise through any of the \noun{Menlops} themes in this article but
stems from the \noun{Powheg }method in general, moreover, all studies
to date in this context suggest that its effects are negligible \cite{Hamilton:2008pd,Hamilton:2009za,Hamilton:2010mb}.
In any case we note that such effects could be included for the processes
studied here by modest extensions of the methods in Refs.~\cite{Hamilton:2008pd,Hamilton:2009ne}.

\subsection{\noun{Menlops} improved simulation of vector boson plus jet production\label{sub:vj_menlops}}

In this section we aim to promote the \noun{Powheg }simulation of
vector boson plus jet production to a \noun{Menlops }one. Whereas,
up to now in the literature, the \noun{Menlops }acronym has stood
for \noun{Nlops }simulations enhanced by higher multiplicity tree
level matrix elements, now we extend it to cover also the inclusion
of lower multiplicity ones; in this case by resummation and soft-collinear
factorisation. By way of distinguishing the\noun{ }simulation here
from that of Sect.~\ref{sub:v_menlops} we append the suffix $0$,
referring to it as \noun{Menlops$^{\text{0}}$} with a double-hard
emission cross section denoted $d\sigma_{0}$. The naming derives
from the fact that the distributions resulting from this simulation
correspond to the limit where the merging scale to be introduced in
Sect.~\ref{sub:v-vj_menlops} is set to zero.

The cross section for vector boson production develops large logarithms,
$\sim\log\, Q^{2}/p_{{\scriptscriptstyle \mathrm{T}}}^{2}$, at all
orders in perturbation theory, as the boson's transverse momentum
becomes small, $Q$ being the off-shell boson mass. In this region
fixed order perturbation theory then becomes unstable, with leading
order and next-to-leading order predictions for the vector boson $p_{{\scriptscriptstyle \mathrm{T}}}$
spectrum diverging to plus and minus infinity respectively, as $p_{{\scriptscriptstyle \mathrm{T}}}\rightarrow0$.
While the \noun{Powheg-Vj} programs exhibit precisely the latter unphysical
behavior, the same large logarithms are summed to all orders in $\alpha_{{\scriptscriptstyle \mathrm{S}}}$
in the \noun{Powheg-V }programs, leading to a stable and physical
description of the Sudakov region and, hence, also inclusive observables.

Thus, here we begin by clarifying the differences between the \noun{Powheg-V}
and \noun{Powheg-Vj }programs in the low transverse momentum region,
in order to better understand when the NLO computation of vector boson
plus jet production can be expected to begin to break down and with
a view to `mapping' the behavior of the former onto that of the latter
when this occurs. We point out that while the role of the renormalization
and factorization scales may appear prosaic, it is central to the
discussion here, hence, it is often referred to in the following. 

In Ref.~\cite{Nason:2006hfa} the authors established the equivalence
of the resummed vector boson transverse momentum spectrum of Ellis
and Veseli \cite{Ellis:1997ii} (based on the DDT formalism \cite{Dokshitzer:1978hw})
and the \noun{Powheg }hardest emission cross section Eq.~\ref{eq:V_hardest_emission_xsec}.
Specifically, it was shown how the \noun{Powheg }Sudakov form factor
for divector boson production relates to the next-to-leading log transverse
momentum space Sudakov form factor in Ref.~\cite{Ellis:1997ii}.
The same proof trivially holds here for vector boson production since
the final-state is still colorless. In particular, neglecting terms
beyond NLL accuracy it was proven that \begin{equation}
\Delta\left(\bm{\Phi}_{{\scriptscriptstyle V}},p_{{\scriptscriptstyle \mathrm{T}}}\right)\simeq\frac{f_{q/h_{\oplus}}\left(\bar{x}_{\oplus},p_{{\scriptscriptstyle \mathrm{T}}}\right)}{f_{q/h_{\oplus}}\left(\bar{x}_{\oplus},Q\right)}\frac{f_{\bar{q}/h_{\ominus}}\left(\bar{x}_{\ominus},p_{{\scriptscriptstyle \mathrm{T}}}\right)}{f_{\bar{q}/h_{\ominus}}\left(\bar{x}_{\ominus},Q\right)}\,\exp\left[\mathcal{T}_{{\scriptscriptstyle \mathrm{NLL}}}\left(p_{{\scriptscriptstyle \mathrm{T}}},Q\right)\right],\label{eq:Nason_Ridolfi_Sudakov}\end{equation}
where $\bar{x}_{\splusminus}$ and $Q$ are evaluated in terms of
the underlying Born kinematics, $\bm{\Phi}_{{\scriptscriptstyle V}}$,
and $\mathcal{T}_{{\scriptscriptstyle \mathrm{NLL}}}$ is the exponent
of the next-to-leading log DDT Sudakov form factor

\begin{equation}
\mathcal{T}_{{\scriptscriptstyle \mathrm{NLL}}}\left(p_{{\scriptscriptstyle \mathrm{T}}},Q\right)=-\int_{p_{{\scriptscriptstyle \mathrm{T}}}^{2}}^{Q^{2}}\frac{dk_{{\scriptscriptstyle \mathrm{T}}}^{2}}{k_{{\scriptscriptstyle \mathrm{T}}}^{2}}\,\frac{A\left(\alpha_{{\scriptscriptstyle \mathrm{S}}}\left(k_{{\scriptscriptstyle \mathrm{T}}}\right)\right)}{2\pi}\,2C_{{\scriptscriptstyle \mathrm{F}}}\left(\ln\frac{Q^{2}}{k_{{\scriptscriptstyle \mathrm{T}}}^{2}}-\frac{3}{2}\right)\,,\label{eq:T_DDT_Ellis_Veseli}\end{equation}
with $C_{{\scriptscriptstyle F}}$ and $C_{{\scriptscriptstyle A}}$
the usual Casimir factors. The function $A\left(\alpha_{{\scriptscriptstyle \mathrm{S}}}\right)$
is defined as \begin{align}
A\left(\alpha_{{\scriptscriptstyle \mathrm{S}}}\right) & =\alpha_{{\scriptscriptstyle \mathrm{S}}}\left(1+\frac{\alpha_{{\scriptscriptstyle \mathrm{S}}}}{2\pi}\, K\right)\,, & K & =\left(\frac{67}{18}-\frac{\pi^{2}}{6}\right)C_{{\scriptscriptstyle A}}-\frac{10}{9}\, n_{f}T_{{\scriptscriptstyle R}}.\label{eq:A_and_K}\end{align}

Substituting the expression for the Sudakov form factor in Eq.~\ref{eq:Nason_Ridolfi_Sudakov}
into the \noun{Powheg-V }hardest emission cross section, Eq.~\ref{eq:V_hardest_emission_xsec},
and noting that the factorization scale there in $\bar{B}$ is $Q$,
while in $R/B$ it is $p_{{\scriptscriptstyle \mathrm{T}}}$, all
PDF factors in the combination $\bar{B}/B\times\Delta$ cancel. Hence,
to NLL accuracy one has 

\begin{equation}
d\sigma_{{\scriptscriptstyle V}}=R\left(\bm{\Phi}_{{\scriptscriptstyle Vj}}\right)\,\exp\left[\mathcal{T}_{{\scriptscriptstyle \mathrm{NLL}}}\left(p_{{\scriptscriptstyle \mathrm{T}}},Q\right)\right]\, d\bm{\Phi}_{{\scriptscriptstyle V}}\, d\bm{\Phi}_{{\scriptscriptstyle j_{1}}}\,,\label{eq:DDT_xsec_in_POWHEG_V_language}\end{equation}
where we-reiterate that the renormalization and factorization scales
in $R$ here are set to the $p_{{\scriptscriptstyle \mathrm{T}}}$
of the W / Z boson.%
\footnote{For brevity we have omitted the unresolved emission term $\Delta\left(\bm{\Phi}_{{\scriptscriptstyle V}},\, p_{_{\mathrm{T}}}^{\mathrm{min}}\right)$
from Eq.~\ref{eq:V_hardest_emission_xsec} here, the effects of which
are negligible, giving rise to a fraction $<1\,\%$ of the events,
in the non-perturbative region $p_{{\scriptscriptstyle \mathrm{T}}}<p_{{\scriptscriptstyle \mathrm{T}}}^{\mathrm{min}}$.%
} We also remark that in order to achieve NLL accuracy in the \noun{Powheg}
Sudakov form factors $\alpha_{{\scriptscriptstyle \mathrm{S}}}$ is
replaced by $A\left(\alpha_{{\scriptscriptstyle \mathrm{S}}}\right)$
in the real cross section therein and, hence, in Eq.~\ref{eq:DDT_xsec_in_POWHEG_V_language}
\cite{Nason:2006hfa,Frixione:2007vw}. Lastly, we note that in writing
Eq.~\ref{eq:DDT_xsec_in_POWHEG_V_language} we have taken the ratio
$\bar{B}/B$ as being equal to that of the associated PDFs, which
is a legitimate replacement at this level of approximation \cite{Nason:2006hfa}. 

Using unitarity to integrate out the $\bm{\Phi}_{j_{2}}$ phase space,
the analogous expression to Eq.~\ref{eq:DDT_xsec_in_POWHEG_V_language}
in the vector boson plus jet \noun{Nlops}\emph{ }is simply \begin{equation}
d\sigma_{{\scriptscriptstyle Vj}}=\bar{B}\left(\bm{\Phi}_{{\scriptscriptstyle Vj}}\right)\, d\bm{\Phi}_{{\scriptscriptstyle V}}\, d\bm{\Phi}_{{\scriptscriptstyle j_{1}}}\,.\label{eq:DDT_xsec_in_POWHEG_VJ_language}\end{equation}
While $\bar{B}(\bm{\Phi}_{{\scriptscriptstyle Vj}})$ may be just
the fixed order, NLO, generalization of $R(\bm{\Phi}_{{\scriptscriptstyle Vj}})$,
since the renormalization and factorization scales in it are also
set to the $p_{{\scriptscriptstyle \mathrm{T}}}$ of the vector boson,
it bears a much closer resemblance to the DDT / Ellis-Veseli resummed
cross section (and hence Eq.~\ref{eq:DDT_xsec_in_POWHEG_V_language})
than it would if, say, the vector boson mass was used to set these
scales. Thus, some significant portion of the DDT resummation of large
logarithms, associated to PDF and coupling constant evolution, is
present in the \noun{Powheg-Vj }programs (Eq.~\ref{eq:DDT_xsec_in_POWHEG_VJ_language})
by virtue of this scale choice. Moreover, given the correspondence
in the scale choices, the NLO corrections in \ref{eq:DDT_xsec_in_POWHEG_VJ_language}
must also reproduce, at least at the leading log level, the first
order expansion of $\exp[\mathcal{T}_{{\scriptscriptstyle \mathrm{NLL}}}]$
in the resummed cross section in Eq.~\ref{eq:DDT_xsec_in_POWHEG_V_language}.

Hence, provided that the higher order terms in the DDT Sudakov form
factor (\emph{not} the \noun{Powheg} Sudakov form factor) are genuinely
$\mathcal{O}(\alpha_{{\scriptscriptstyle \mathrm{S}}}^{2})$, \emph{i.e.
}provided $\exp[\mathcal{T}_{{\scriptscriptstyle \mathrm{NLL}}}]\gtrsim1-\alpha_{{\scriptscriptstyle \mathrm{S}}}$,
the fixed order cross section in Eq.~\ref{eq:DDT_xsec_in_POWHEG_VJ_language}
will not be invalidated by the effects of large logs at higher orders.
This inequality translates directly, albeit roughly, into a lower
bound on the underlying Born transverse momentum: $\tilde{p}_{{\scriptscriptstyle \mathrm{T}}}\gtrsim14\,\mathrm{GeV}$.
The plots of the vector boson $p_{{\scriptscriptstyle \mathrm{T}}}$
spectra following this discussion are consistent with this estimate.
Let us make clear that this bound is lower than that which would be
obtained were different renormalization / factorization scale choices
made in Eq.~\ref{eq:DDT_xsec_in_POWHEG_VJ_language} (see \emph{e.g.
}Fig.~1 of Ref\emph{.}~\cite{Bozzi:2010xn}) \emph{i.e.} logarithmically
enhanced corrections are already in effect to some extent above this
threshold but they are offset by the resummation implicit in the scale
choices.

Explicitly, we have taken $\tilde{p}_{{\scriptscriptstyle \mathrm{T}}}=14\,\mathrm{GeV}$
in Z production and $\tilde{p}_{{\scriptscriptstyle \mathrm{T}}}=12.5\,\mathrm{GeV}$
in W production \emph{throughout}. While the values inferred from
the aforementioned inequality should give a conservative estimate
for $\tilde{p}_{{\scriptscriptstyle \mathrm{T}}}$, to put things
on a more solid footing, one should ideally check that $\tilde{p}_{{\scriptscriptstyle \mathrm{T}}}$
is chosen above, yet close to, the point where the fixed order and
resummed spectra just start to diverge. The values quoted here fulfill
the latter criterion (see Fig.~\ref{fig:menlops_0_validation_sudakov_peak_1});
in practice they should be adequate for all vector boson production
applications. 

Below $\tilde{p}_{{\scriptscriptstyle \mathrm{T}}}$ missing logarithmically
enhanced terms in the DDT Sudakov form factor begin to render the
NLO corrections to vector boson plus jet production invalid by themselves.
Just above $\tilde{p}_{{\scriptscriptstyle \mathrm{T}}}$ the combination
of the NLO corrections and renormalization and factorization scale
choices serve to well model the onset of large logarithmic Sudakov
effects --- we take the fact that the \noun{Powheg-V }and \noun{Powheg-Vj
}transverse momentum spectra in Fig.~\ref{fig:menlops_0_validation_sudakov_peak_1}
agree to within 5\% in this region to be good evidence of that. 

Having noted the various scale dependencies and the projected domain
of validity of the vector boson plus jet NLO computation, our \noun{Menlops}
strategy for enhancing the \noun{Powheg-Vj }simulations reduces to
cutting off the NLO contributions to $\bar{B}(\bm{\Phi}_{{\scriptscriptstyle Vj}})$
where they start to become compromised by the neglect of large logarithms
at higher orders, $p_{{\scriptscriptstyle \mathrm{T,1}}}\lesssim\tilde{p}_{{\scriptscriptstyle \mathrm{T}}}$,
and instead resumming those same Sudakov logs, in precisely the same
way as is done in the \noun{Powheg-V }programs. Thus, we write the
modified \noun{Menlops$^{\text{0}}$} cross section as \begin{equation}
d\sigma_{0}=\mathcal{P}(p_{{\scriptscriptstyle \mathrm{T},1}})\, d\sigma_{{\scriptscriptstyle Vj}}^{{\scriptscriptstyle \mathrm{NLL}}}+\left(1-\mathcal{P}(p_{{\scriptscriptstyle \mathrm{T},1}})\right)\, d\sigma_{{\scriptscriptstyle Vj}}\,;\label{eq:menlops_0_xsec}\end{equation}
where $\mathcal{P}(p_{{\scriptscriptstyle \mathrm{T},1}})$ is the
following switching function,%
\footnote{The width of the switching threshold is determined by $\epsilon$,
which we have taken to be 2.5 GeV throughout --- the fact that the
switching threshold has a finite width allows for small $\mathcal{O}\left(1\%\right)$
differences between $d\sigma_{{\scriptscriptstyle Vj}}^{{\scriptscriptstyle \mathrm{NLL}}}$
and $d\sigma_{{\scriptscriptstyle Vj}}$, in the vicinity of $p_{{\scriptscriptstyle \mathrm{T}}}^{{\scriptscriptstyle \mathrm{merge}}}$,
to be smoothed out.%
} \begin{eqnarray}
\mathcal{P}(p_{{\scriptscriptstyle \mathrm{T},1}}) & = & \left\{ \begin{array}{ll}
1 & p_{{\scriptscriptstyle \mathrm{T},1}}<\tilde{p}_{{\scriptscriptstyle \mathrm{T}}}-\epsilon\\
\frac{1}{2}\sin\left(\frac{\pi}{2\epsilon}\,(\tilde{p}_{{\scriptscriptstyle \mathrm{T}}}-p_{{\scriptscriptstyle \mathrm{T},1}})\right)+\frac{1}{2}\,\,\,\, & \tilde{p}_{{\scriptscriptstyle \mathrm{T}}}-\epsilon<p_{{\scriptscriptstyle \mathrm{T},1}}<\tilde{p}_{{\scriptscriptstyle \mathrm{T}}}+\epsilon\\
0 & p_{{\scriptscriptstyle \mathrm{T},1}}>\tilde{p}_{{\scriptscriptstyle \mathrm{T}}}+\epsilon\,;\end{array}\right.\label{eq:switching_function}\end{eqnarray}
$d\sigma_{{\scriptscriptstyle Vj}}$ is the unmodified NLO hardest
emission cross section of the \noun{Powheg-Vj }programs (Eq.~\ref{eq:Vj_hardest_emission_xsec})
and $d\sigma_{{\scriptscriptstyle Vj}}^{{\scriptscriptstyle \mathrm{NLL}}}$
a reweighting of $d\sigma_{{\scriptscriptstyle Vj}}$, at leading
order ($\bar{B}(\bm{\Phi}_{{\scriptscriptstyle Vj}})\rightarrow R(\bm{\Phi}_{{\scriptscriptstyle Vj}})$),
intended to replicate the $\bm{\Phi}_{{\scriptscriptstyle Vj}}$ dependence
of the \noun{Powheg-V }programs:

\begin{eqnarray}
d\sigma_{{\scriptscriptstyle Vj}}^{{\scriptscriptstyle \mathrm{NLL}}} & = & \left.\mathcal{K}\, B\left(\bm{\Phi}_{{\scriptscriptstyle V}}\right)\right|_{\mu_{F}=m_{{\scriptscriptstyle V}}}\, d\bm{\Phi}_{{\scriptscriptstyle V}}\,\left[\,\frac{R\left(\bm{\Phi}_{{\scriptscriptstyle Vj}}\right)}{B\left(\bm{\Phi}_{{\scriptscriptstyle V}}\right)}\,\delta\left(k_{_{\mathrm{T}}}\left(\bm{\Phi}_{{\scriptscriptstyle Vj}}\right)-p_{{\scriptscriptstyle \mathrm{T},1}}\right)\,\Delta\left(\bm{\Phi}_{{\scriptscriptstyle V}},\, p_{{\scriptscriptstyle \mathrm{T},1}}\right)\, d\bm{\Phi}_{{\scriptscriptstyle j_{1}}}\, dp_{{\scriptscriptstyle \mathrm{T,1}}}\right.\nonumber \\
 &  & \phantom{BLANK}\times\left\{ \,\,\Delta\left(\bm{\Phi}_{{\scriptscriptstyle Vj}},\, p_{_{\mathrm{T}}}^{\mathrm{min}}\right)\phantom{+\frac{R\left(\bm{\Phi}_{{\scriptscriptstyle Vj}}\right)}{B\left(\bm{\Phi}_{{\scriptscriptstyle V}}\right)}\,\delta\left(k_{_{\mathrm{T}}}\left(\bm{\Phi}_{{\scriptscriptstyle Vj}}\right)-p_{{\scriptscriptstyle \mathrm{T}}}\right)\,\Delta\left(\bm{\Phi}_{{\scriptscriptstyle V}},\, p_{{\scriptscriptstyle \mathrm{T}_{1}}}\right)\, d\bm{\Phi}_{{\scriptscriptstyle j_{1}}}\, dp_{{\scriptscriptstyle \mathrm{T}},1}}\right.\label{eq:menlops_0_xsec_low}\\
 &  & \phantom{BLANK}\phantom{\times}\left.\left.\,+\,\Delta\left(\bm{\Phi}_{{\scriptscriptstyle V}j},\, p_{{\scriptscriptstyle \mathrm{T},2}}\right)\,\sum_{\alpha}\,\frac{R^{\alpha}\left(\bm{\Phi}_{{\scriptscriptstyle Vjj}}^{\alpha}\right)}{B\left(\bm{\Phi}_{{\scriptscriptstyle Vj}}\right)}\,\delta\left(k_{_{\mathrm{T}}}^{\alpha}\left(\bm{\Phi}_{{\scriptscriptstyle Vjj}}^{\alpha}\right)-p_{{\scriptscriptstyle \mathrm{T},2}}\right)\, d\bm{\Phi}_{{\scriptscriptstyle j_{2}}}\, dp_{{\scriptscriptstyle \mathrm{T},2}}\right\} \right]\,.\nonumber \end{eqnarray}
This being so, when evaluating the \emph{leading order} single emission
cross section, $R(\bm{\Phi}_{{\scriptscriptstyle Vj}})$, in Eq.~\ref{eq:menlops_0_xsec_low},
we replace $\alpha_{{\scriptscriptstyle \mathrm{S}}}\rightarrow A\left(\alpha_{{\scriptscriptstyle \mathrm{S}}}\right)$,
as when generating radiation in the \noun{Powheg-V }programs. Furthermore,
as in that case, $R(\bm{\Phi}_{{\scriptscriptstyle Vj}})$ is multiplied
by a ratio of PDFs, in the form \begin{equation}
\left.B\left(\bm{\Phi}_{{\scriptscriptstyle V}}\right)\right|_{\mu_{F}=m_{{\scriptscriptstyle V}}}/\left.B\left(\bm{\Phi}_{{\scriptscriptstyle V}}\right)\right|_{\mu_{F}=p_{{\scriptscriptstyle \mathrm{T},1}}}\,.\label{eq:B_over_B_PDF_ratio}\end{equation}
Note that the latter ratio comprises a tower of large logarithms,
the same large logarithmic content as the ratio $\bar{B}/B$ in the
conventional \noun{Powheg-W / -Z} cross section formulae. While we
do not claim that the description of fully inclusive observables is
better than leading order here, given the guiding principle amounts
to trying to match, as far as practically possible, the $\bm{\Phi}_{{\scriptscriptstyle Vj}}$
dependence of $d\sigma_{{\scriptscriptstyle V}}$, we can at least
correct the overall normalization using a \emph{K}-factor, $\mathcal{K}=\bar{\left\langle B\,\right\rangle }/\left\langle B\right\rangle $,
for the vector boson production process (determined with $\mu_{R}=\mu_{F}=m_{{\scriptscriptstyle V}}$
throughout). Lastly, we include the \noun{Powheg }Sudakov form factor
$\Delta(\bm{\Phi}_{{\scriptscriptstyle V}},\, p_{{\scriptscriptstyle \mathrm{T,1}}})$
leading to a double-hard emission cross section much like that of
the \noun{Menlops$^{\text{\ensuremath{\infty}}}$} simulation (Eq.~\ref{eq:menlops_inf_xsec}),
with the vector boson production hardest emission cross section recovered
up to the replacement $\bar{B}\rightarrow\mathcal{K}\, B$ on integrating
out $\bm{\Phi}_{j_{2}}$.

All of the modifications and reweightings above are carried out by
straightforward modifications to the $\bar{B}(\bm{\Phi}_{{\scriptscriptstyle Vj}})$
function, except for the addition of the Sudakov form factor. With
the Monte Carlo generation cut on $p_{{\scriptscriptstyle \mathrm{T},1}}$
set to that defining unresolvable radiation in the \noun{Powheg~Box
}programs ($p_{{\scriptscriptstyle \mathrm{T}}}^{{\scriptscriptstyle \mathrm{min}}}=0.8\,\mathrm{GeV}$),
the modified event generation process including Sudakov form factor
suppression proceeds as follows:
\begin{enumerate}
\item A vector boson plus jet configuration, $\bm{\Phi}_{{\scriptscriptstyle Vj}}$,
is generated according to Eq.~\ref{eq:menlops_0_xsec} but with the
term $\Delta\left(\bm{\Phi}_{{\scriptscriptstyle V}},\, p_{{\scriptscriptstyle \mathrm{T,1}}}\right)$
in $d\sigma_{{\scriptscriptstyle Vj}}^{{\scriptscriptstyle \mathrm{NLL}}}$
replaced\emph{ }by\emph{ }one\emph{.}
\item A random number $\mathcal{R}$ is generated and if $\mathcal{R}>\mathcal{P}(p_{{\scriptscriptstyle \mathrm{T},1}})$
the event generation continues as in the unmodified \noun{Powheg-Vj
}program at step 7.
\item The inverse mapping $\bm{\Phi}_{{\scriptscriptstyle Vj}}\rightarrow\{\bm{\Phi}_{{\scriptscriptstyle V}},\bm{\Phi}_{{\scriptscriptstyle j_{1}}}\}$
is performed exactly and unambiguously.
\item Kinematics for vector boson production are assembled from $\bm{\Phi}_{{\scriptscriptstyle V}}$.
\item The \noun{Powheg-V} radiation generating code is used to generate
a `ghost' emission, $\bm{\Phi}_{{\scriptscriptstyle j_{1}}}^{\prime}$,
with respect to $\bm{\Phi}_{{\scriptscriptstyle V}}$.
\item If $p_{{\scriptscriptstyle \mathrm{T}}}^{\prime}>p_{{\scriptscriptstyle \mathrm{T},1}}$
, where $p_{{\scriptscriptstyle \mathrm{T}}}^{\prime}=k_{_{\mathrm{T}}}\left(\bm{\Phi}_{{\scriptscriptstyle V}},\,\bm{\Phi}_{{\scriptscriptstyle j_{1}}}^{\prime}\right)$
and $p_{{\scriptscriptstyle \mathrm{T},1}}=k_{_{\mathrm{T}}}\left(\bm{\Phi}_{{\scriptscriptstyle V}},\,\bm{\Phi}_{{\scriptscriptstyle j_{1}}}\right)$,
the vector boson plus jet configuration, $\bm{\Phi}_{{\scriptscriptstyle Vj}}$,
is rejected and the event generation is restarted at step 1.%
\footnote{Given the interpretation of the Sudakov form factor as a no-emission
probability, events with kinematics $\bm{\Phi}_{{\scriptscriptstyle V}},$
$p_{{\scriptscriptstyle \mathrm{T},1}}$ then pass this veto with
probability $\Delta\left(\bm{\Phi}_{{\scriptscriptstyle V}},\, p_{{\scriptscriptstyle \mathrm{T},1}}\right)$,
thus generating the Sudakov suppression term that was set to one in
the `crude' distribution in step 1. %
}
\item Radiation is generated from the $\bm{\Phi}_{{\scriptscriptstyle Vj}}$
state exactly as in the \noun{Powheg-Vj }program, completing the \noun{Powheg}
event.
\end{enumerate}
Although generating the vector boson plus jet configurations at such
low values of $p_{{\scriptscriptstyle \mathrm{T}}}$ and then vetoing
them in large numbers may seem computationally expensive, the actual
efficiency associated with the rejection procedure above (with the
generation cut set to $0.8\,\mathrm{GeV}$) is not so costly, the
acceptance rate being a little under 20\%.

When all modifications are considered together, the full event generation
procedure actually proceeds much faster than normal. The switching
off of the NLO corrections, when the $p_{{\scriptscriptstyle \mathrm{T}}}$
of the underlying Born configuration dips below $\sim14\,\mathrm{GeV}$,
means one only needs to evaluate relatively simple matrix elements
there, moreover, the fact that the distribution is then manifestly
positive in this region, greatly reduces the number of negative weight
events and hence the number of \emph{folded-integrations} \cite{Nason:2007vt,Alioli:2010qp}
required to remove them. We have used a single folding of the real
emission phase space $\bm{\Phi}_{j_{1}}$ and $\bm{\Phi}_{j_{2}}$
(a so-called 2-1-1 folding in the notation of Ref.~\cite{Alioli:2010qp}),
simply out of a desire to test the programming apparatus under general
conditions, rather than any inclination to remove negative weights.
In the event samples created with the unmodified \noun{Powheg-Vj}
programs used in this study, the fraction of negative weight events
was found to be $\sim20\,\%$, while in all other \noun{Powheg-V}
and \noun{Menlops }samples it was below $\sim0.5\%$. Although the
computational speed here proved to be more than adequate for the purposes
of this study, it is clear that the efficiency of the Sudakov suppression
procedure can be improved without too much effort.

We remark that once the $\tilde{p}_{{\scriptscriptstyle \mathrm{T}}}$
threshold is crossed and the NLO corrections in $\bar{B}(\bm{\Phi}_{{\scriptscriptstyle Vj}})$
are turned off, the basic idea behind the procedure here is really
no different to the Sudakov suppression applied to matrix elements
in the CKKW(-L) and MLM tree-level merging methods. On a basic level,
the whole approach here is a direct application of those same principles.

Unlike the general tree-level merging techniques, the \noun{Powheg}
formalism, since it aims at\noun{ NLO }accuracy\noun{,} is defined
and realized with a high degree of technical precision, affording
more control over the implementation here. Note, in particular, that
the inverse mapping of the phase space, $\bm{\Phi}_{{\scriptscriptstyle Vj}}\rightarrow\{\bm{\Phi}_{{\scriptscriptstyle V}},\bm{\Phi}_{{\scriptscriptstyle j_{1}}}\}\,$,
is clearly and unambiguously defined; this follows from the fact that
the mapping covers the full single emission phase space, without \emph{dead
zones}, without the need for any kind of emitter-spectator assignment,
and because it is, in this case, uniquely attributable to initial-state
radiation. The clear specification of the formalism, combined with
this phase space inversion, underpins all of the modifications described
here, ultimately allowing us to reconstruct, up to the replacement
$\bar{B}\rightarrow\mathcal{K}\, B$, \emph{exactly} the same distribution
for $\bm{\Phi}_{{\scriptscriptstyle V}}$ and $\bm{\Phi}_{j_{1}}$
as in the \noun{Powheg-W }and \noun{Powheg-Z }programs. This exactness
of the\noun{ }implementation should reveal itself in a number of ways,
as in the case of Sect.~\ref{sub:v_menlops}. In particular, predictions
for inclusive W / Z production should closely trace their genuine
NLO \noun{Powheg-V} counterparts, the only source of differences being
$\bar{B}\rightarrow\mathcal{K}\, B$, also the large-log resummation
in regions sensitive to Sudakov effects should maintain its quality.

For completeness we note that in the \noun{Powheg-V }simulations some
events occur with non-radiative kinematics, $\bm{\Phi}_{{\scriptscriptstyle V}}$
only, being physically interpreted as ones in which the emitted radiation
is unresolvable and in a region dominated by non-perturbative dynamics.
These events are neglected in Eq.~\ref{eq:menlops_0_xsec_low}, comprising
just a fraction $\Delta(p_{{\scriptscriptstyle \mathrm{T}}}^{\mathrm{min}})<1\,\%$
of the total sample in the \noun{Powheg-V }programs --- a negligible
contribution given that the accuracy aimed at for fully inclusive
observables in this section is just leading order.

\begin{figure}[H]
\begin{centering}
\includegraphics[width=0.4\textwidth]{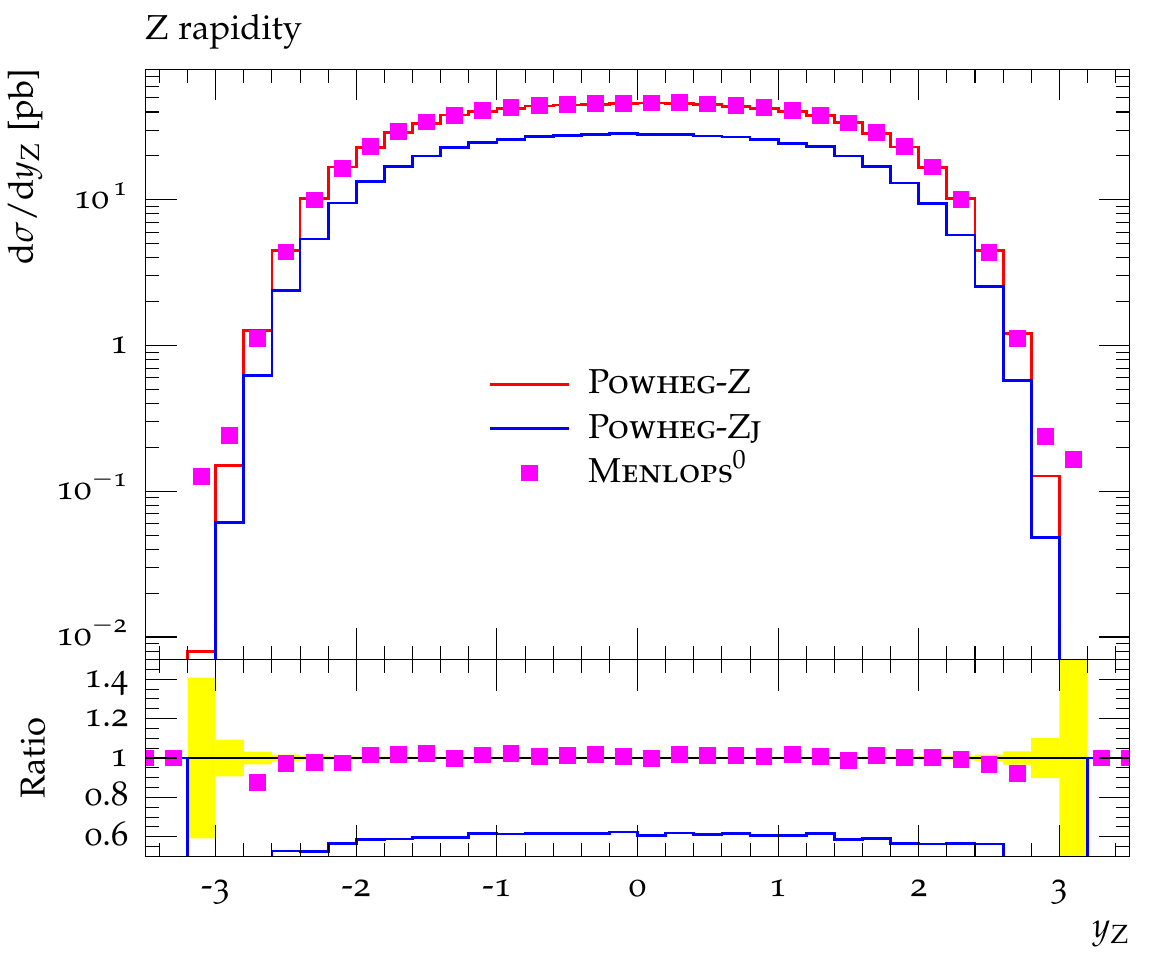}\hfill{}\includegraphics[width=0.4\textwidth]{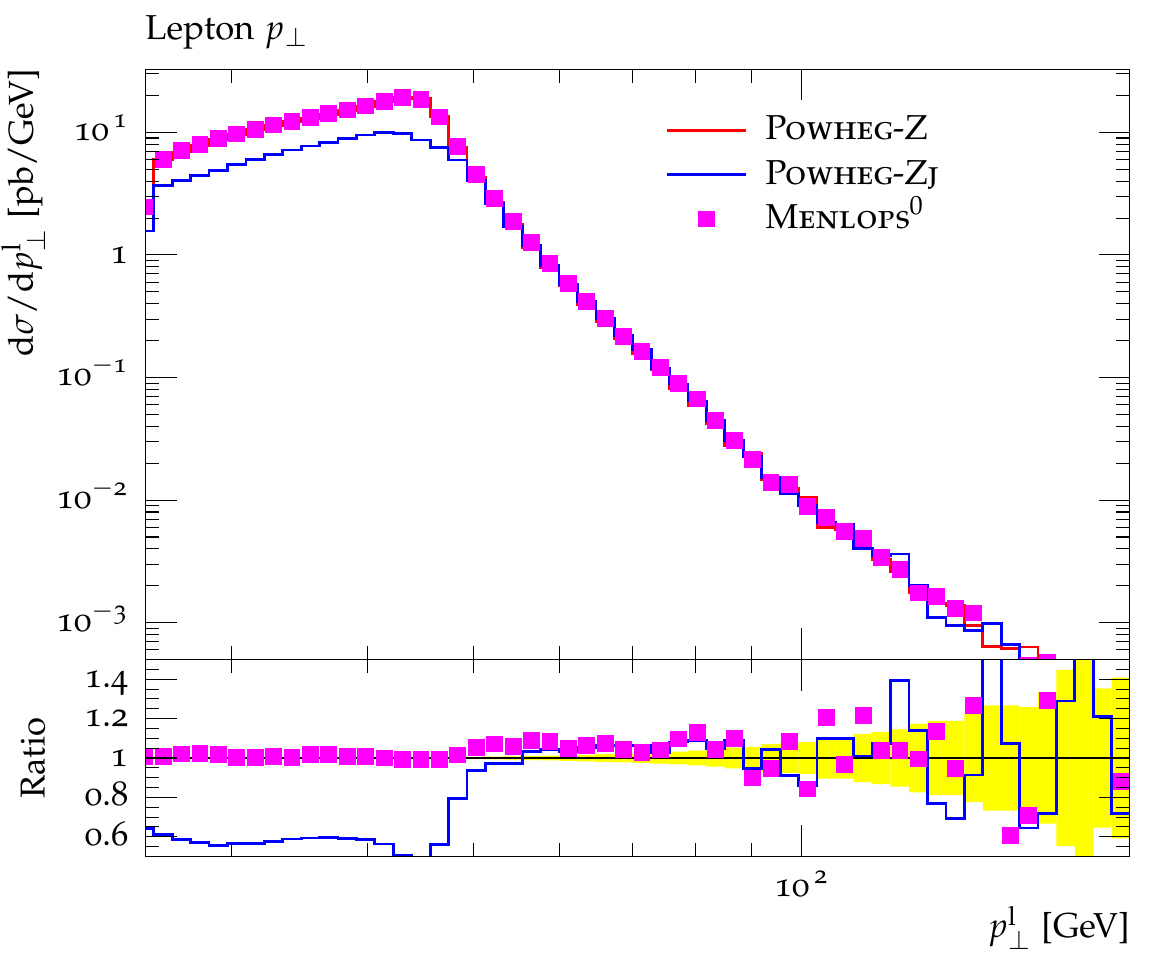}
\par\end{centering}

\begin{centering}
\includegraphics[width=0.4\textwidth]{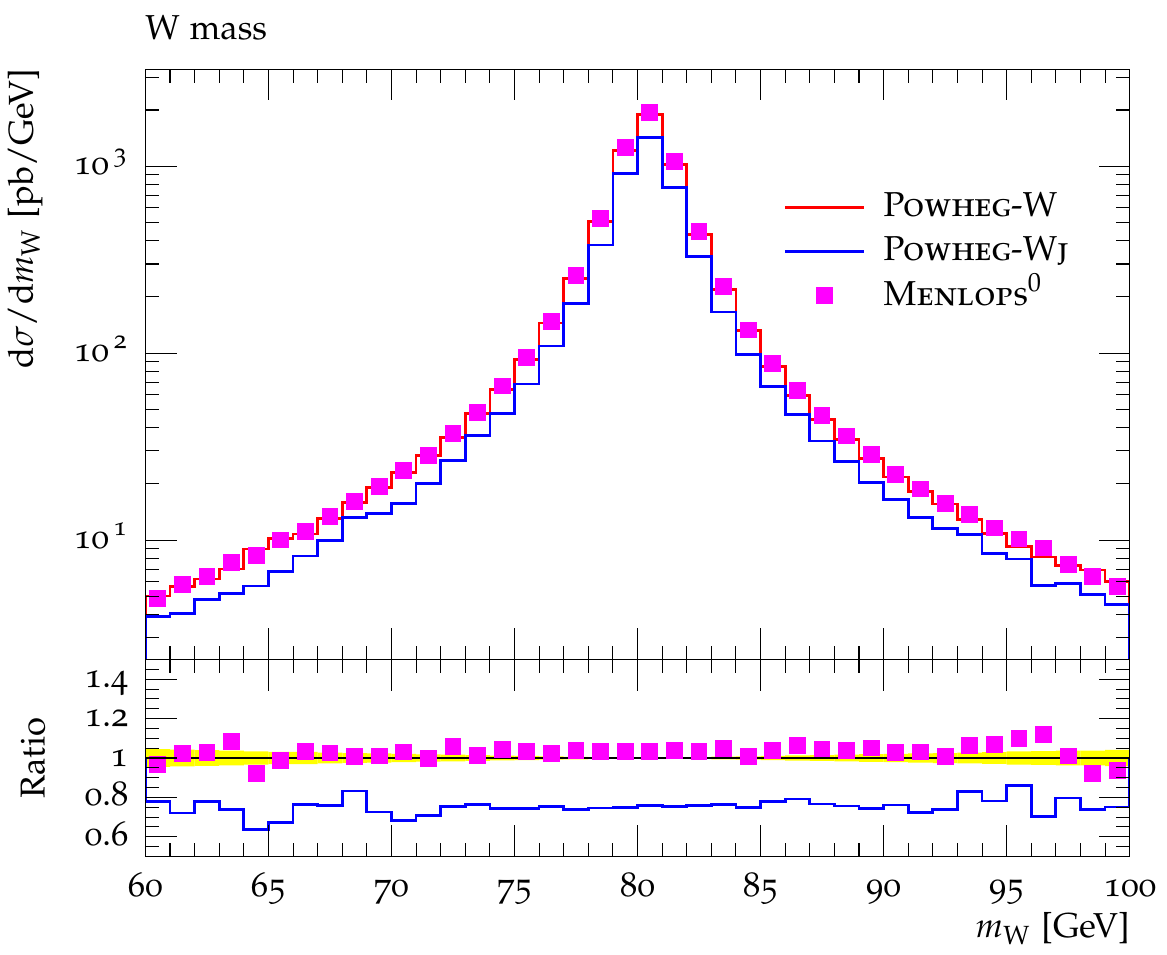}\hfill{}\includegraphics[width=0.4\textwidth]{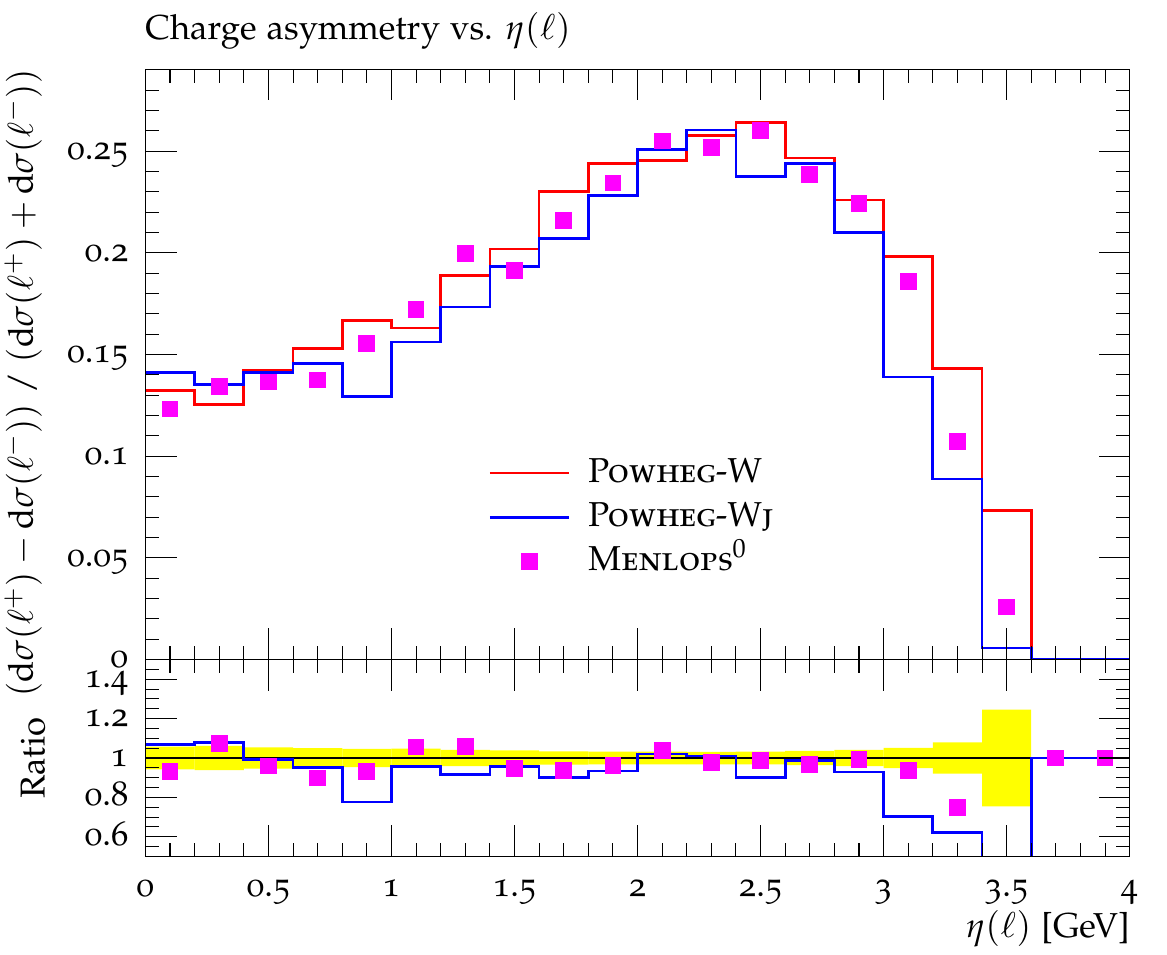}
\par\end{centering}

\vspace{5mm}

\caption{In this figure we compare predictions for inclusive observables obtained
using the \noun{Powheg }simulations of vector boson production (red)
and jet-associated vector boson production (blue), to those of the
\noun{Menlops$^{\text{0}}$} vector boson production simulation (magenta
squares). Plots in the upper half of the figure correspond to Z production
in $\sqrt{s}=1.96\,\mathrm{TeV}$ $p\bar{p}$ collisions, while those
in the lower half are for W production in $\sqrt{s}=7\,\mathrm{TeV}$
$pp$ reactions.}

\label{fig:menlops_0_validation_fully_inclusive} 
\end{figure}

\begin{figure}[H]
\begin{centering}
\includegraphics[width=0.4\textwidth]{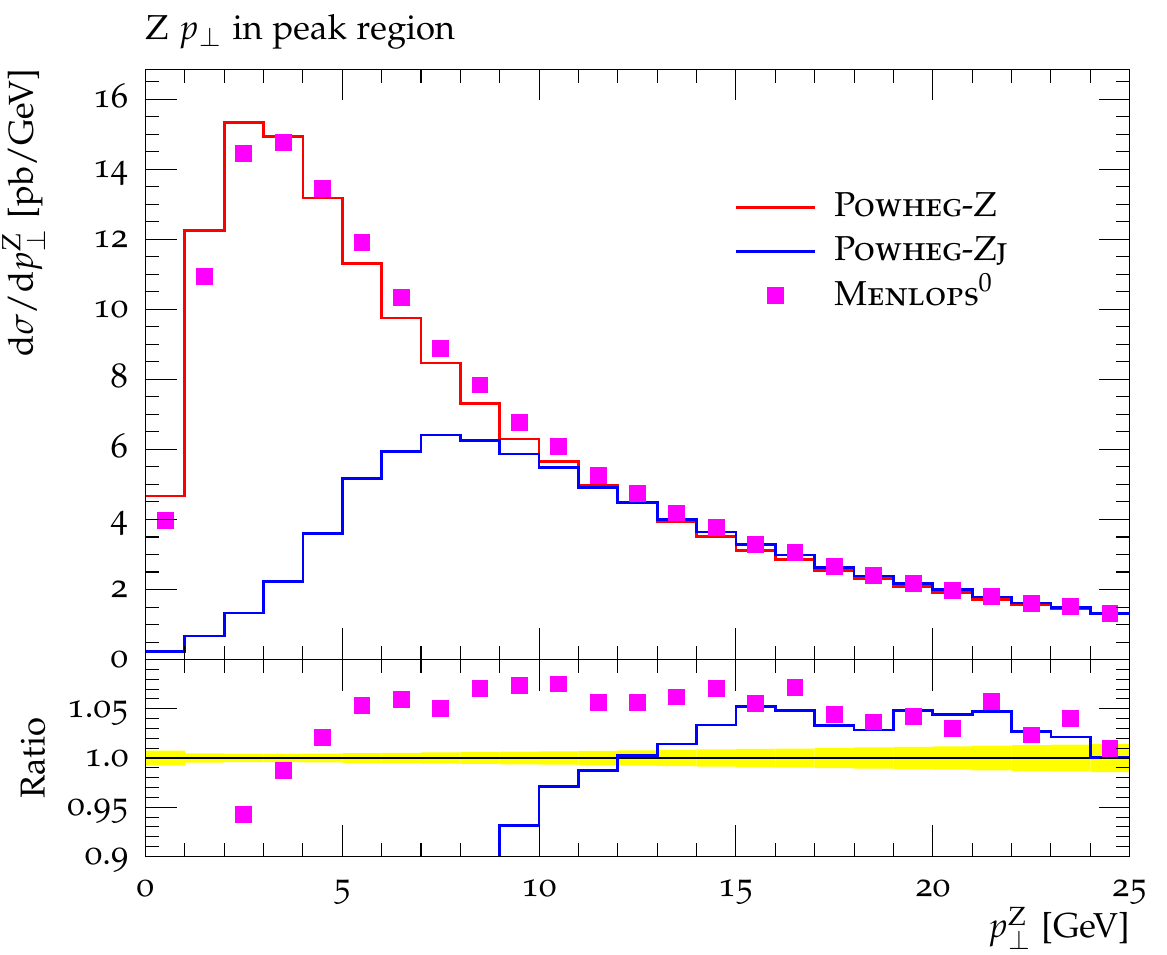}\hfill{}\includegraphics[width=0.4\textwidth]{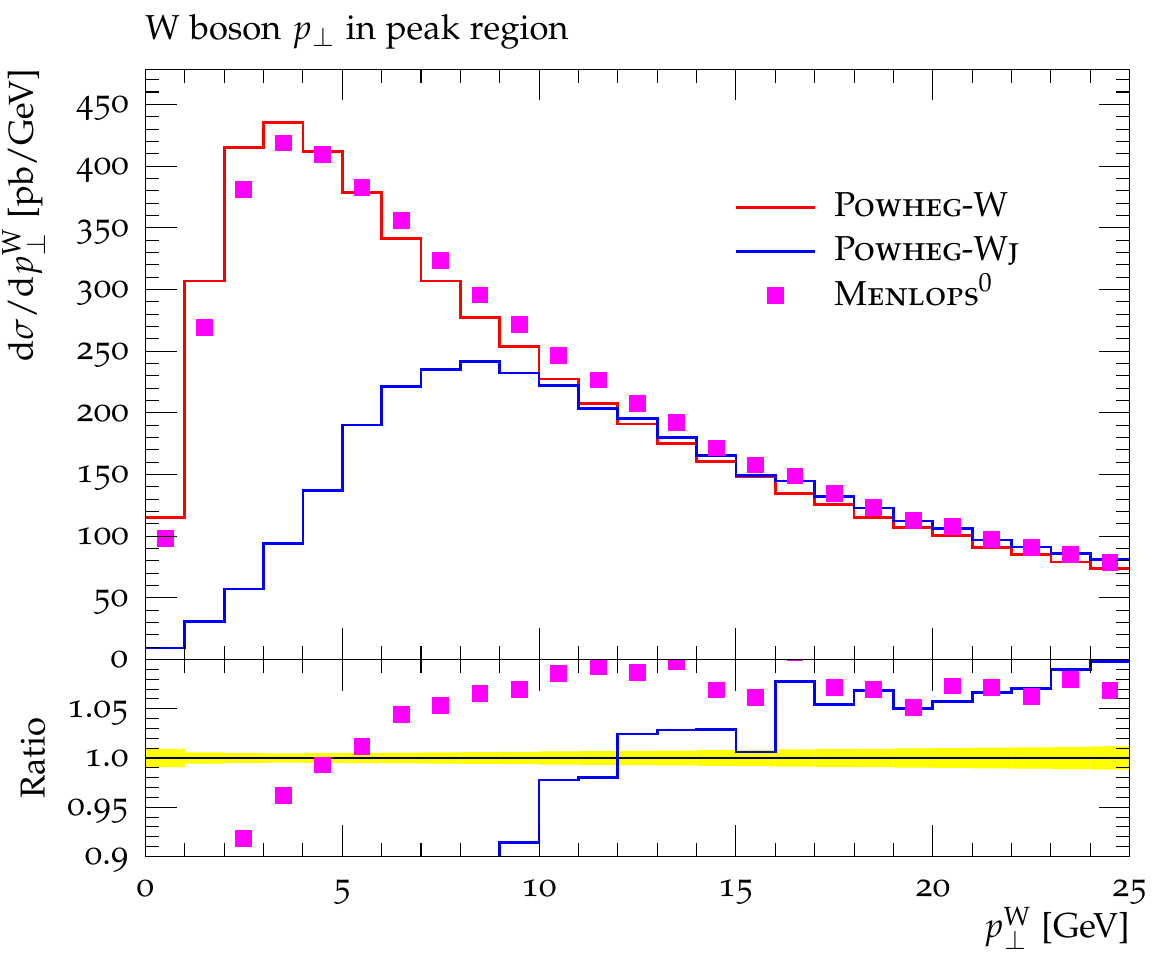}
\par\end{centering}

\caption{Here we show\noun{ Powheg }vector boson (red) vector boson plus jet
(blue) and \noun{Menlops$^{0}$} (magenta squares) predictions for
the Sudakov peak region of the vector boson transverse momentum spectrum,
in Z production at the Tevatron (left) and W production at the LHC
(right). The results shown here have been obtained by showering the
initial \emph{bare} events, stored in Les Houches event files, with
\noun{Pythia 8.150}; hadronization and multiple parton interactions
are not simulated. }

\label{fig:menlops_0_validation_sudakov_peak_1} 
\end{figure}

As in section~\ref{sub:v_menlops}, here we begin to demonstrate
that the \noun{Menlops$^{\text{0}}$ }method\noun{ }has been faithfully
implemented according to Eq.~\ref{eq:menlops_0_xsec_low} by first
considering fully inclusive observables in Fig.~ \ref{fig:menlops_0_validation_fully_inclusive}.
Immediately one can see, as expected, that the vector boson plus jet
production programs give markedly different predictions to the \noun{Powheg-V
}and \noun{Menlops$^{\text{0}}$ }simulations. This should not come
as any surprise, since the behavior of the NLO calculation underlying
the former becomes unphysical when the observables include contributions
from the Sudakov region. Recall that the vector boson plus jet programs
include an unphysical cut-off on the $p_{{\scriptscriptstyle \mathrm{T}}}$
of the underlying Born kinematics, to screen the associated collinear
singularity, without which event generation is not possible. We have
taken this Monte Carlo generation cut to be $5\,\mathrm{GeV}$ here.
Although we have not shown it explicitly, taking a different value
of the cut for this type of observable will produce significant variations
in the results here, in fact by making the cut sufficiently small
($\mathrm{2-3\, GeV}$) all of the predictions become negative \cite{Alioli:2010qp}.
On the contrary, since the large logarithms associated with the same
Sudakov region are resummed in Eq.~\ref{eq:menlops_0_xsec_low},
the \noun{Menlops$^{\text{0}}$ }formulation exhibits no such unphysical
behavior. In fact, in this case, the technical cut has been taken
down to that used in generating radiation in the vector boson production
programs, $0.9\,\mathrm{GeV}$, moreover, the dependence of the predictions
upon the cut is physical. Pleasingly, we see that the level of agreement
in the predictions of the \noun{Powheg} vector boson production and
the \noun{Menlops$^{\text{0}}$ }simulations is by-and-large very
good, with the two rarely deviating by more than 10\%; recall that
the overall normalization is to a large extent fixed to the NLO total
cross section by the constant factor, $\mathcal{K}$, in the low-$p_{{\scriptscriptstyle \mathrm{T}}}$
region.\emph{ }

Continuing as in section~\ref{sub:v_menlops}, we next look to check
the implementation of resummation in the Sudakov region, where the
transverse momentum of the vector boson becomes small in Eq.~\ref{eq:menlops_0_xsec_low}.
As noted earlier, the similarities of Eq.~\ref{eq:menlops_0_xsec_low}
with the double-hard emission cross section in Eq.~\ref{eq:menlops_inf_xsec}
of Sect.~\ref{eq:V_hardest_emission_xsec} are apparent, moreover,
the surrounding discussion on resummation there applies here too,
without modification. Here again, in figure \ref{fig:menlops_0_validation_sudakov_peak_1},
one can see good agreement between the \noun{Menlops$^{0}$} and \noun{Powheg-V
}predictions, with deviations being limited to about 10\%, which we
take as confirmation that the accuracy of the Sudakov resummation
in either case is the same. 

The \noun{Powheg-Vj }prediction is shown in blue in Fig.~\ref{fig:menlops_0_validation_sudakov_peak_1},
the fact that it fails to describe the low $p_{{\scriptscriptstyle \mathrm{T}}}$
end of spectrum is no surprise, as we have already cited it as a motivating
factor for the modifications proposed here. We have chosen to show
it since we believe it substantiates our earlier statements regarding
when the NLO calculation can be expected to diverge from the resummed
predictions, $\sim14\,\mathrm{GeV}$. Note that this value for when
one can expect the resummed and fixed order predictions to exhibit
deviations is noticeably lower than that which one might have inferred
from Fig.~1 of Ref.~\cite{Bozzi:2010xn}. Note, however, that in
the case of Ref.~\cite{Bozzi:2010xn} the renormalization and factorization
scales are set to the Z boson mass, whereas the \noun{Powheg-Vj }program
uses the $p_{{\scriptscriptstyle \mathrm{T}}}$ of the Z boson in
the underlying Born configuration and is thus in closer correspondence
with the DDT formalism. 

\begin{figure}[H]
\begin{centering}
\includegraphics[width=0.4\textwidth]{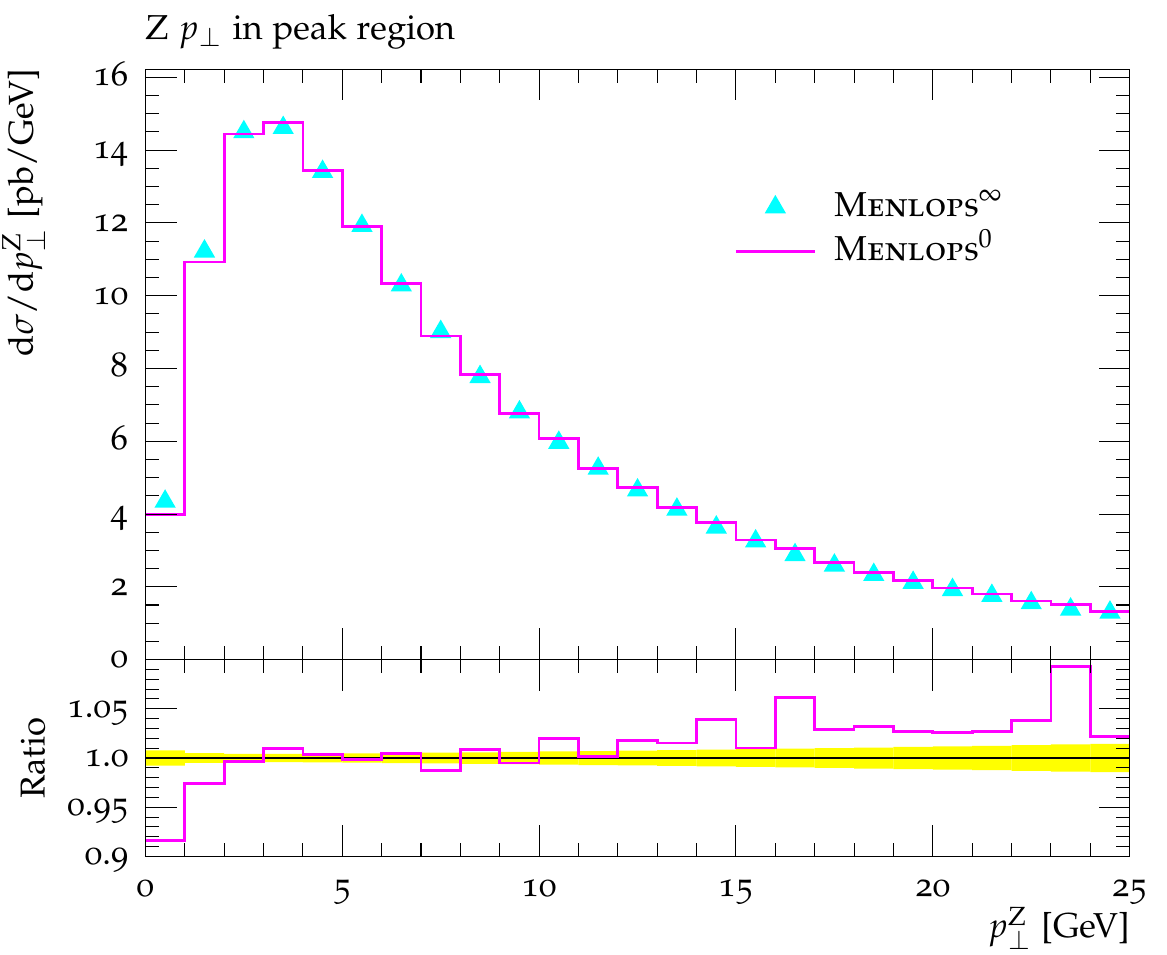}\hfill{}\includegraphics[width=0.4\textwidth]{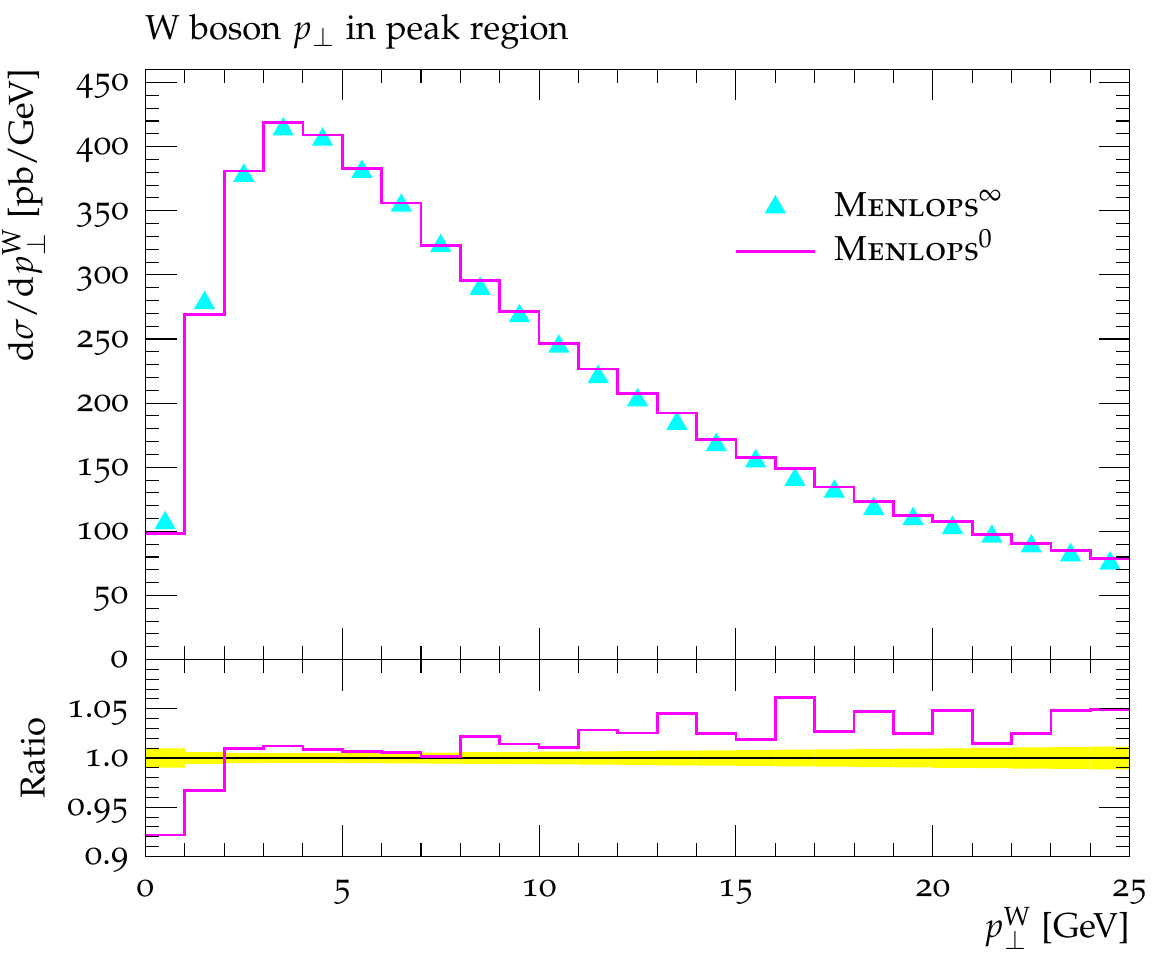}
\par\end{centering}

\caption{\noun{Menlops$^{0}$} (magenta) and \noun{Menlops$^{\infty}$} (cyan
triangles) predictions for the low transverse momentum region in Z
production at the Tevatron (left) and W production at the LHC (right).
The results shown here include the effects of multiple emissions through
parton showering; the simulation of hadronization and multiple parton
interactions has been omitted. These plots serve to prove that the
non-trivial implementation of Sudakov suppression effects has been
carried out properly in the \noun{Menlops$^{0}$} case.}

\label{fig:menlops_0_validation_sudakov_peak_2} 
\end{figure}

In order to prevent the distributions obscuring one another and due
to its importance in the context of the modifications being discussed
in this section, in figure \ref{fig:menlops_0_validation_sudakov_peak_2}
we show the same plots again but this time comparing the \noun{Menlops$^{\text{0}}$
}and \noun{Menlops$^{\text{\ensuremath{\infty}}}$ }output. If the
hardest emission cross section formulae have been implemented precisely
according to Eqs.~\ref{eq:menlops_inf_xsec} and \ref{eq:menlops_0_xsec_low}
the two sets of predictions should be essentially identical in the
peak region, as is seen to be the case. 

\begin{figure}[H]
\begin{centering}
\includegraphics[width=0.4\textwidth]{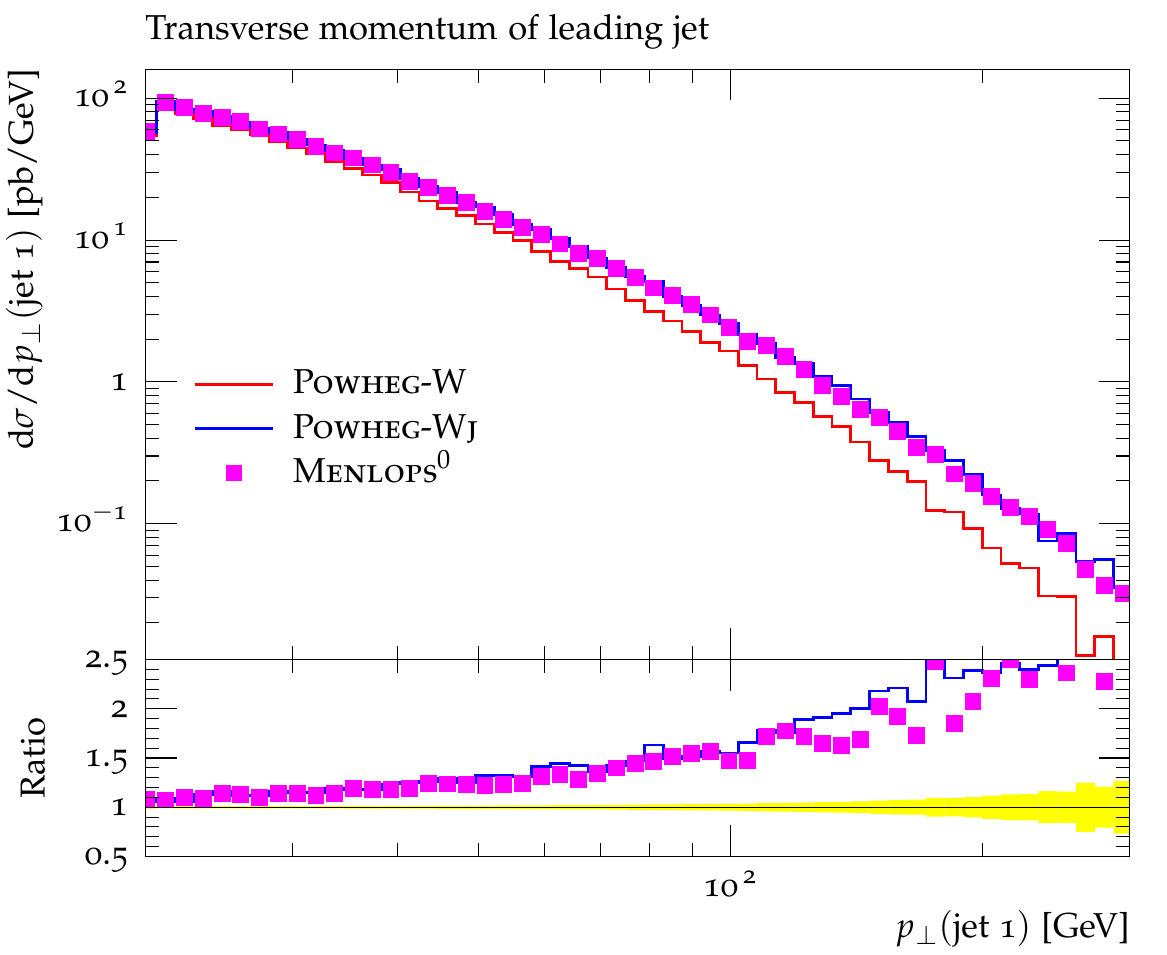}\hfill{}\includegraphics[width=0.4\textwidth]{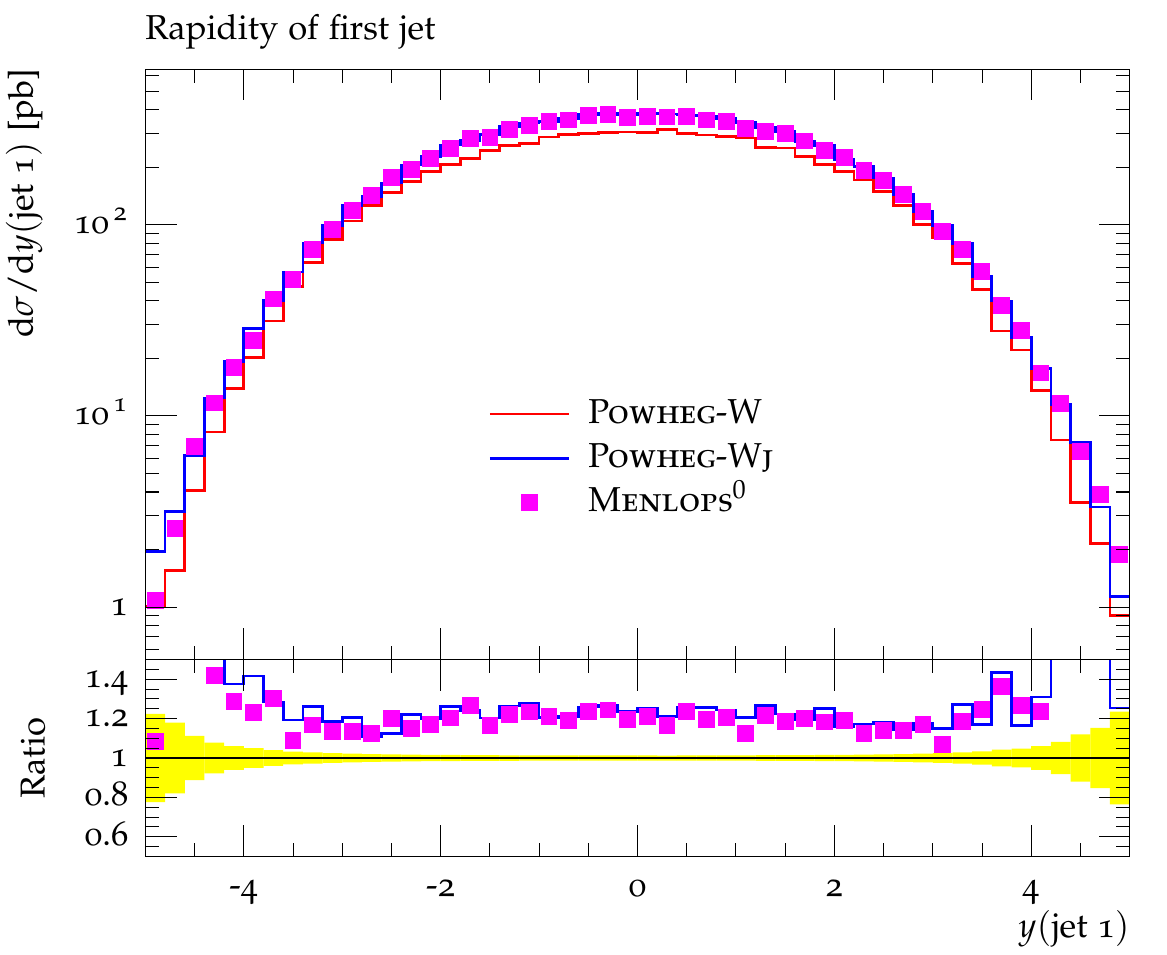}
\par\end{centering}

\caption{\noun{Powheg-W }(red), \noun{Powheg-Wj }(blue) and \noun{Menlops$^{0}$}
(magenta squares) predictions for the leading jet transverse momentum
and rapidity spectra for current LHC $pp$ beam energies.}

\label{fig:menlops_0_validation_leading_jet} 
\end{figure}

In figure \ref{fig:menlops_0_validation_leading_jet} we compare the
predictions of the leading jet transverse momentum and rapidity distributions
among the \noun{Powheg-W, Powheg-Wj} and \noun{Menlops$^{\text{\ensuremath{\infty}}}$
}approaches\noun{.} As expected, since these distributions are inclusive
with respect to the jet-associated vector boson production process
and since they do not receive contributions from regions of phase
space in which the W boson has a low $p_{{\scriptscriptstyle \mathrm{T}}}$,
the \noun{Powheg-Wj} and \noun{Menlops$^{\text{0}}$} are indistinguishable
from one another. In the latter two simulations a small $10-20\,\%$
NLO enhancement of the total vector boson plus jet cross section is
discernible from the low-$p_{{\scriptscriptstyle \mathrm{T}}}$ end
of the transverse momentum spectrum and again, more clearly, in the
rapidity spectrum. The jet transverse momentum spectrum clearly shows
the NLO enhancement increasing with the $p_{{\scriptscriptstyle \mathrm{T}}}$
\cite{Rubin:2010xp}. Recall that a similar but slightly less robust,
enhancement was visible in the leading jet transverse momentum spectrum
in the \noun{Menlops$^{\infty}$ }case (Fig.~\ref{fig:menlops_infty_validation_ge_1j}).
This is not unexpected since in that case the simulation should offer
the same leading order description of \emph{two}-\emph{jet} events
(whose rate is underestimated by the single emission \noun{Powheg-V}
programs) which serve to harden the leading jet $p_{{\scriptscriptstyle \mathrm{T}}}$
spectrum. 

\begin{figure}[H]
\begin{centering}
\includegraphics[width=0.4\textwidth]{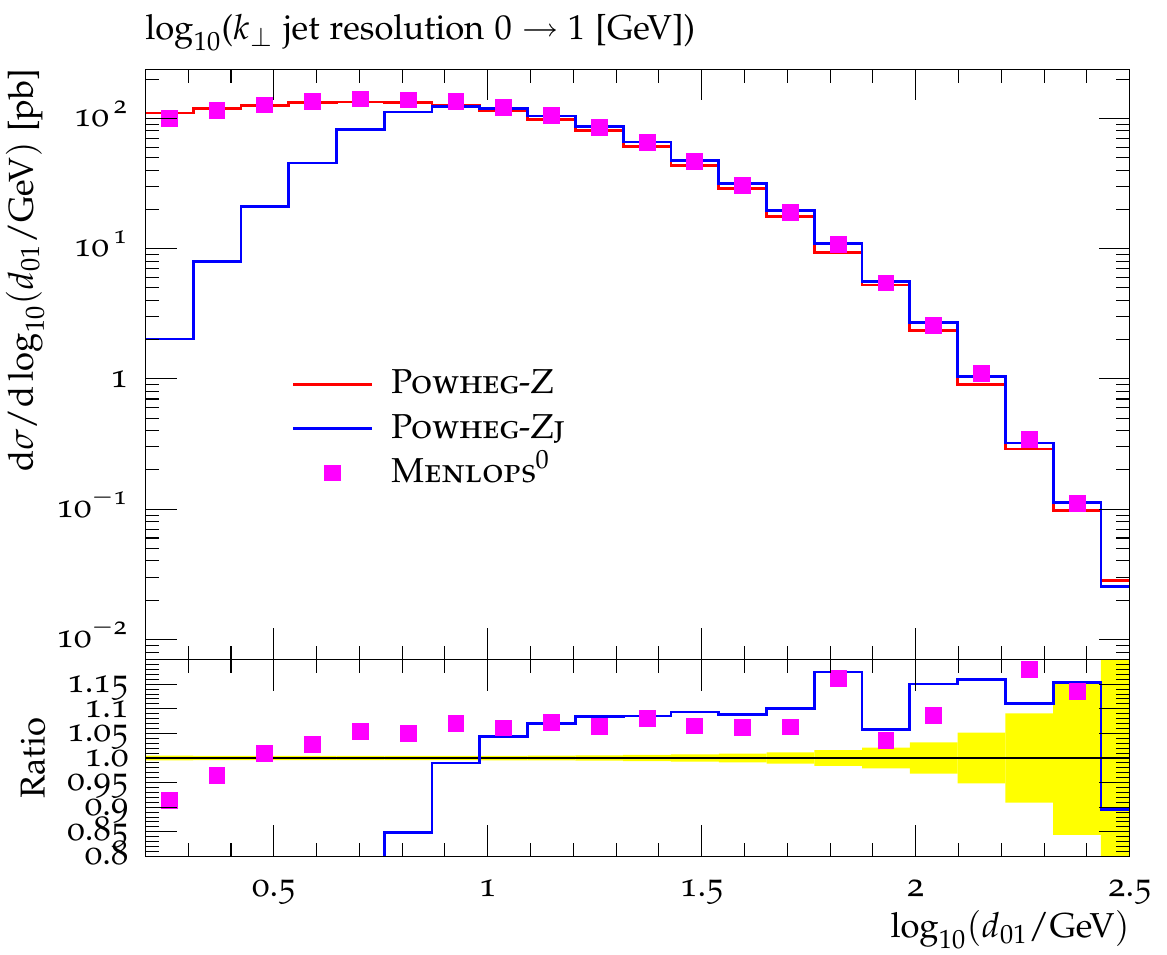}\hfill{}\includegraphics[width=0.4\textwidth]{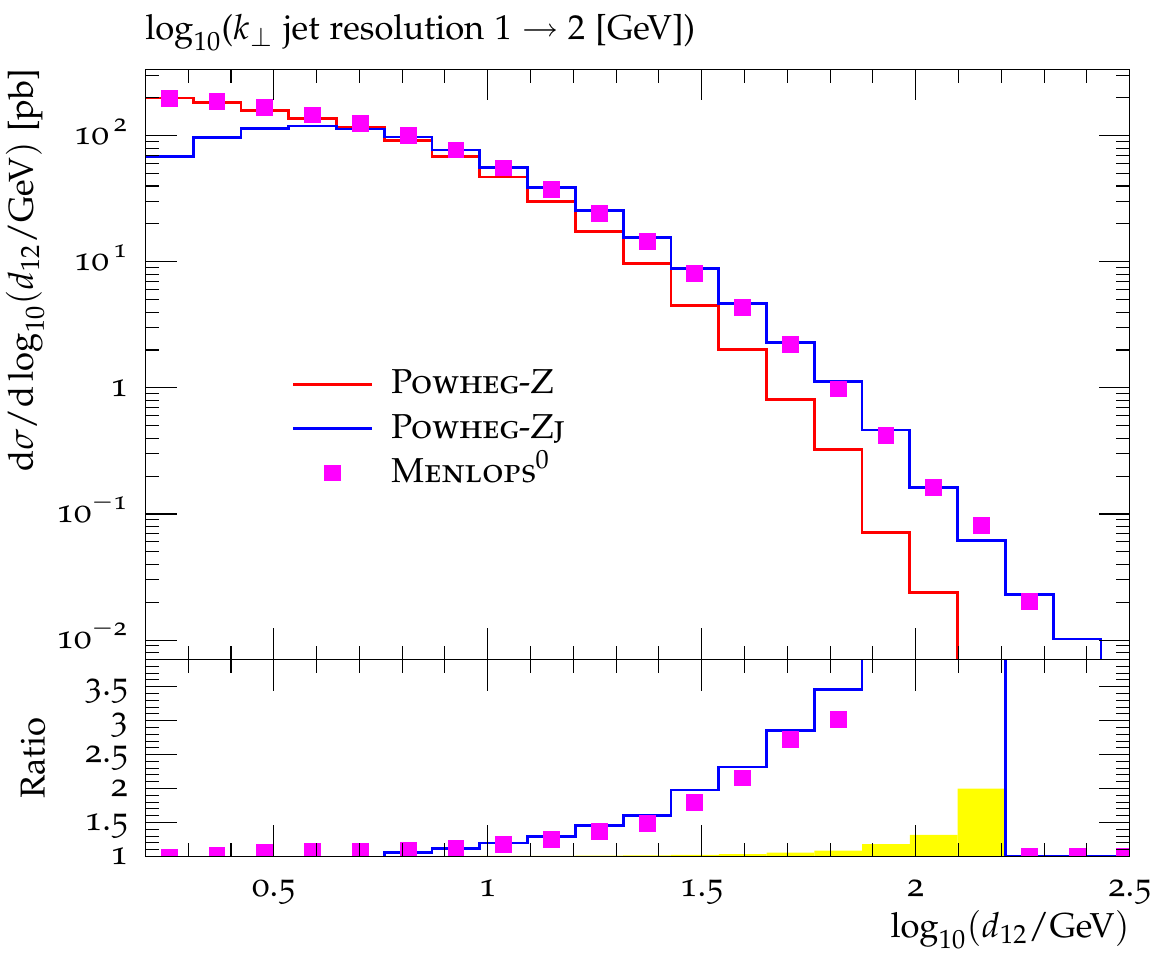}
\par\end{centering}

\begin{centering}
\includegraphics[width=0.4\textwidth]{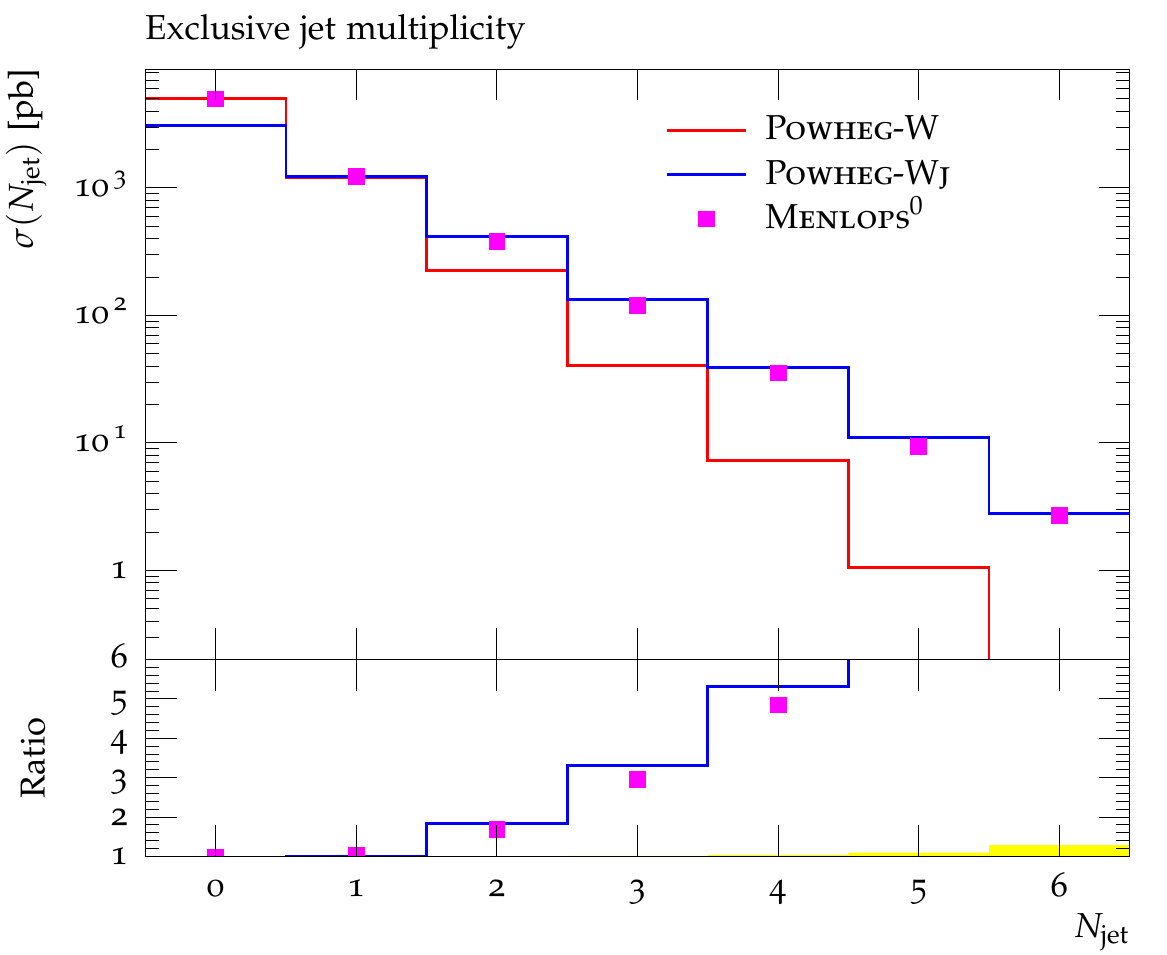}\hfill{}\includegraphics[width=0.4\textwidth]{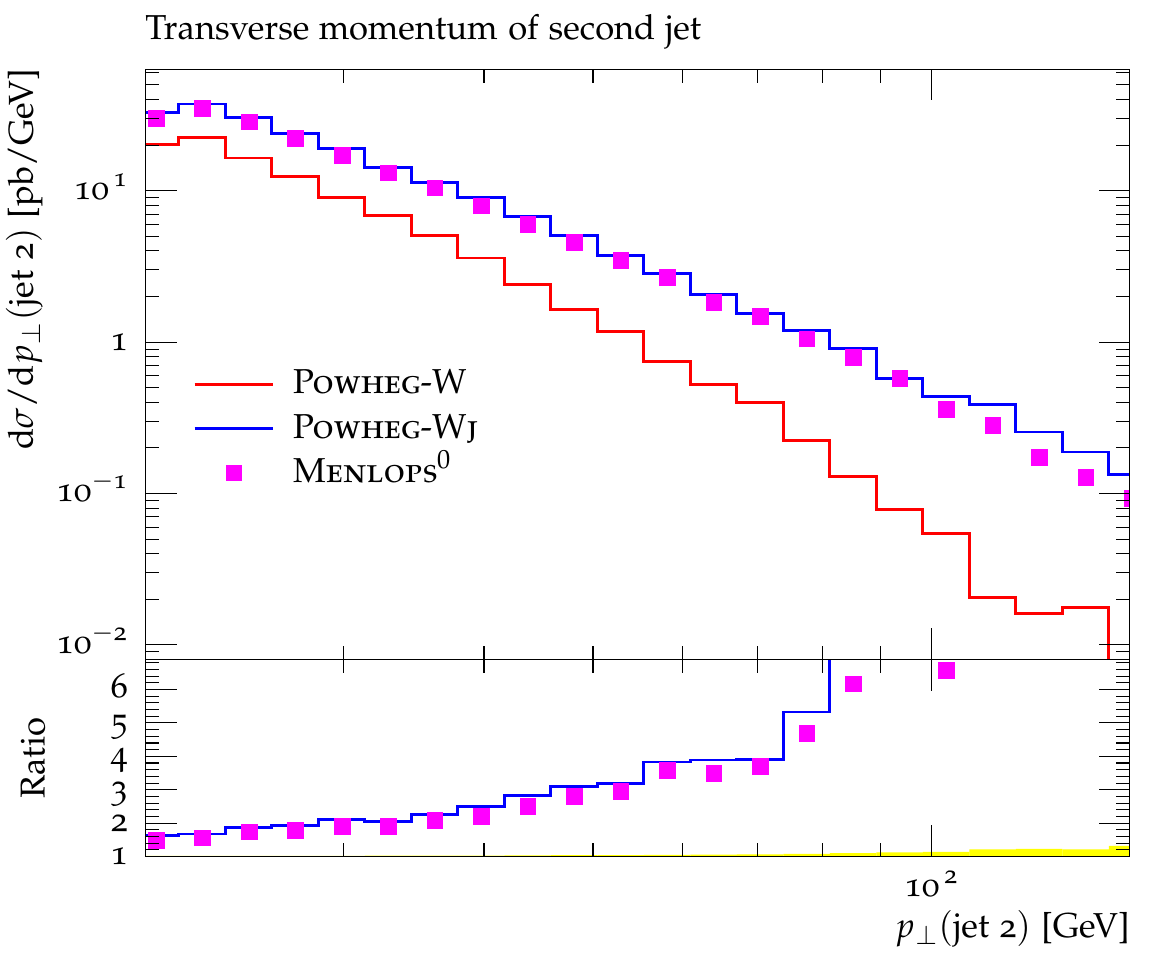}
\par\end{centering}

\caption{Observables sensitive to the emission of at least one hard jet. \noun{Powheg-}V,
\noun{Powheg-Vj} and \noun{Menlops$^{0}$} results are shown in red,
blue and magenta respectively. As in all earlier plots, the sub-plot
displays the ratio of the various predictions with respect to that
given by the first entry in the legend; the inset yellow band corresponds
to the statistical errors on the reference data.}

\label{fig:menlops_0_validation_ge_1j} 
\end{figure}

Figure \ref{fig:menlops_0_validation_ge_1j} focuses more on observables
sensitive to the emission of multiple hard jets. The trends shown
here are readily understandable in terms of the basic features and
frailties of the various generators. In the upper right-hand plot
the differential $0\rightarrow1$-jet rate is presented, physically
representing the exclusive $k_{{\scriptscriptstyle \mathrm{T}}}$-jet
clustering scale at which a 0-jet event becomes resolved as a 1-jet
one. Here one can see good agreement among all three approaches except
in the low transverse momentum / Sudakov region where the Monte Carlo
generation cut and the lack of any large-log resummation in the \noun{Powheg-Zj
}program becomes evident with respect to the other two resummed predictions.
The same feature is apparent in the differential $1\rightarrow2$-jet
rate, for the same reason. In this case, however, at the high $p_{{\scriptscriptstyle \mathrm{T}}}$
end of the spectrum, the NLO and partial NLO corrections to the jet-associated
vector boson production cross section reveal themselves, leading to
an excess of the\noun{ Powheg-Zj }and \noun{Menlops$^{0}$ }results
over, what is for this observable, in this region of phase space,
only a LO prediction from the \noun{Powheg-Z }program. The lower plots,
displaying the exclusive jet multiplicities and transverse momentum
spectrum of the next-to-leading jet, show the now-familiar picture
of the \noun{Powheg-Vj }and \noun{Menlops$^{0}$ }simulations predicting
more radiation in the high transverse momentum regions of phase space,
where, by construction, they agree exactly with one another. Before
moving on though we draw attention to the one place where these two
predictions do not agree in the lower two plots \emph{viz} the 0-jet
multiplicity bin; there the \noun{Menlops$^{0}$} prediction correctly
interpolates to that of the vector boson production \noun{Nlops}. 

\subsection{Merging \noun{Menlops} samples\label{sub:v-vj_menlops}}

The \noun{Menlops }enhancement described in Sect.~\ref{sub:v_menlops}
augments the \noun{Powheg Nlops }simulation of vector boson production
freely with next-to-leading order real emission matrix elements, while
that of Sect.~\ref{sub:vj_menlops} includes, through the factorisation
theorem and Sudakov resummation, lower order matrix elements in the
vector boson plus jet simulation. Neither simulation therefore includes
a full set of NLO corrections to both\emph{ }vector boson production
or jet-associated vector boson production. 

A practical, if seemingly crude, way to improve on the predictions
of either \noun{Menlops }simulation,\noun{ }is to simply combine the
event samples they produce appropriately, such that results for both
fully inclusive observables and also observables inclusive with respect
to the vector boson plus jet final-state, are accurate at the NLO
level. To this end we follow a similar line of reasoning to Ref.~\cite{Hamilton:2010wh},
in particular, the discussion in Sect.~3.

In Ref.~\cite{Hamilton:2010wh} an approximation to the exact \noun{Menlops
}method is implemented where events from \noun{Nlops }and \noun{Meps
}samples are filtered according to their jet multiplicity, using the
exclusive $k_{{\scriptscriptstyle \mathrm{T}}}$-jet clustering algorithm,
at the so-called \noun{Menlops }merging scale, and arranged in 0-jet,
1-jet and $\geqslant$2-jet sub-samples. The \noun{Nlops }sub-samples
are generally considered to offer a better description of the 0-jet
and 1-jet events than their \noun{Meps }counterparts since both simulations
possess the same tree-level matrix elements, yet, at least for the
0-jet class, the \noun{Powheg }simulation also includes virtual corrections.
Conversely, the \noun{Meps }simulations contain higher multiplicity
matrix elements that the \noun{Nlops} simulations do not, so the $\geqslant$2-jet
sample is regarded as having a better description there. Thus, in
the pragmatic implementation of Ref.~\cite{Hamilton:2010wh}, the
final \noun{Menlops }samples are constructed by complementing events
from the 0- and 1-jet \noun{Powheg }sub-samples with those in the
$\geqslant$2-jet \noun{Meps }one, in carefully specified proportions;
the key consideration being that the unitary violating \noun{Meps}
component should be restricted so as not to diminish the NLO accuracy
of inclusive observables. 

Here the situation is somewhat analogous. Rather than implement a
multi-particle phase space partition according to a conventional jet
measure, we use the transverse momentum of the vector boson plus jet
kinematics, at the level of the bare (pre-shower) events. As noted
in Ref.~\cite{Alioli:2010xa}, in principle one can assemble an infrared
safe jet algorithm from the radiation phase space mappings underlying
the \noun{Powheg~Box} implementation \cite{Alioli:2010xd}, hence
the partitioning here is in effect much like that of a Durham-type
jet algorithm. From a theoretical point of view though, it is more
appealing to base the partition on the underlying Born kinematics,
since these are more directly interpretable in terms of the hardest
emission cross section formulae, Eqs.~\ref{eq:menlops_inf_xsec},
\ref{eq:menlops_0_xsec_low}. To create the doubly NLO accurate event
sample we then populate the region $p_{{\scriptscriptstyle \mathrm{T,1}}}<p_{{\scriptscriptstyle \mathrm{T}}}^{{\scriptscriptstyle \mathrm{merge}}}$
using the \noun{Menlops$^{\text{\ensuremath{\infty}}}$} program and
the region above $p_{{\scriptscriptstyle \mathrm{T}}}^{{\scriptscriptstyle \mathrm{merge}}}$
using the \noun{Menlops$^{\text{0}}$ }simulation --- we adjust the
number of events taken from either sample appropriately, to relieve
(small) differences in their event weights, giving uniform $\pm1$
weighted events in the final sample.

Plainly, with no modifications to the component \noun{Nlops} programs,
the dependence on the unphysical phase space partition is such that
as it tends to high values one recovers the single vector boson \noun{Nlops}
results, having, we note, only a leading-log, soft-collinear, approximation
for the next-to-leading jet. Conversely, as the $p_{{\scriptscriptstyle \mathrm{T}}}$
defining the merging scale tends to low values, the associated vector
boson production \noun{Nlops} is recovered, for which the description
of fully inclusive observables is (negatively) divergent and unphysical.
Hence by merging instead the output of the \noun{Menlops} simulations
we mitigate the unwanted scale dependence to the same extent as in
the CKKW method.

In order to preserve the NLO accuracy of inclusive observables in
Ref.~\cite{Hamilton:2010wh} it was important that the merging scale
be bounded from below, to prevent too much of the tree level \noun{Meps
}component entering and spoiling it, specifically, it was stipulated
that the \noun{Meps }events form no more than a fraction $\mathcal{O}(\alpha_{{\scriptscriptstyle \mathrm{S}}})$
of the total. Equally, in the present case, while the merging scale
dependence is lessened by combining \noun{Menlops }improved samples,
if we require the sample to be doubly NLO accurate, in the sense indicated
by the second paragraph in this section, it must be subject to restrictions.
Here, as the merging scale is decreased towards zero, the \noun{Menlops$^{\text{0}}$
}sample is recovered, which is strictly only LO accurate in the description
of fully inclusive observables (modulo the NLO normalization\emph{
K-}factor), hence, the key point is again to bound the merging scale
from below. Following the same logic as Ref.~\cite{Hamilton:2010wh},
it is clear that NLO accuracy for inclusive observables will be preserved,
up to fluctuations of NNLO significance, if one restricts the fraction
of \noun{Menlops$^{\text{0}}$ }events in the sample to be less than
$\mathcal{O}(\alpha_{{\scriptscriptstyle \mathrm{S}}})$.

In practice, it is almost certainly the case that very much bolder
merging scale choices, taking in a much greater \noun{Menlops$^{\text{0}}$
}fraction would in fact not cause deviations in the predictions for
inclusive observables. This point is clearly made in Fig.~\ref{fig:menlops_0_validation_fully_inclusive},
where there is a remarkable level of agreement between the latter
and the genuinely NLO \noun{Powheg-V }predictions. This is not something
to be deceived by though. Whereas the \noun{Menlops$^{\text{0}}$
}kinematics, $\bm{\Phi}_{{\scriptscriptstyle V}}$, there are distributed
according to a \emph{K-}factor enhanced leading order distribution,
$B(\bm{\Phi}_{{\scriptscriptstyle V}})$, the \noun{Powheg-V} predictions
include full NLO correlations through $\bar{B}(\bm{\Phi}_{{\scriptscriptstyle V}})$.
Therefore, while it may be the case that the combination of the \emph{K-}factor
and $B(\bm{\Phi}_{{\scriptscriptstyle V}})$ work well, a more complete
study than this ought to expose the leading order sensitivity to changes
in the renormalization and factorization scales. In order to have
a truly doubly NLO accurate sample the lower bound on the merging
scale should be respected.

We have enforced this criterion in building our merged samples: in
the case of W production at the LHC, merging the two \noun{Menlops
}samples at a scale $p_{{\scriptscriptstyle \mathrm{T}}}^{{\scriptscriptstyle \mathrm{merge}}}=\mathrm{35\, GeV}$,
leads to a fraction of $11-12\,\%$ of \noun{Menlops$^{\text{0}}$}
events in the final sample%
\footnote{Setting $p_{{\scriptscriptstyle \mathrm{T}}}^{{\scriptscriptstyle \mathrm{merge}}}=20\,\mathrm{GeV}$
in LHC W boson production yields a sample $25\,\%$ of which is \noun{Menlops$^{\text{0}}$}
events; note that the figures in section \ref{sec:Results} indicate
that inclusive predictions are rather insensitive to maximal merging
scale variations (as does Fig.~\ref{fig:menlops_0_validation_fully_inclusive}).%
} and in the case of Z production at the Tevatron $p_{{\scriptscriptstyle \mathrm{T}}}^{{\scriptscriptstyle \mathrm{merge}}}=\mathrm{25\, GeV}$
leads to the same amount. Note that in combining the samples, obviously
the \noun{Menlops$^{\infty}$} and \noun{Menlops$^{\text{0}}$ }cross
sections are preserved in the regions either side of the cut, thus
the final \emph{total} \emph{cross} \emph{section} can differ from
the NLO fully inclusive \noun{Powheg-V / Menlops$^{\infty}$ }results.
In all cases we find the difference between the \noun{Menlops$^{\infty}$
}result and those of the merged samples --- which we refer to as \noun{Menlops$^{25}$
}and \noun{Menlops$^{35}$ }on\noun{ }account\noun{ }of the merging
scales used --- is not greater than 1\,\%. This latter observation
is unremarkable given the fraction of \noun{Menlops$^{\infty}$ }events
allowed and the agreement shown in Fig.~\ref{fig:menlops_0_validation_fully_inclusive}
for inclusive quantities, we merely note it for technical reasons. 

Furthermore, in practical applications, provided that the vector boson
$p_{{\scriptscriptstyle \mathrm{T}}}$ is used as the renormalization
and factorization scale in the fixed order components, the latter
constraint renders the added resummation in the \noun{Menlops$^{0}$}
simulations pointless; so long as the merging scale is not too small
one can obviously use the jet-associated vector boson production program,
unmodified, in the high $p_{{\scriptscriptstyle \mathrm{T}}}$ region.
We therefore employ the \noun{Menlops$^{0}$} simulation in merging
more as a matter of theoretical correctness. In this regard, when
considering the dependency of the predictions on the unphysical merging
scale, the Sudakov suppression factors take on their practical as
well as theoretical significance, in the same way as in the CKKW method.
Actually the merging scale dependence here is substantially reduced
with respect to the CKKW case since, as we shall see, the two component
sub-samples are exactly unitary, moreover, they both include precisely
all the same tree-level matrix elements and exact same resummation.
In particular, from the point of view of the region below $p_{{\scriptscriptstyle \mathrm{T}}}^{{\scriptscriptstyle \mathrm{merge}}}$,
we expect that using the \noun{Powheg-V} simulation instead of its
\noun{Menlops} enhanced version would be notable as a marked dependence
on the merging scale in multi-jet observables. As noted in the introduction,
to lessen the dependence on the merging scale beyond this, would seem
to be equivalent to the task of achieving genuine NNLO-parton shower
matching, something which is beyond the more modest, practical, aims
of this work. 

A final, related, technical point concerns the presence of discontinuities
in the differential jet rates and $p_{{\scriptscriptstyle \mathrm{T}}}$
spectra. We claim a level of matching at least as good as in the CKKW
method, in particular, that all logarithms pertaining to $p_{{\scriptscriptstyle \mathrm{T},1}}$
and hence $p_{{\scriptscriptstyle \mathrm{T}}}^{{\scriptscriptstyle \mathrm{merge}}}$
are resummed. On the other hand, as is by now clear, the same is not
true for the finite, non-logarithmic, NLO (virtual) corrections. This
alone defines the level of ambiguity in the present merging. Below
$p_{{\scriptscriptstyle \mathrm{T}}}^{{\scriptscriptstyle \mathrm{merge}}}$
we have full NLO corrections to vector boson production observables
only, while above it there are only full NLO corrections to vector
boson plus jet production. Of course, one expects and finds that the
NLO corrections to the latter enhance jet rates and $p_{{\scriptscriptstyle \mathrm{T}}}$
spectra with respect to the former by $\mathcal{O}(\alpha_{{\scriptscriptstyle \mathrm{S}}})$
at intermediate / high $p_{{\scriptscriptstyle \mathrm{T}}}^{{\scriptscriptstyle \mathrm{merge}}}$
--- these are, after all, the differences we hope to take into account
by merging --- leading to the possibility that small discontinuities
may occur around $p_{{\scriptscriptstyle \mathrm{T}}}^{{\scriptscriptstyle \mathrm{merge}}}$
on merging the two samples. Note, however, that the exact value of
the merging scale should not be taken too seriously, it is wrong,
for instance, to imagine that the quality of either component sample
instantly degrades either side of $p_{{\scriptscriptstyle \mathrm{T}}}^{{\scriptscriptstyle \mathrm{merge}}}$;
there is certainly some blurring in that regard. This being the case,
to smoothen any small NLO differences occurring between the different
jet rates in the vicinity of the merging partition, we do not implement
an immediate step-function cut-off at $p_{{\scriptscriptstyle \mathrm{T}}}^{{\scriptscriptstyle \mathrm{merge}}}$,
but rather a smooth sinusoidal damping function of the form Eq.~\ref{eq:switching_function},
interpolating between $0$ and $1$ in the region $\pm5\,\mathrm{GeV}$
about $p_{{\scriptscriptstyle \mathrm{T}}}^{{\scriptscriptstyle \mathrm{merge}}}$.

~

~

\begin{figure}[H]
\begin{centering}
\includegraphics[width=0.4\textwidth]{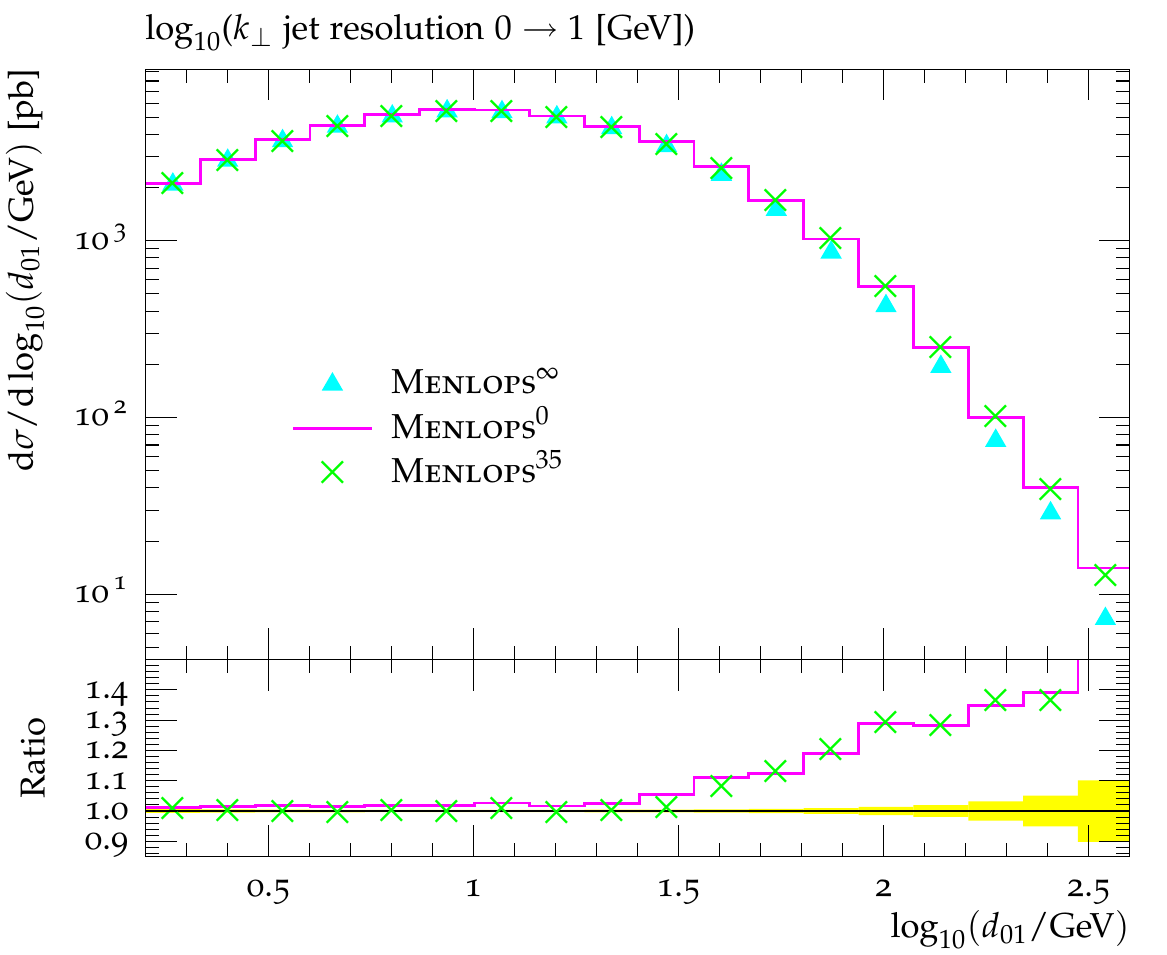}\hfill{}\includegraphics[width=0.4\textwidth]{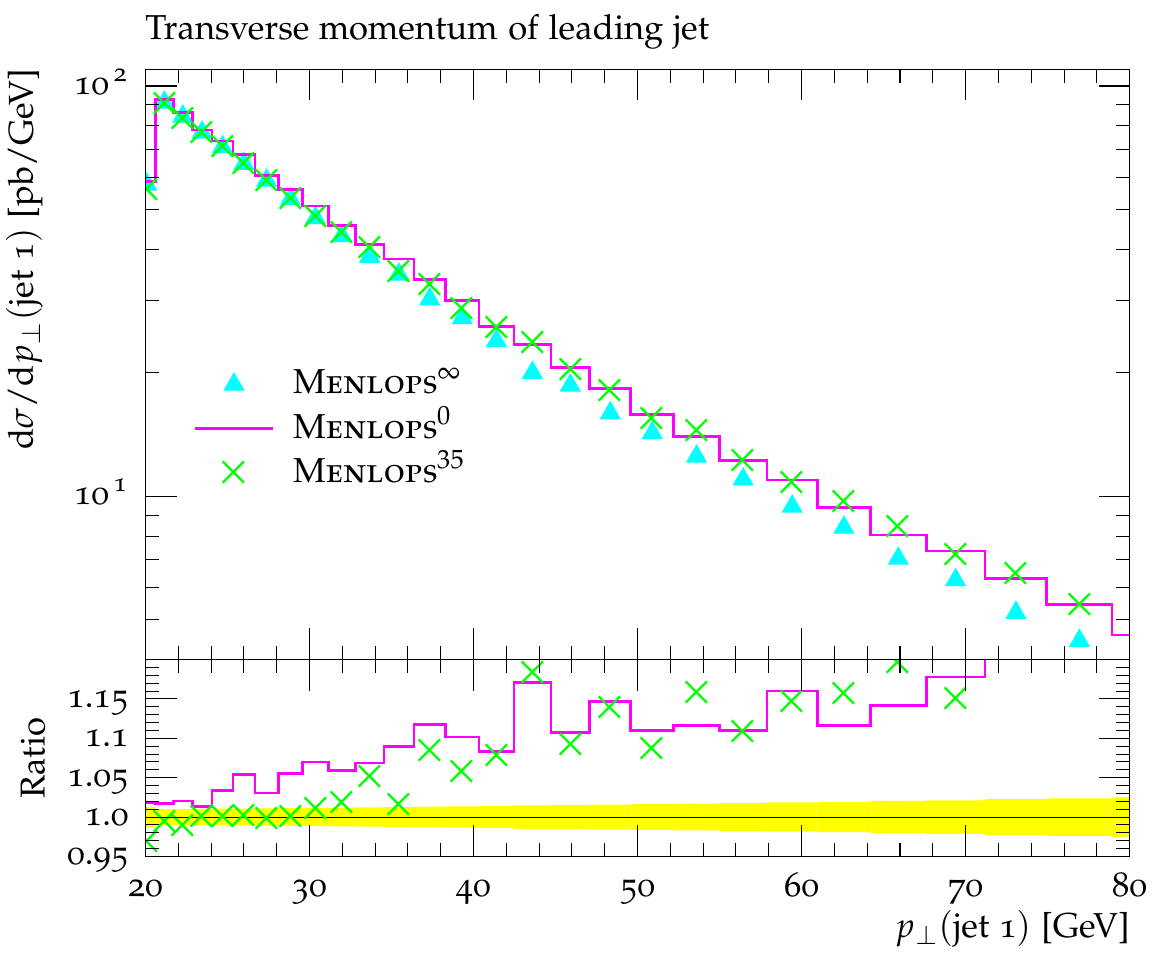}
\par\end{centering}

\caption{The 0- to 1-jet differential jet rate (left) and leading jet transverse
momentum spectrum (right), in W boson production at the LHC. In each
figure we show, as green crosses, the prediction from the merging
of the component \noun{Menlops$^{\text{0}}$ }and \noun{Menlops$^{\text{\ensuremath{\infty}}}$
}samples, with the merging scale set to $p_{{\scriptscriptstyle \mathrm{T}}}^{{\scriptscriptstyle \mathrm{merge}}}=35\,\mathrm{GeV}$.
The magenta lines and cyan triangles, respectively, show the corresponding
predictions from the component \noun{Menlops$^{0}$ }and\noun{ Menlops$^{\infty}$}
samples.\noun{ }This collider setup and these distributions were selected
on the basis that they show the matching ambiguity between the two
samples in its worst light, specifically, the difference between the
magenta and cyan distributions at $p_{{\scriptscriptstyle \mathrm{T}}}^{{\scriptscriptstyle \mathrm{merge}}}$.}

\label{fig:menlops_v-vj_validation} 
\end{figure}

By way of showing the worst that can be expected in terms of such
discontinuities at the merging scale, we show in figure \ref{fig:menlops_v-vj_validation}
the 0- to 1-jet differential jet rate and the leading jet $p_{{\scriptscriptstyle \mathrm{T}}}$
spectrum, which we found to be most sensitive. Here one can see the
\noun{Menlops$^{\text{0}}$ }and \noun{Menlops$^{\text{\ensuremath{\infty}}}$
}predictions\noun{ }disagree at the level of $\sim5\,\%$ in the vicinity
of the 35 GeV merging scale, with the\noun{ Menlops$^{\text{35}}$
}merged prediction interpolating between them. Due to the size of
the differences probed here the Monte Carlo statistical errors make
the interpolation difficult to see, we suffice to say that in this
regard we have used here event samples comprising of well over one
million events. Other distributions such as the vector boson $p_{{\scriptscriptstyle \mathrm{T}}}$
spectrum exhibit less sensitivity / ambiguity in the region of the
cut, moreover, the magnitude of the NLO corrections at the Tevatron
is smaller, hence, there \noun{Menlops$^{\text{0}}$ }and \noun{Menlops$^{\text{\ensuremath{\infty}}}$
}results are more-or-less indistinguishable in the low $p_{{\scriptscriptstyle \mathrm{T}}}$
region near $p_{{\scriptscriptstyle \mathrm{T}}}^{{\scriptscriptstyle \mathrm{merge}}}$.\noun{
}Many other distributions from all of the \noun{Menlops }samples will
now be studied in comparison to data in Sect.~\ref{sec:Results}.

\section{Tevatron and LHC predictions\label{sec:Results}}

\noindent Here we confront the\noun{ }native \noun{Powheg~Box }programs,
their \textsc{Menlops} enhancements and the merged \noun{Menlops}
samples, with Tevatron and early LHC data. The analyses presented
in this section are therefore carried out having developed the events
output by the \noun{Powheg }and \noun{Menlops} simulations to the
hadron level and included multiple interaction effects. To this end
we have used the \noun{Pythia }8.150 program, following closely the
recommendations set out in Ref.~\cite{Corke:2010zj} but otherwise
maintaining the default settings.%
\footnote{In particular we used the so-called \noun{Pythia} \emph{power}-\emph{shower}
option which is understood to reduce the presence of radiation gaps. %
} We have interfaced \noun{Pythia }to the \noun{HepMC} package \cite{Dobbs:2001ck}
and hence on to the \noun{Rivet} framework \cite{Buckley:2010ar}
to perform the experimental analyses. The selection of results is
intended to be representative.

We have endeavored to keep the colouring of the predictions the same
here as in earlier plots in the article. As in all preceding histograms,
each sub-plot shows the ratio of each type of prediction to that placed
first in the legend, here the experimental data. Also, as before,
in each case the shaded yellow band depicts the projection of the
errors on the reference data into the same ratio distribution.

\noindent \smallskip{}

\subsection{Comparison to Tevatron data\label{sub:Tevatron_comparison}}

\begin{figure}[H]
\begin{centering}
\includegraphics[width=0.4\textwidth]{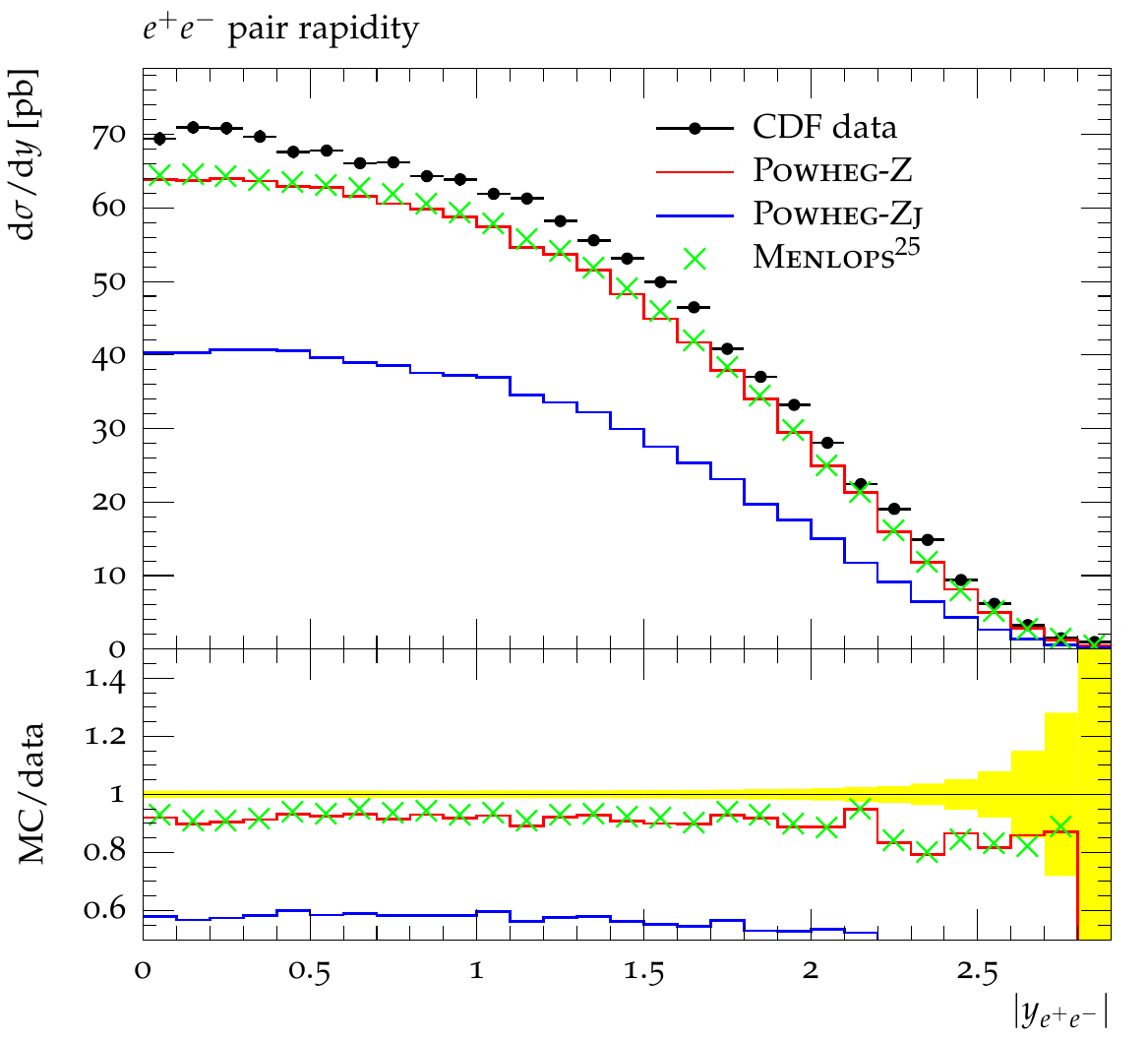}\hfill{}\includegraphics[width=0.4\textwidth]{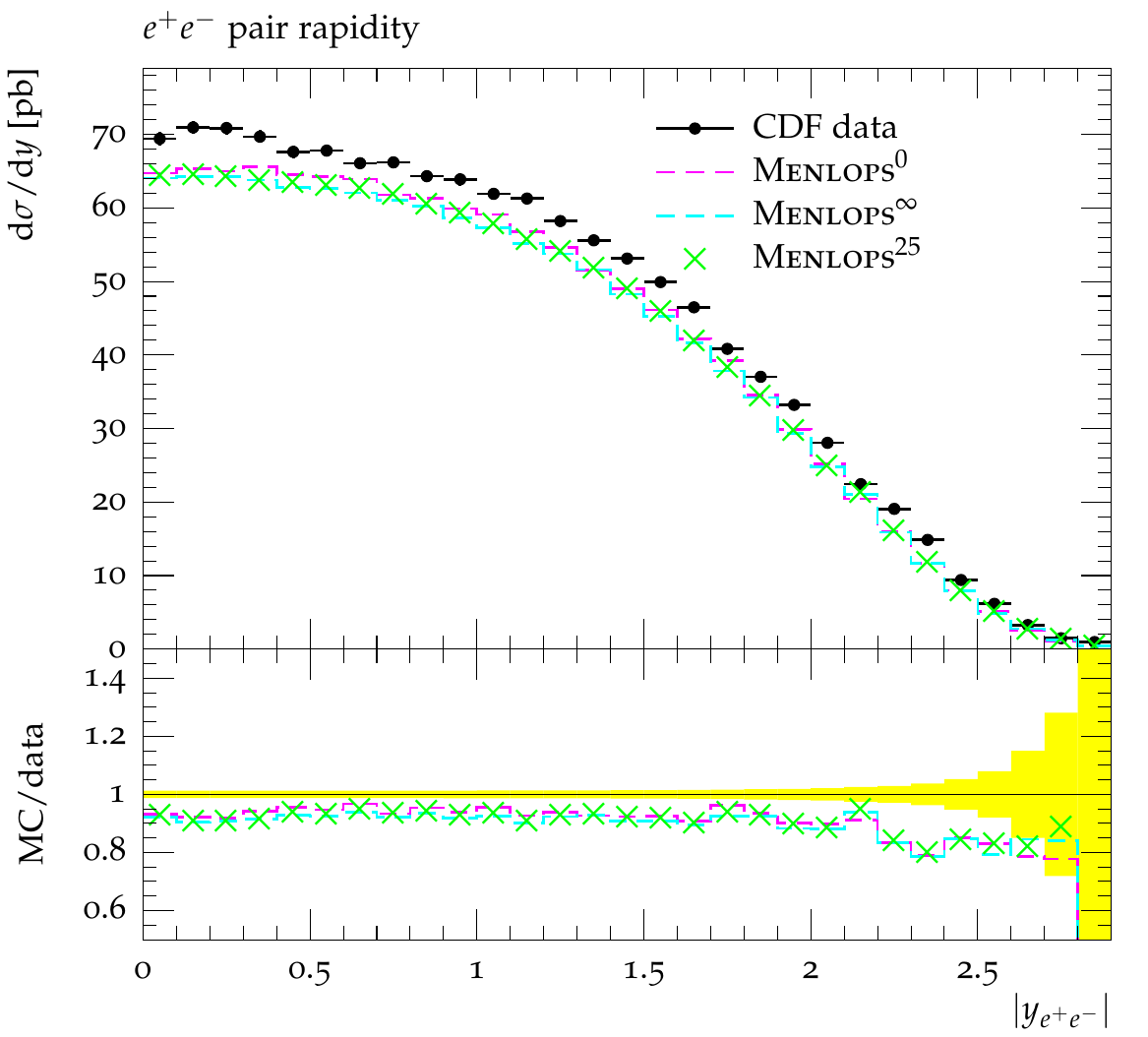}
\par\end{centering}

\caption{On the left-hand side we show the inclusive Z boson rapidity spectrum
as measured by the CDF collaboration at the Tevatron \cite{Aaltonen:2010zza},
superimposed on the data are \noun{Powheg }predictions from the vector
boson production and jet-associated production programs, in red and
blue lines respectively, as well as our \noun{Menlops} prediction
obtained by merging\noun{ Menlops$^{\text{0}}$} and\noun{ Menlops$^{\text{\ensuremath{\infty}}}$
}samples at a merging scale of $25\,\mathrm{GeV}$ (Sect.~\ref{sub:v-vj_menlops}).
On the right we show the effects of varying the same merging scale
between logical extremes ($0\leftrightarrow\infty$) displaying the
pure \noun{Menlops$^{\text{0}}$} and\noun{ Menlops$^{\text{\ensuremath{\infty}}}$}
predictions alongside that shown in the left-hand plot. As throughout
the article, the yellow band signifies the projection of the errors
on the reference data (always the first entry in the legend) into
the MC / data ratio.}

\label{fig:tevatron_inclusive_Z_ee_rapidity} 
\end{figure}

In figure \ref{fig:tevatron_inclusive_Z_ee_rapidity} we show the
Z boson rapidity spectrum as measured by the CDF collaboration at
the Tevatron \cite{Aaltonen:2010zza}. On the left \noun{Powheg-Z,
Powheg-Zj }and merged \noun{Menlops }predictions are compared to data,
while on the right the \noun{Menlops$^{\text{0}}$ }and \noun{Menlops$^{\text{\ensuremath{\infty}}}$
}component samples are shown along with the same merged \noun{Menlops$^{\text{25}}$
}sample ($p_{{\scriptscriptstyle \mathrm{T}}}^{{\scriptscriptstyle \mathrm{merge}}}=25\,\mathrm{GeV}$).

Given the inclusivity of the measurement here (Fig.~\ref{fig:tevatron_inclusive_Z_ee_rapidity})
the \noun{Powheg-Zj} prediction unsurprisingly fails to describe the
data; we present it for completeness and to illustrate the improvement
rendered to it by the \noun{Menlops$^{\text{0}}$} enhancement described
in Sect.~\ref{sub:vj_menlops}, as shown in magenta on the right-hand
plot. The \noun{Powheg-Z }and merged \noun{Menlops }sample predictions
(red line and green crosses) offer a good description of the shape
of the distribution while their overall normalization is negatively
offset from the data by about 10\,\%, an acceptable disagreement
for an NLO computation and one echoed by the other, independent, theoretical
predictions given in Ref.~\cite{Aaltonen:2010zza}. 

From the point of view of validating the \noun{Menlops }procedure,
we draw attention to the fact that the result of merging the \noun{Menlops
}samples is in complete agreement with the \noun{Powheg-Z }one for
this fully inclusive observable. From the same perspective, the stability
of the predictions against extremal variations in the merging scale,
in the right-hand plot, is remarkable. While the level of agreement
between the three \noun{Menlops} predictions is striking we do not
wish to over emphasize it; by virtue of the rules applied to the composition
of the merged sample, the \noun{Menlops$^{\text{\ensuremath{\infty}}}$
}and \noun{Menlops$^{\text{25}}$} results should be in near perfect
agreement, whereas, in principle, one could have expected some marginal
deformations of the shape of the \noun{Menlops$^{\text{0}}$} result
with respect to them, similar in size to those seen between NLO and
\emph{K-}factor adjusted LO predictions. We cannot claim that the
excellent level of agreement between the \noun{Menlops$^{\text{0}}$}
results and the other two \noun{Menlops }ones is not, to some extent,
coincidental --- as noted in Sect.~\ref{sub:vj_menlops}, the true
leading order nature of the shape of fully inclusive \noun{Menlops$^{\text{0}}$}
predictions would reveal itself as a sensitivity to renormalization
and factorization scale variations, while the genuine NLO quality
of the other two \noun{Menlops }samples would show them to be suitably
insensitive. 

The two key points to take away here are: firstly, that the \noun{Menlops$^{\text{0}}$}
enhancement of the \noun{Powheg-Zj }program means that it now gives
a physical prediction for this inclusive observable where it did not
before, moreover, one that has an uncanny resemblance to the fully
NLO predictions; the second important point is that the \noun{Menlops$^{\text{\ensuremath{\infty}}}$
}and \noun{Menlops$^{\text{25}}$ }results are in complete agreement
with that of the \noun{NLO Powheg-Z} program, from which they should
have inherited this prediction.

\begin{figure}[H]
\begin{centering}
\includegraphics[width=0.4\textwidth]{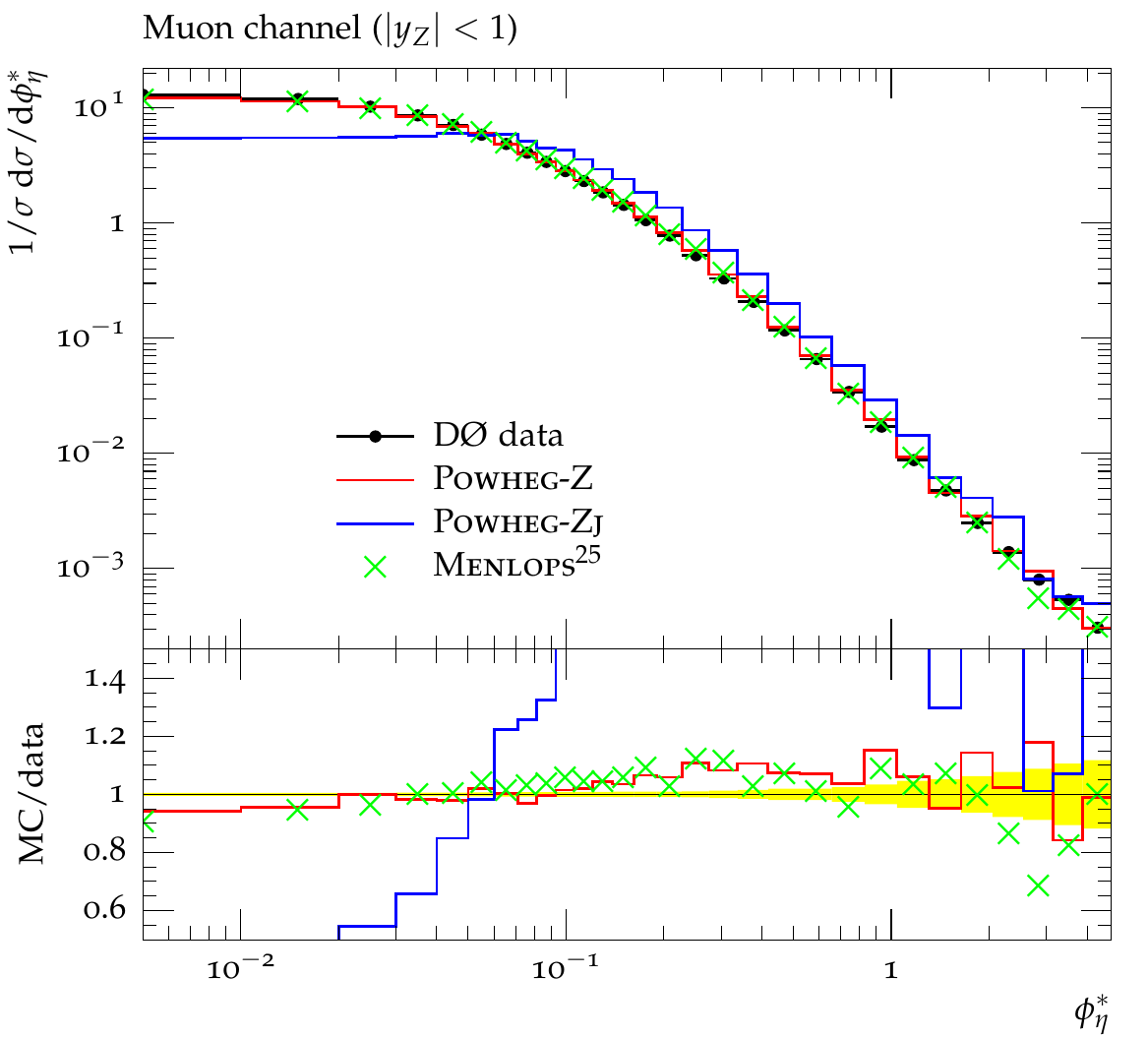}\hfill{}\includegraphics[width=0.4\textwidth]{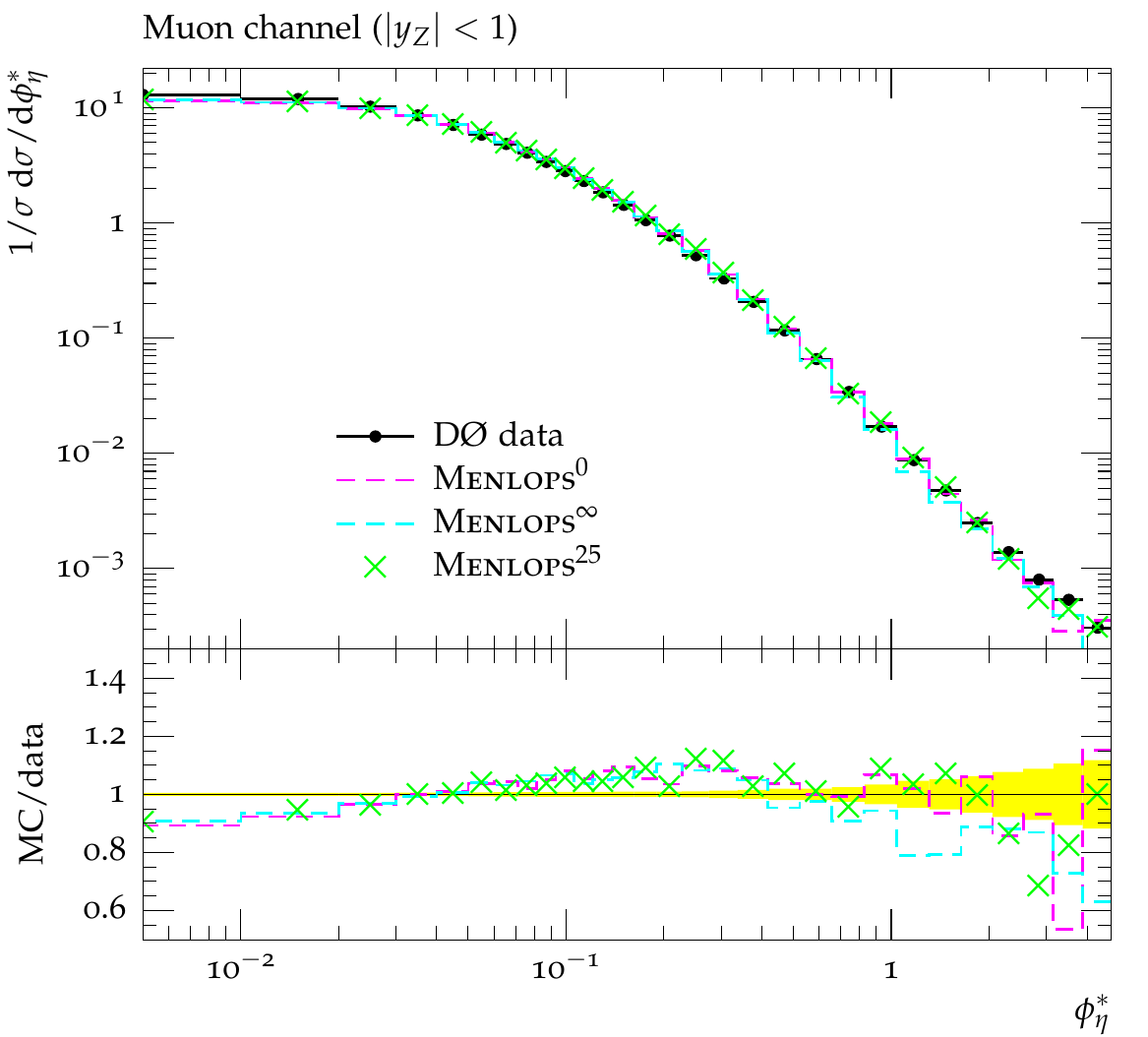}
\par\end{centering}

\begin{centering}
\includegraphics[width=0.4\textwidth]{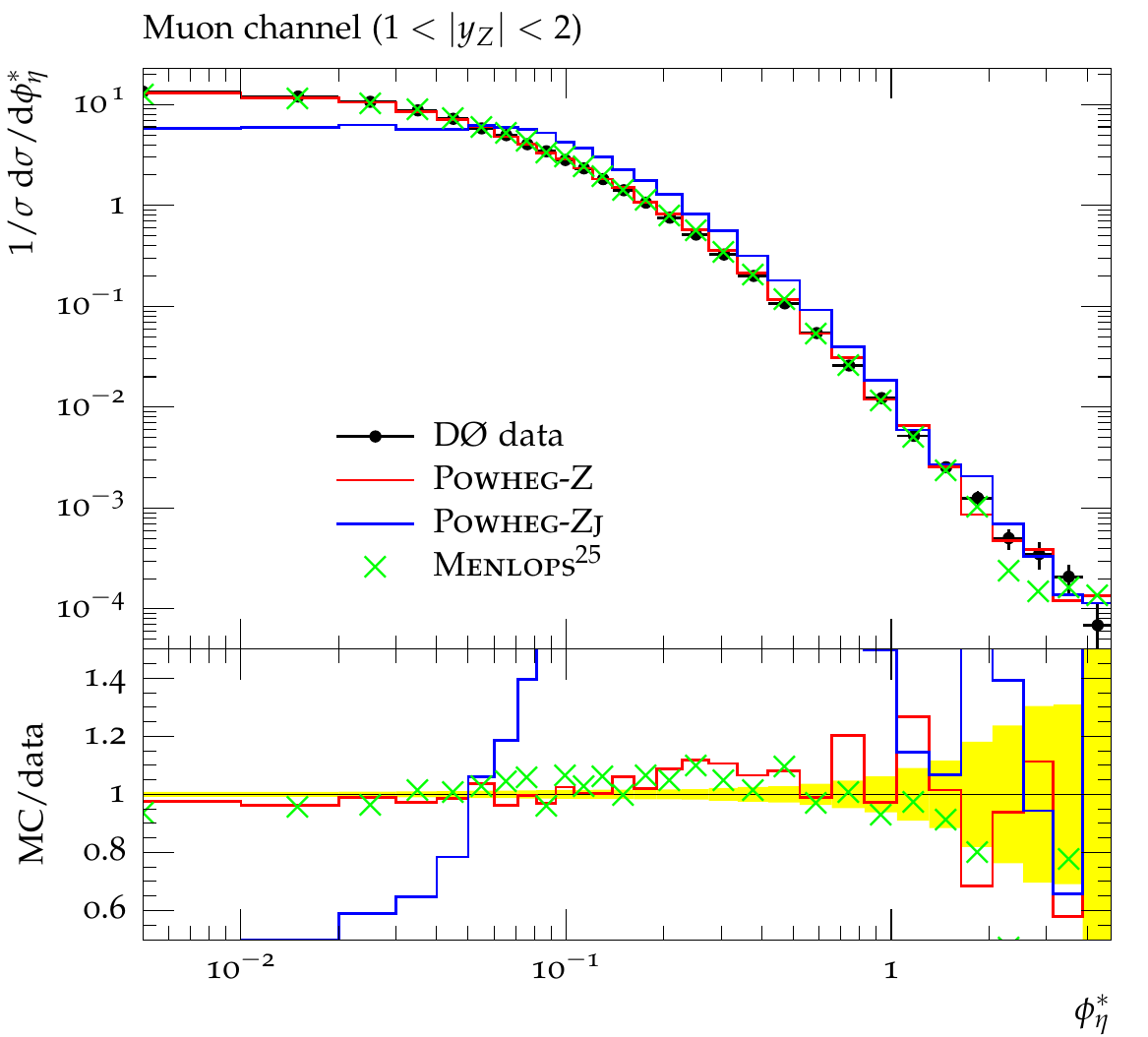}\hfill{}\includegraphics[width=0.4\textwidth]{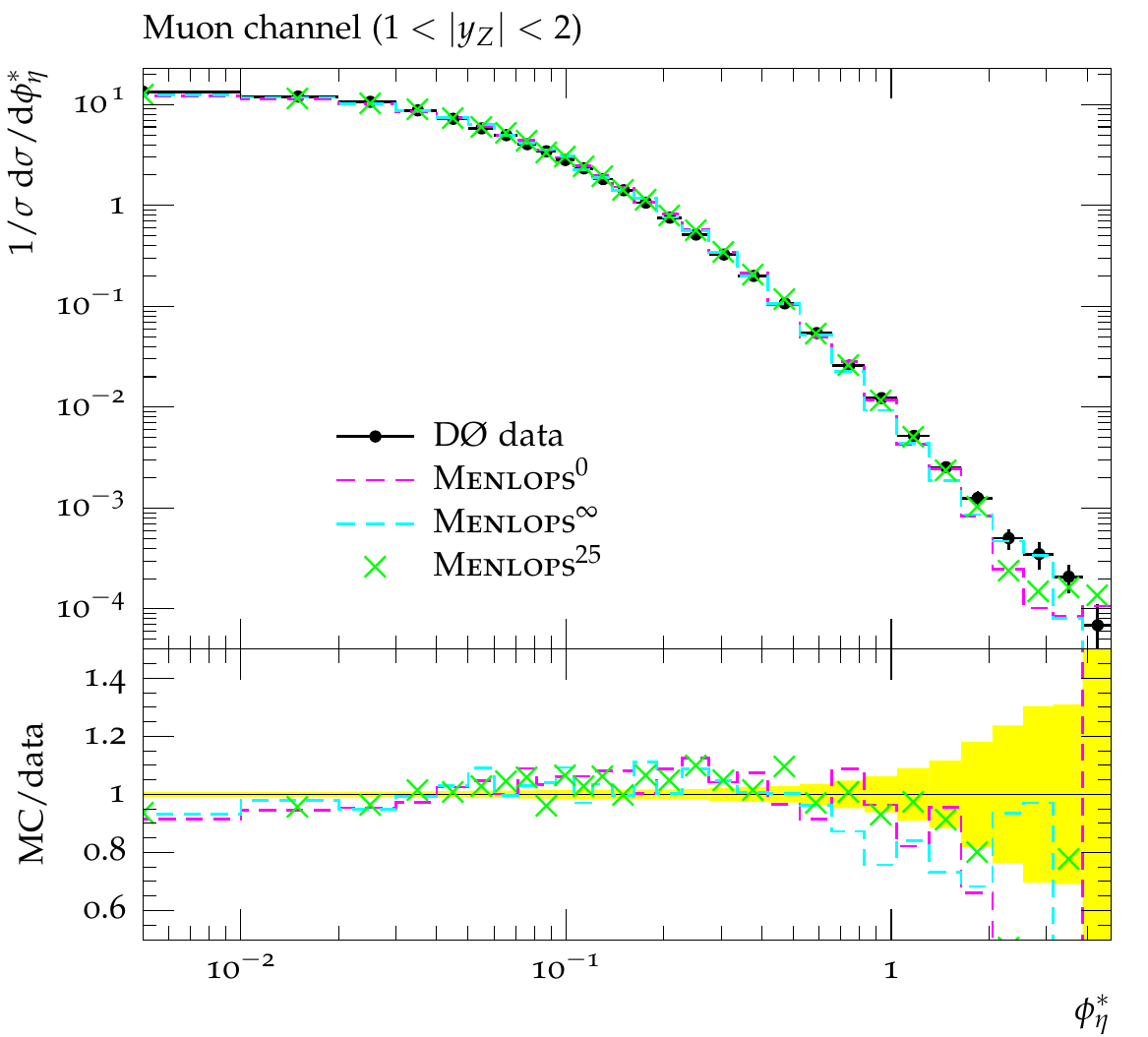}
\par\end{centering}

\caption{\noun{Powheg }and \noun{Menlops }predictions overlaid on D\O~measurements
of the variable $\phi_{\eta}^{*}=\tan\left(\phi_{{\scriptscriptstyle \mathrm{acop}}}/2\right)\sin\left(\theta_{\eta}^{*}\right)$
\cite{Abazov:2010mk}, with $\phi_{{\scriptscriptstyle \mathrm{acop}}}$
defined as $\pi$ minus the azimuthal separation of the Z decay products
and $\theta_{\eta}^{*}$ a measure of the scattering angle of the
leptons with respect to the proton beam in the dimuon rest frame.
Specifically, $\cos\theta_{\eta}^{*}=\tanh\left[\left(\eta^{-}-\eta^{+}\right)/2\right]$,
where $\eta^{+/-}$ denotes the pseudorapidity of $\mu^{+/-}$. In
Ref.~\cite{Abazov:2010mk} it is noted that the angular nature of
$\phi_{{\scriptscriptstyle \mathrm{acop}}}^{*}$ and $\phi_{\eta}^{*}$
means they can be measured more precisely than quantities more dependent
on the lepton momenta as a whole. All histograms here are normalized
to unity.}

\label{fig:tevatron_inclusive_Z_mm_phi*} 
\end{figure}

Figure \ref{fig:tevatron_inclusive_Z_mm_phi*} shows the same set
of comparisons with respect to a recent D\O~measurement \cite{Abazov:2010mk}
of an inclusive quantity, $\phi_{\eta}^{*}$ , said to be closely
related to the Z boson transverse momentum: $\phi_{\eta}^{*}=\tan\left(\phi_{{\scriptscriptstyle \mathrm{acop}}}/2\right)\sin\left(\theta_{\eta}^{*}\right)$
, with $\phi_{{\scriptscriptstyle \mathrm{acop}}}$ defined as $\pi$
minus the azimuthal separation of the muonic decay products and $\theta_{\eta}^{*}$
a measure of the scattering angle of the dimuons with respect to the
proton beam in their rest frame. From the definition one can see that
when the final-state leptons are back-to-back in azimuth, as would
be the case if there were no additional radiation in the final state,
$\phi_{\eta}^{*}$ tends to zero. Hence the region $\phi_{\eta}^{*}=0$
is highly correlated with the Sudakov region in the Z boson transverse
momentum spectrum.

It follows that, despite the plots being normalized to unit area,
the \noun{Powheg-Zj }prediction undershoots the data for $\phi_{\eta}^{*}\rightarrow0$,
while the others offer reasonable agreement there. The fact that this
area of the plots is associated with the Sudakov peak region also
accounts for the small experimental errors there. By-and-large all
of the conclusions drawn in regard to the Z boson rapidity spectrum,
in particular the \noun{Menlops }predictions, apply here unchanged.
What is worth noting in addition, is the binning of the prediction
according to the rapidity of the Z boson. The Sudakov form factor,
amongst others resumming large-logs in the \noun{Menlops}$^{\text{0}}$
improved \noun{Powheg-Zj }simulation, depends on this, hence, it is
interesting and reassuring to see that the low $\phi_{\eta}^{*}$
region is described equally well by this program in the $\left|y_{{\scriptscriptstyle \mathrm{Z}}}\right|<1$
bin as in the $1<\left|y_{{\scriptscriptstyle \mathrm{Z}}}\right|<2$
one.

\begin{figure}[H]
\begin{centering}
\includegraphics[width=0.4\textwidth]{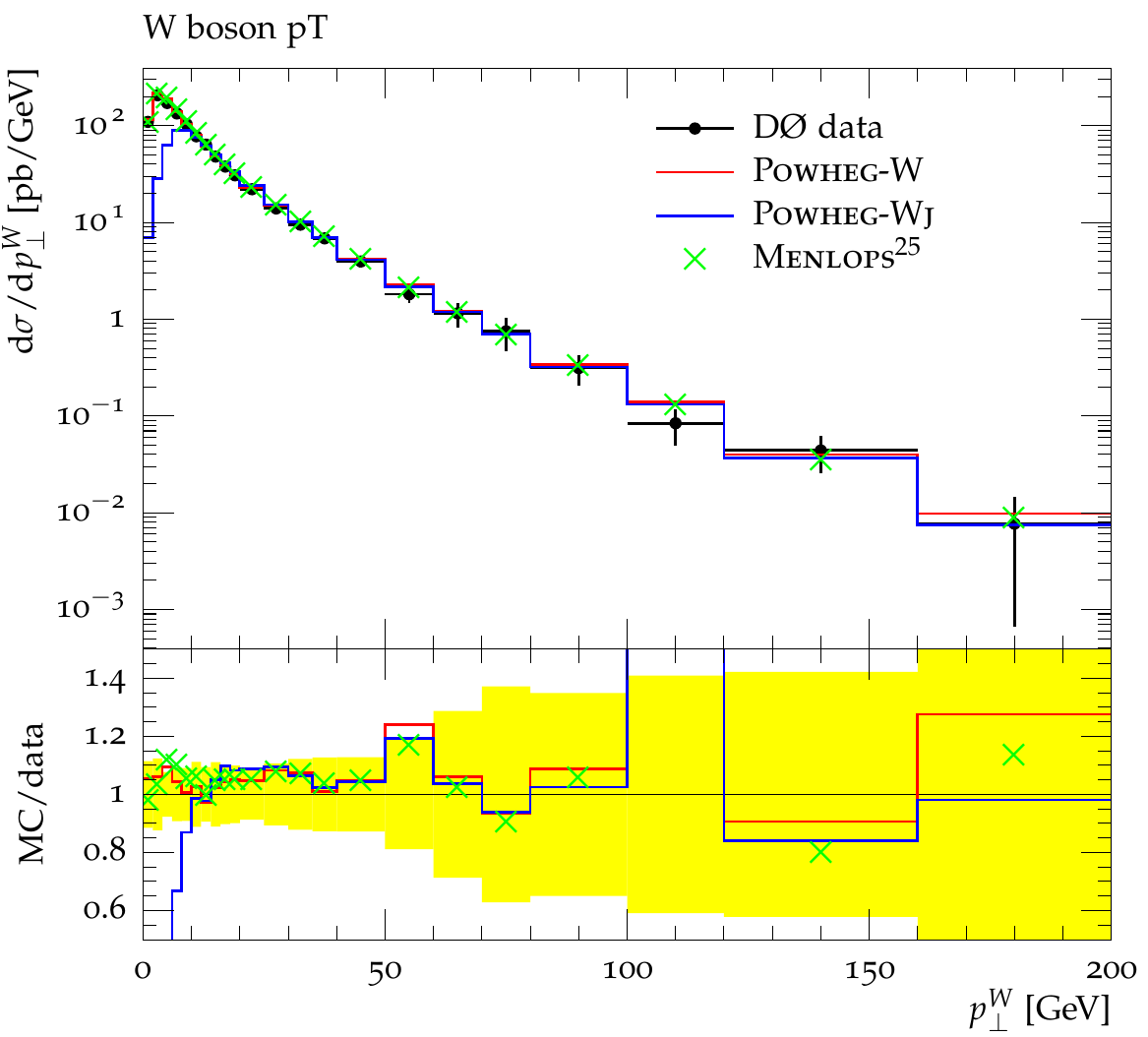}\hfill{}\includegraphics[width=0.4\textwidth]{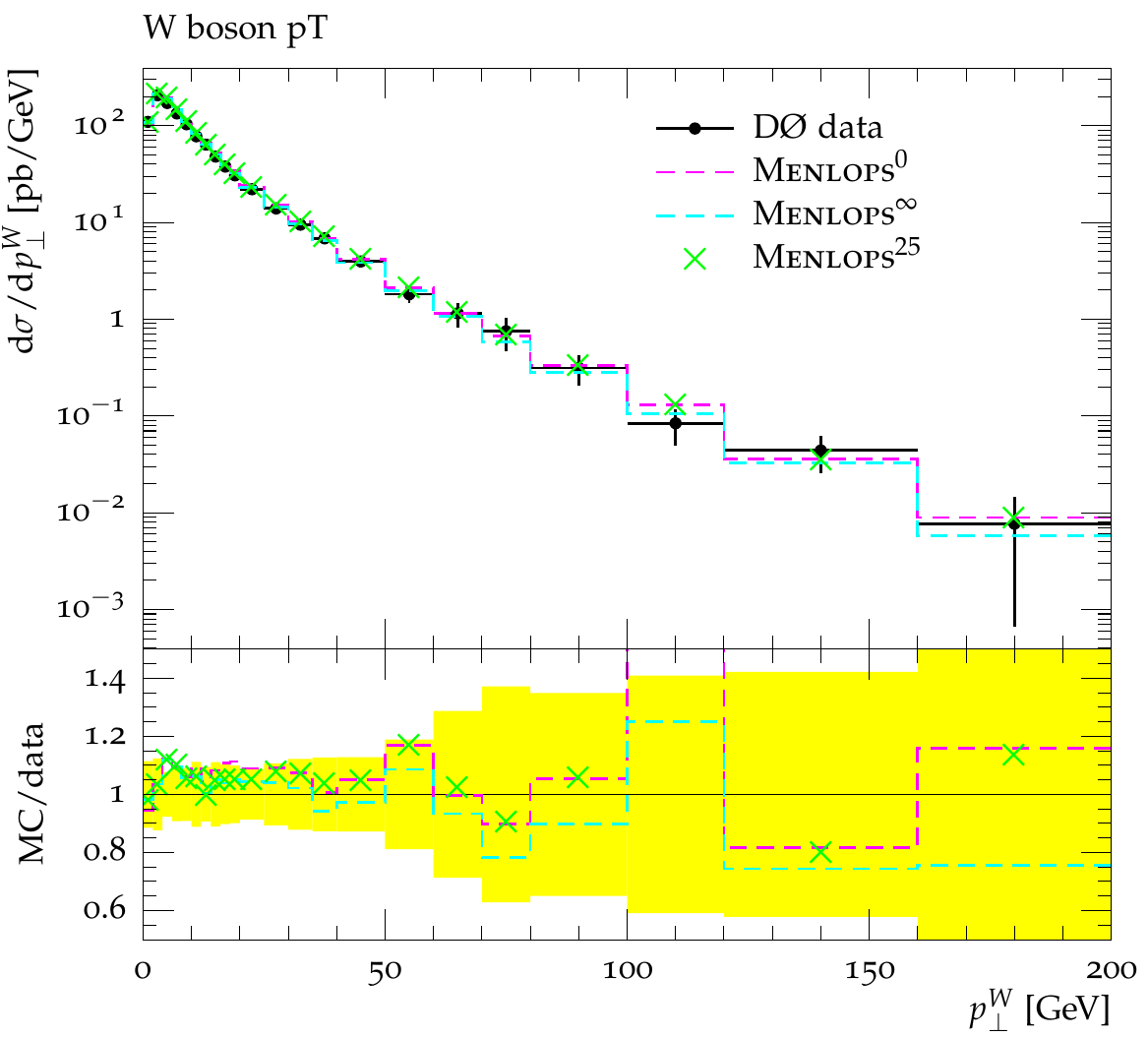}
\par\end{centering}

\caption{The data here correspond to a D\O~measurement of the W boson $p_{{\scriptscriptstyle \mathrm{T}}}$
spectrum, in the electron channel \cite{Abbott:2000xv}. Superimposed
on the left are results from \noun{Powheg Nlops }simulations of W
production and W plus jet production, as well as a \noun{Menlops }prediction.
The \noun{Menlops$^{\text{25}}$} event sample was formed by merging
\noun{Menlops$^{\text{0}}$ }and \noun{Menlops$^{\text{\ensuremath{\infty}}}$}
simulations at an underlying Born $p_{{\scriptscriptstyle \mathrm{T}}}$
of $25\,\mathrm{GeV}$. On the right we show the maximum variation
induced in the \noun{Menlops} predictions by changing the merging
scale. }

\label{fig:tevatron_W_enue_pT} 
\end{figure}
\begin{figure}[H]
\begin{centering}
\includegraphics[width=0.4\textwidth]{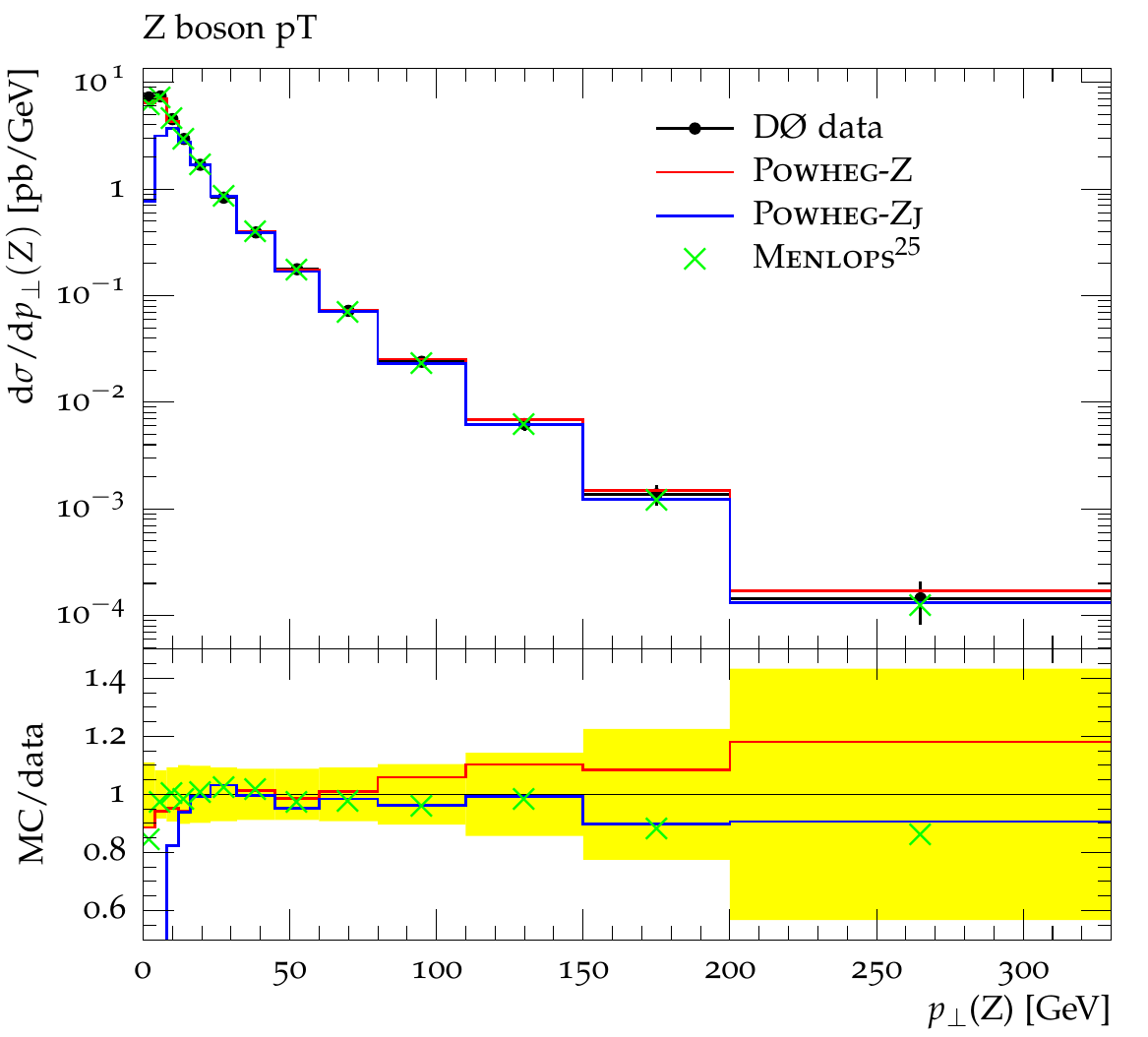}\hfill{}\includegraphics[width=0.4\textwidth]{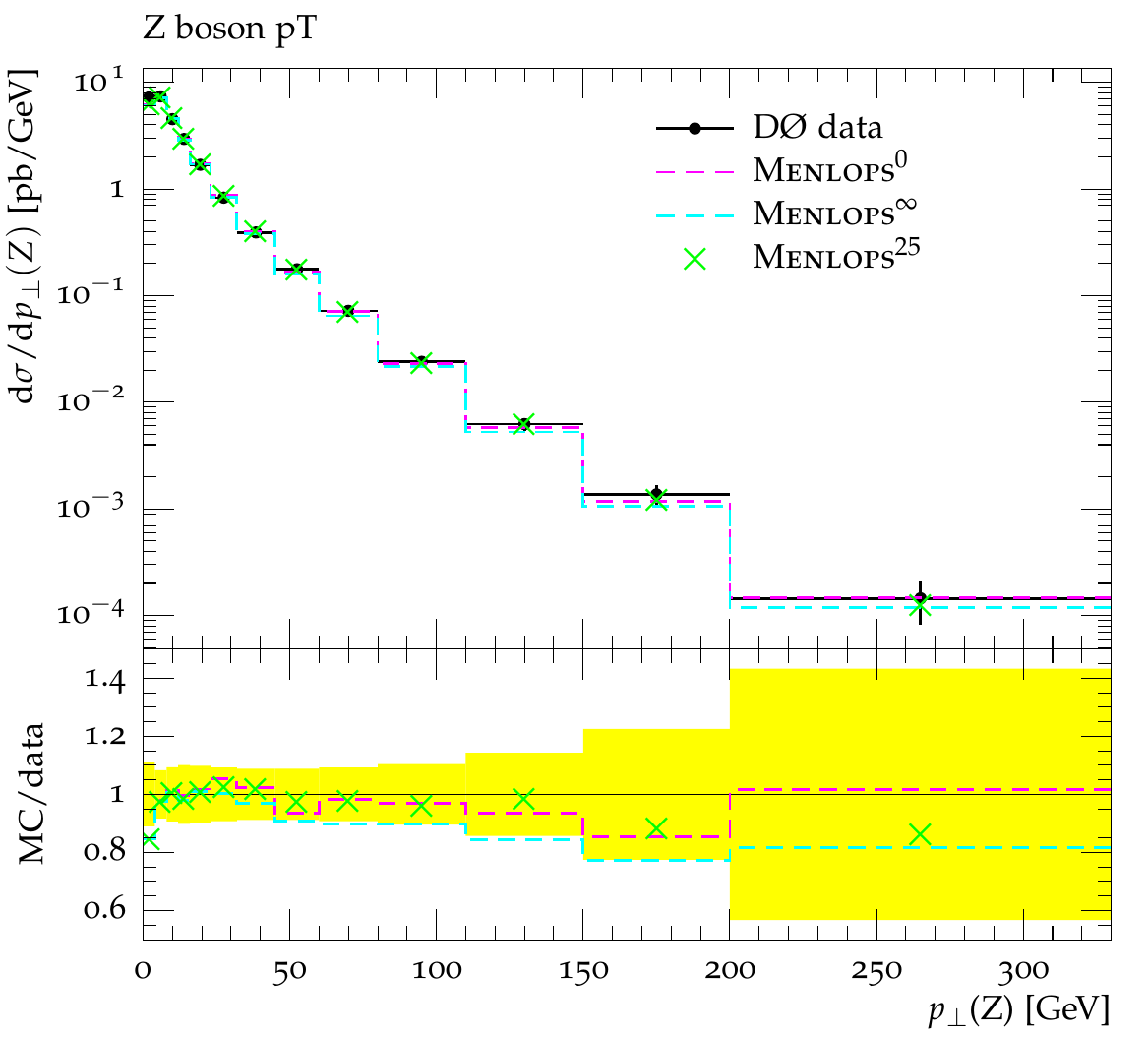}
\par\end{centering}

\caption{On the left we compare \noun{Powheg }simulations of vector boson production
and vector boson plus jet production, as well as a \noun{Menlops }prediction,
to a recent D\O~measurement of the Z boson $p_{{\scriptscriptstyle \mathrm{T}}}$
spectrum, in the dimuon channel \cite{Abazov:2010kn}. The \noun{Menlops}
event sample was formed by merging \noun{Menlops$^{\text{0}}$ }and
\noun{Menlops$^{\text{\ensuremath{\infty}}}$} samples at an underlying
Born $p_{{\scriptscriptstyle \mathrm{T}}}$ of $25\,\mathrm{GeV}$.
The right-hand plot shows the maximum variation in \noun{Menlops}
predictions associated to changing the merging scale. The yellow band
signifies the projection of the errors on the reference data --- the
first entry in the legend --- into the ratio MC / data.}

\label{fig:tevatron_Z_mm_pT} 
\end{figure}

In figures \ref{fig:tevatron_W_enue_pT} and \ref{fig:tevatron_Z_mm_pT}
we show the transverse momentum spectra of W and Z bosons as measured
by the Tevatron D\O~collaboration in Refs.~\cite{Abbott:2000xv}
and \cite{Abazov:2010kn} respectively. In all cases the agreement
of the various Monte Carlo predictions is quite pleasing, the only
exception being the \noun{Powheg-Wj }and \noun{Powheg-Zj }programs'
failure to describe the low $p_{{\scriptscriptstyle \mathrm{T}}}$
region. As can be seen on the left-hand side of each of the figures,
this problem is remedied well in the \noun{Menlops$^{\text{0}}$ }enhanced
versions, which include the appropriate resummation of large Sudakov
logarithms there (magenta lines). Once again the agreement between
the merged \noun{Menlops }prediction (green crosses) and those of
its two component samples is very good.

As with the case of the \noun{Menlops$^{\text{0}}$ }result for the
Z boson rapidity spectrum, the fact that the \noun{Menlops$^{\text{\ensuremath{\infty}}}$
}simulation replicates the high-$p_{{\scriptscriptstyle \mathrm{T}}}$
behavior of the \noun{Powheg-Vj }and other \noun{Menlops} predictions
in this region is noteworthy, since its description there is only
partially next-to-leading order --- through the inclusion of the double
real emission matrix elements and associated Sudakov suppression ---
the others include full NLO corrections to this observable throughout
the high $p_{{\scriptscriptstyle \mathrm{T}}}$ region. While it is
the case that the high-$p_{{\scriptscriptstyle \mathrm{T}}}$ description
of the \noun{Menlops$^{\text{\ensuremath{\infty}}}$} program appears
in good agreement with its fully NLO counterparts, one expects that
it exhibits a little more renormalization and factorization scale
sensitivity there.

Here the first main point is that the \noun{Menlops }results give
an improved description over the full $p_{{\scriptscriptstyle \mathrm{T}}}$
range with respect to their unenhanced versions, in particular the
description of the low-$p_{{\scriptscriptstyle \mathrm{T}}}$ region
is always physical and the predictions exhibit no more variation among
the improved programs than in the originals (which, for this observable,
was already quite mild). The second important point is that the merged
\noun{Menlops$^{\text{25}}$ }sample can be considered to give a genuine
NLO+NLL accurate prediction all through the spectrum and that here
the sensitivity of the prediction to the exact choice of merging scale
in a `real-life' application should be practically negligible.

\begin{figure}[H]
\begin{centering}
\includegraphics[width=0.4\textwidth]{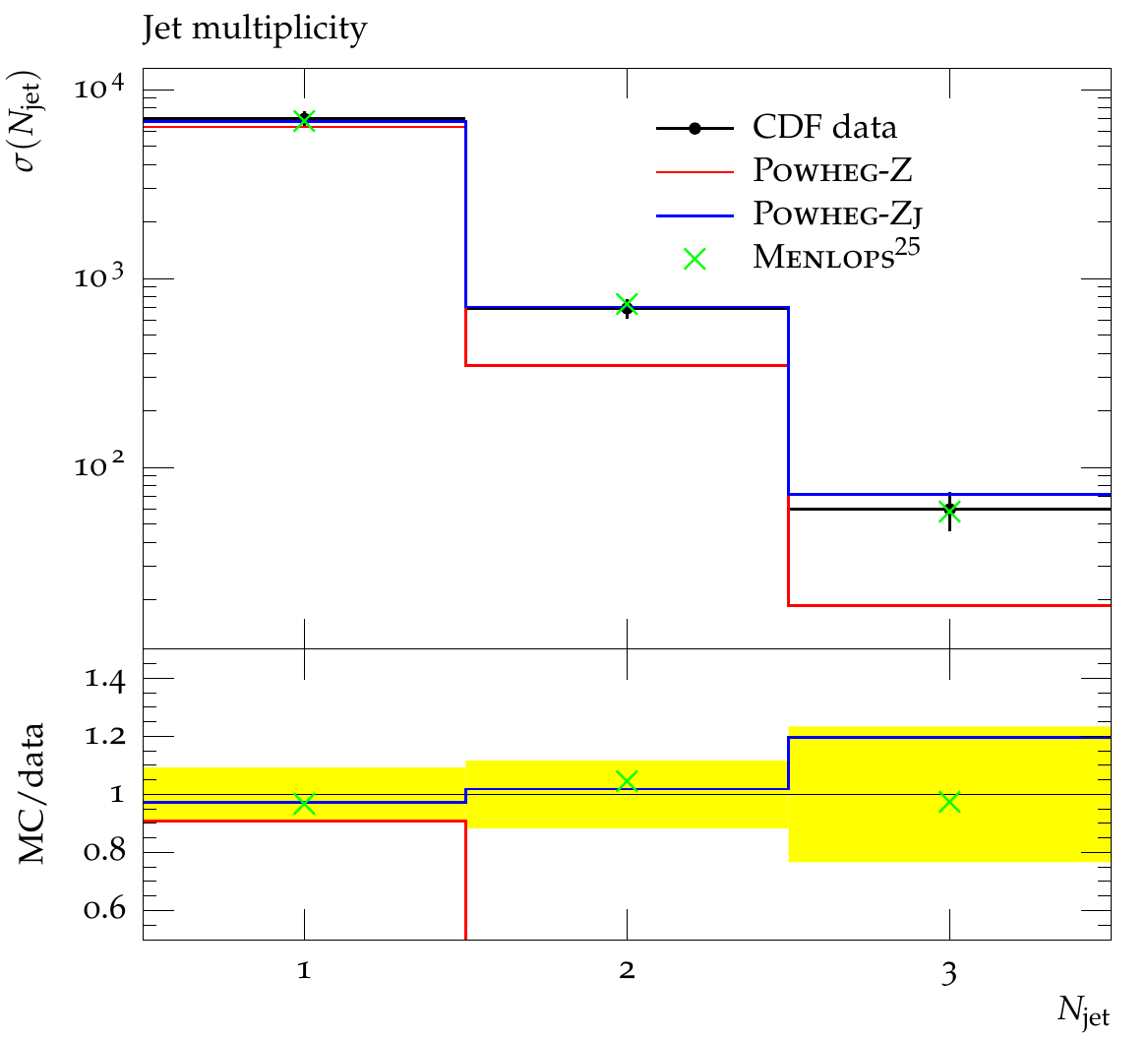}\hfill{}\includegraphics[width=0.4\textwidth]{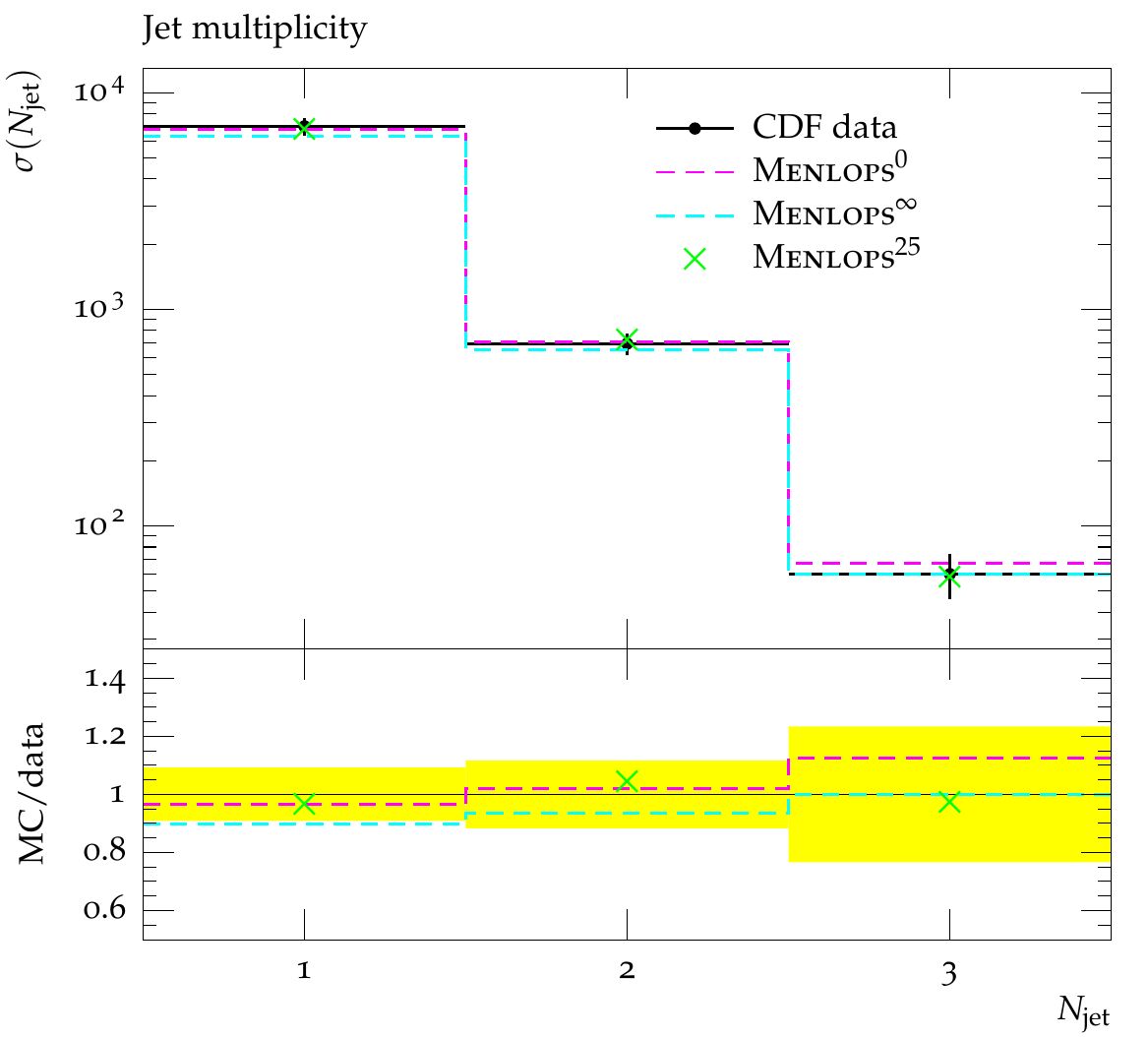}
\par\end{centering}

~

\begin{centering}
\includegraphics[width=0.4\textwidth]{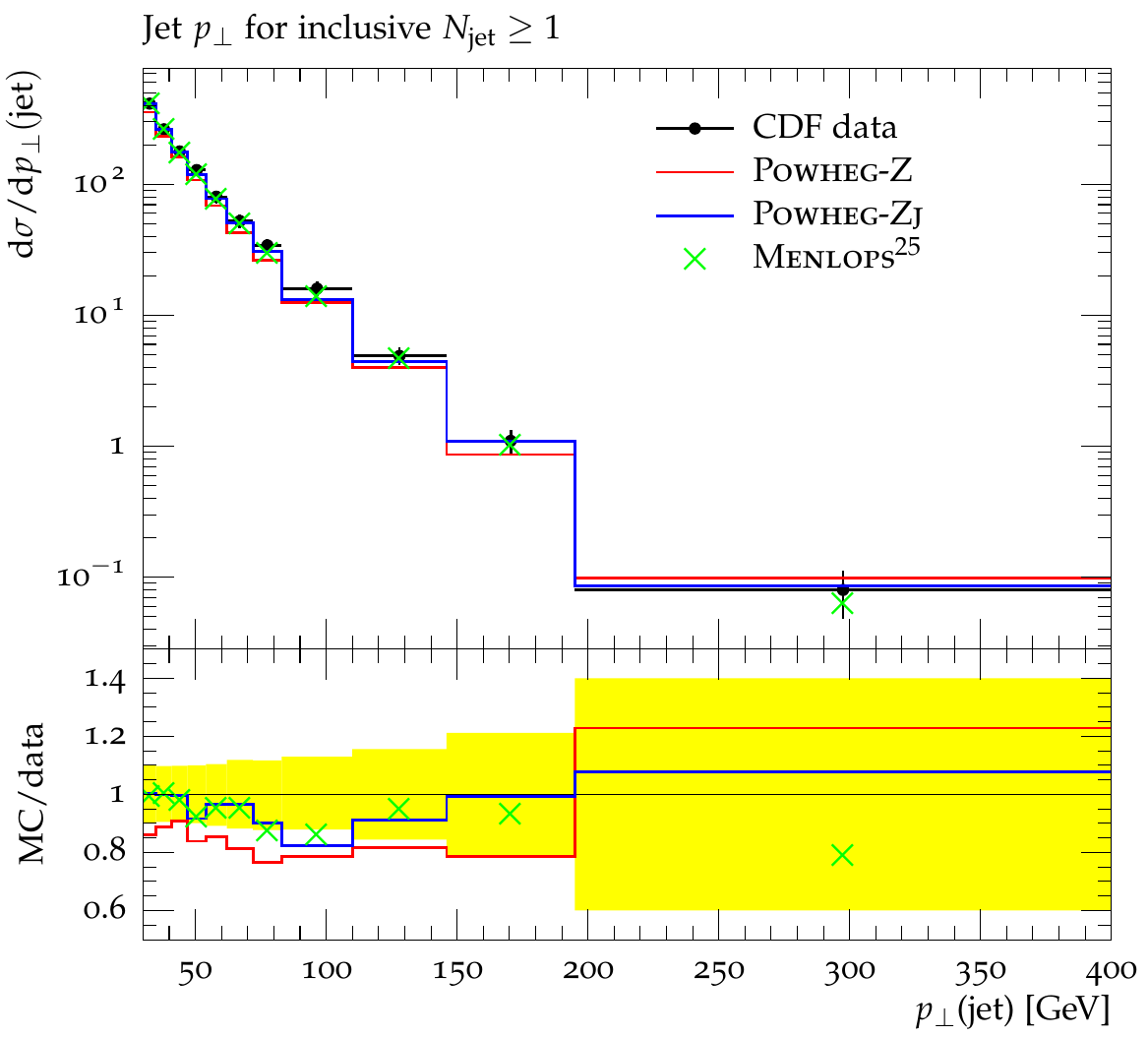}\hfill{}\includegraphics[width=0.4\textwidth]{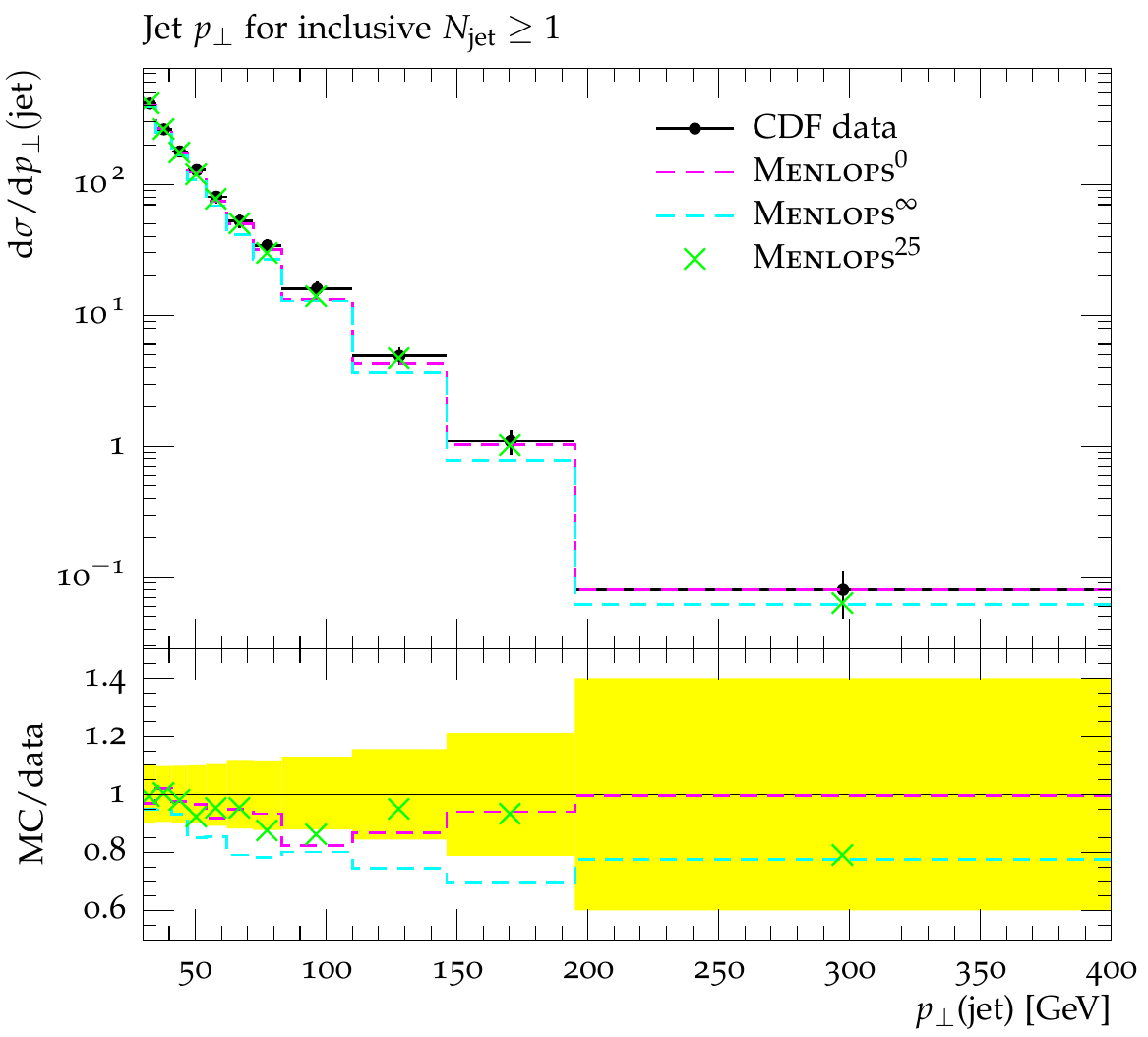}
\par\end{centering}

~

\begin{centering}
\includegraphics[width=0.4\textwidth]{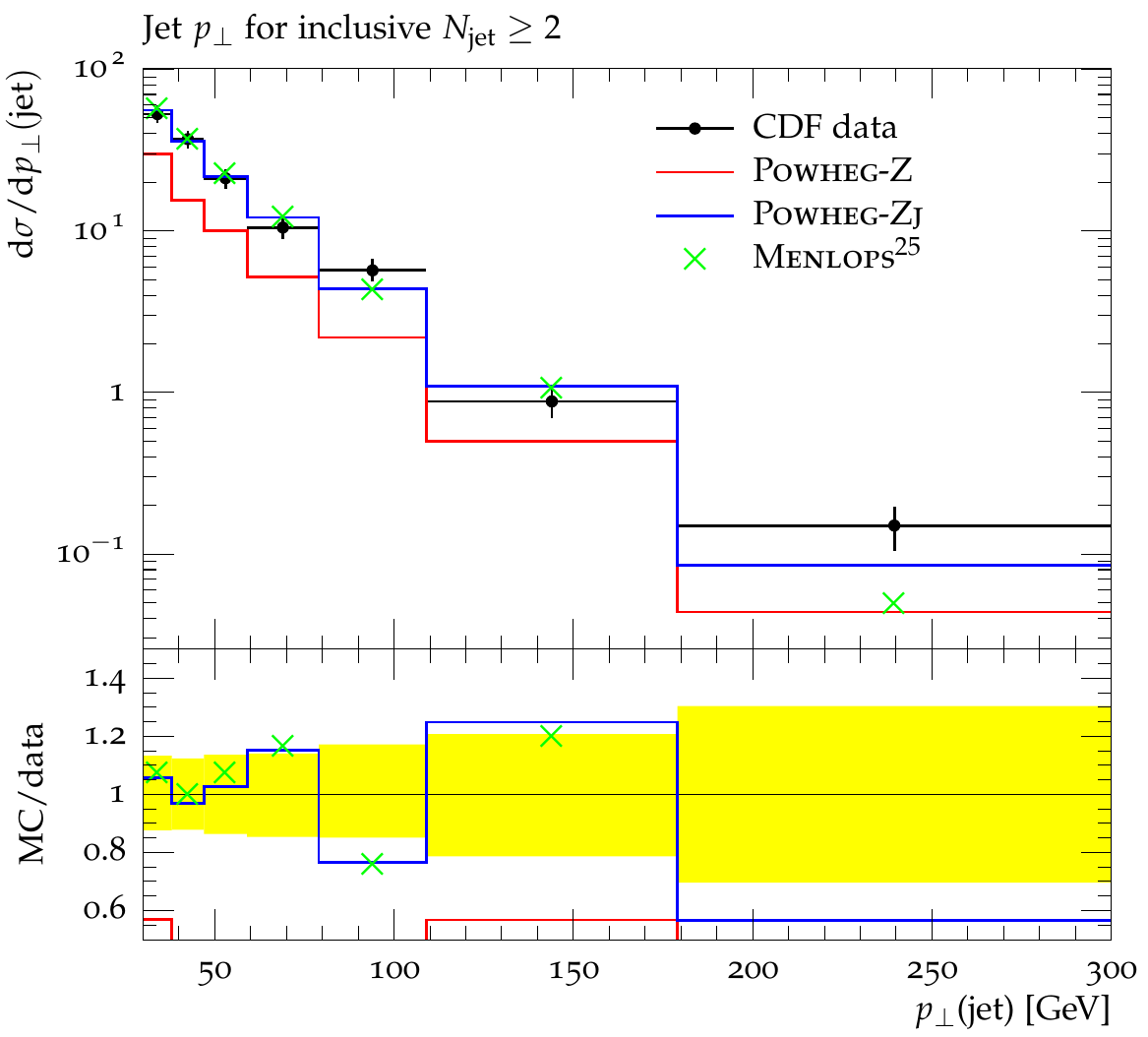}\hfill{}\includegraphics[width=0.4\textwidth]{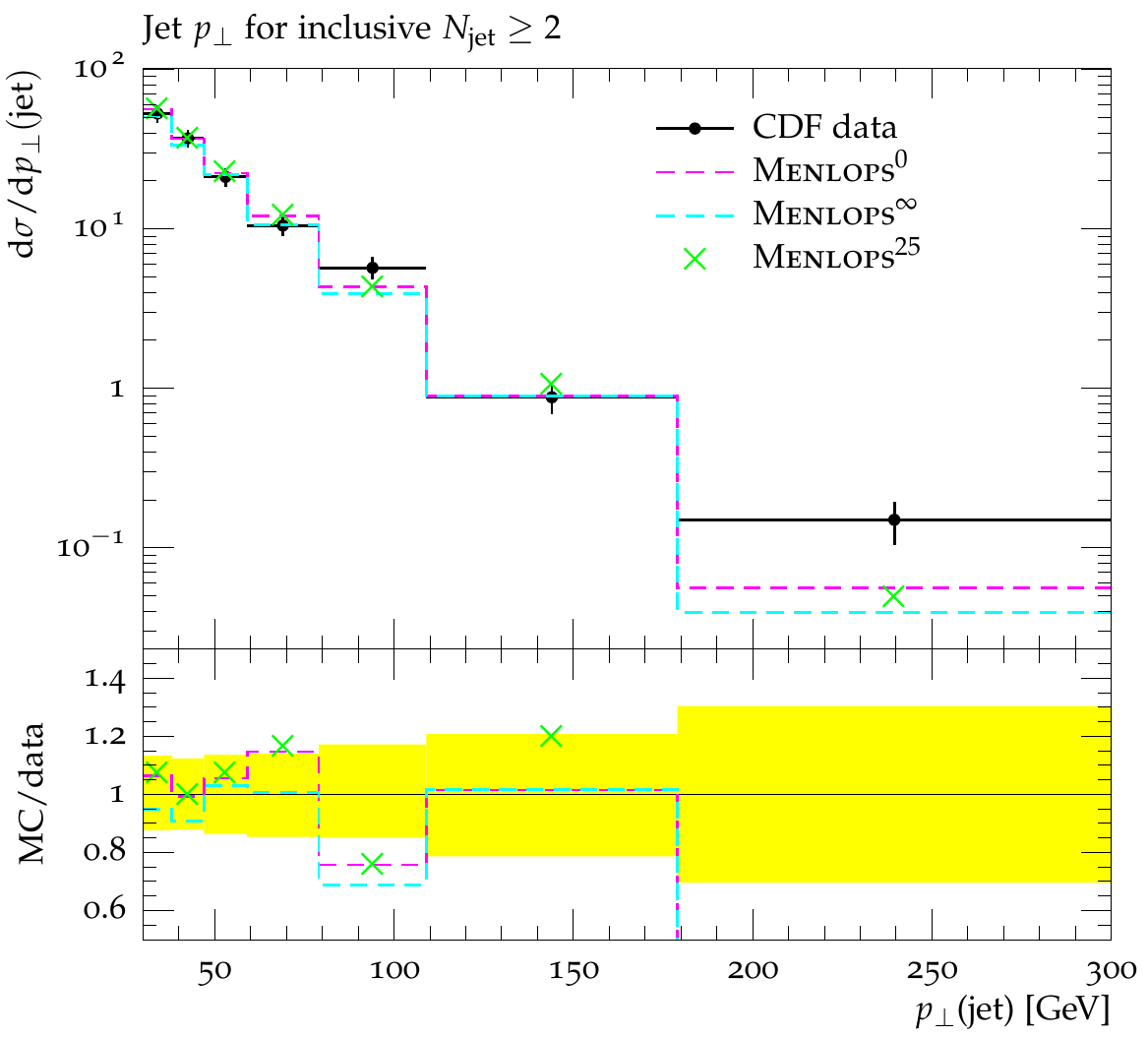}
\par\end{centering}

\caption{In the left-hand column we compare \noun{Powheg}-Z, \noun{Powheg-Zj}
and \noun{Menlops }predictions to CDF measurements of jet multiplicities
and $p_{{\scriptscriptstyle \mathrm{T}}}$ spectra in the dielectron
decay channel \cite{AaltonenT:2007cp}. On the right we repeat the
exercise showing the full range obtained by varying the \noun{Menlops
}merging scale described in Sect.~\ref{sub:v-vj_menlops} from zero
to infinity. }

\label{fig:tevatron_Z_ee_jet_rates_and_pTs} 
\end{figure}

Figure \ref{fig:tevatron_Z_ee_jet_rates_and_pTs} displays CDF measurements
of the 1-, 2- and 3-jet cross sections, as well as inclusive jet $p_{{\scriptscriptstyle \mathrm{T}}}$
spectra underneath. The various Monte Carlo simulation predictions
are relatively straightforward to understand in terms of their constituent
matrix elements. Since all of the observables demand the presence
of at least one jet, the \noun{Powheg-Z }predictions obviously suffer
with respect to all of the others, since it includes only a tree-level
single emission matrix element for Z production, with additional radiation
being generated according to the parton shower approximation. In particular,
the \noun{Powheg-Z} predictions have a characteristically softer radiation
pattern associated with them, manifested here as diminished multi-jet
cross sections and inclusive jet $p_{{\scriptscriptstyle \mathrm{T}}}$
spectra. It is then pleasing to see that the \noun{Menlops$^{\text{\ensuremath{\infty}\ }}$
}improvement (Sect.~\ref{sub:v_menlops}), which generates a secondary
emission, according to the exact double emission matrix element, from
the \noun{Powheg-Z }single emission configurations, gives a set of
predictions in better agreement with experimental data and the other
approaches.\noun{ }On a related note, one can also see the general
trend persisting from the previous figures, whereby the \noun{Menlops-}improved\noun{
}results are in closer agreement with one another than the underlying
\noun{Powheg} programs. 

While the general agreement of the improved approaches with the data
in Fig.~\ref{fig:tevatron_Z_ee_jet_rates_and_pTs} is good, clearly,
in regards to the analysis carried out there, all of the \noun{Menlops
}approaches offer no practical improvements over the description afforded
by the \noun{Powheg-Zj} program; the low $p_{{\scriptscriptstyle \mathrm{T}}}$
region is excluded by the nature of the observables, hence, there
is no need for Sudakov resummation (provided optimal scale choices
are made) or NLO corrections to the underlying inclusive Z production
process. Thus, with this unique application in mind, the advantages
of \noun{Menlops }event samples are rendered moot, only becoming useful
when viewed as a tool to perform many different analyses.Alternatively,
more importantly, when Z production is being considered as a background
rather than a signal process, the distinction between the inclusive
and associated Z plus jet production processes is not often clear
and generally applicable tools are required. It follows that the ultimate
practical significance of these plots (and in fact all plots in this
section) is in verifying the ability of the \noun{Menlops }approach
to capture the best description from the \noun{Nlops }simulations
throughout phase space and to show that the sensitivity of merged
\noun{Menlops }predictions to the merging scale, $p_{{\scriptscriptstyle \mathrm{T}}}^{{\scriptscriptstyle \mathrm{merge}}}$,
is remarkably mild. In particular, comparing the left- and right-hand
columns, that the merging scale sensitivity is minimized when the
component sub-samples are made from \noun{Menlops} improved\noun{
}programs as opposed to unimproved ones.

\begin{figure}[H]
\begin{centering}
\includegraphics[width=0.4\textwidth]{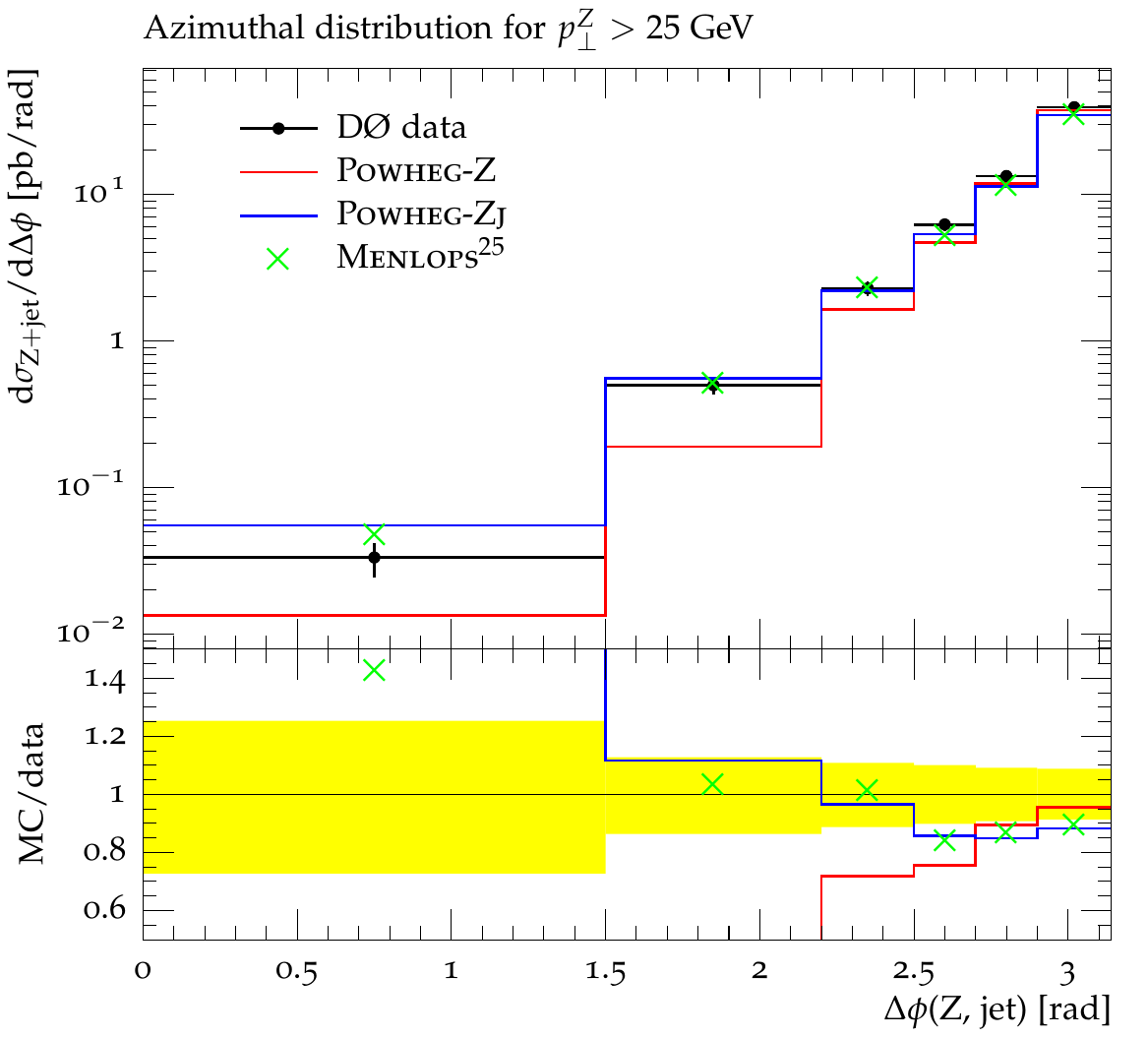}\hfill{}\includegraphics[width=0.4\textwidth]{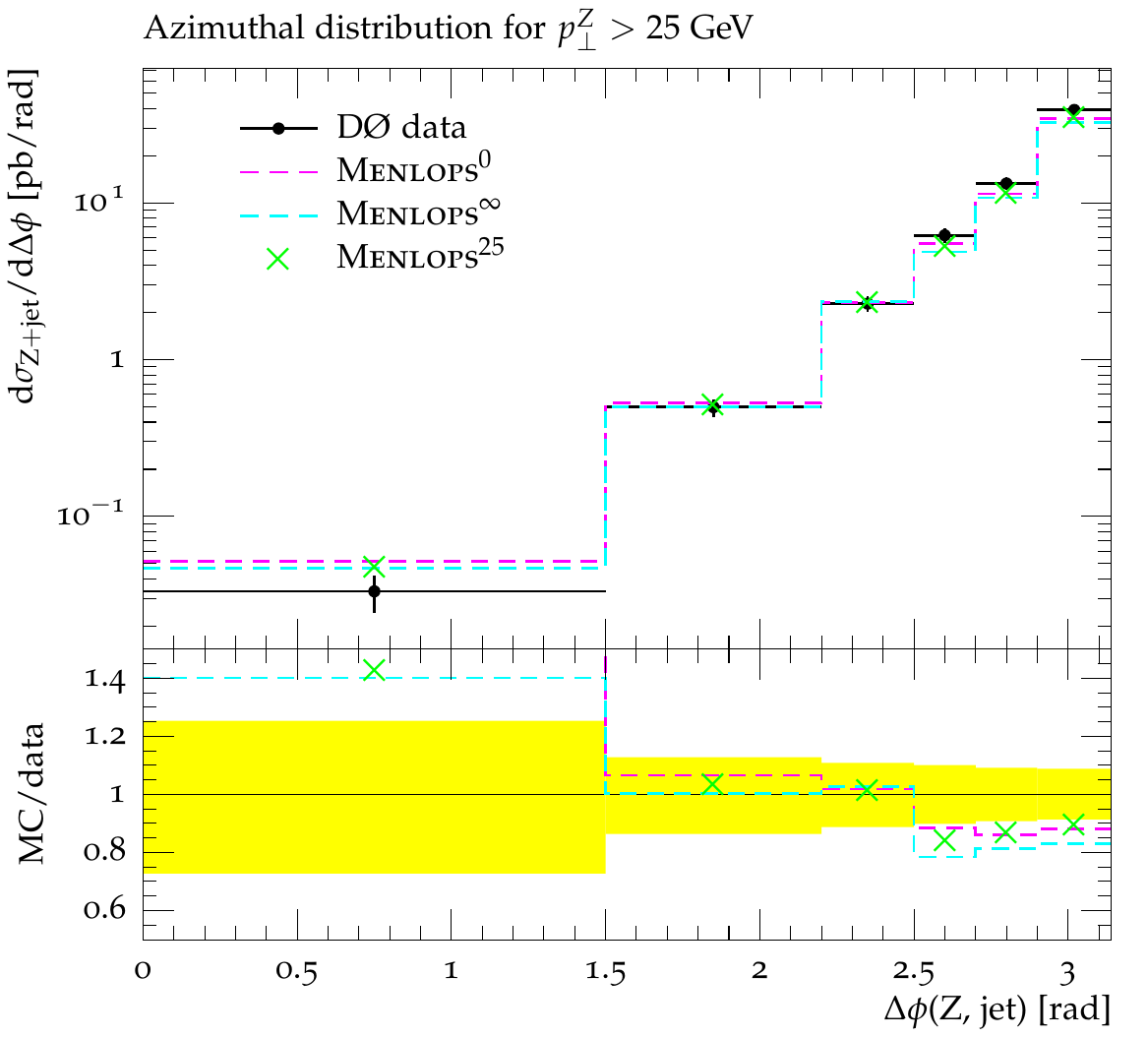}
\par\end{centering}

\begin{centering}
\includegraphics[width=0.4\textwidth]{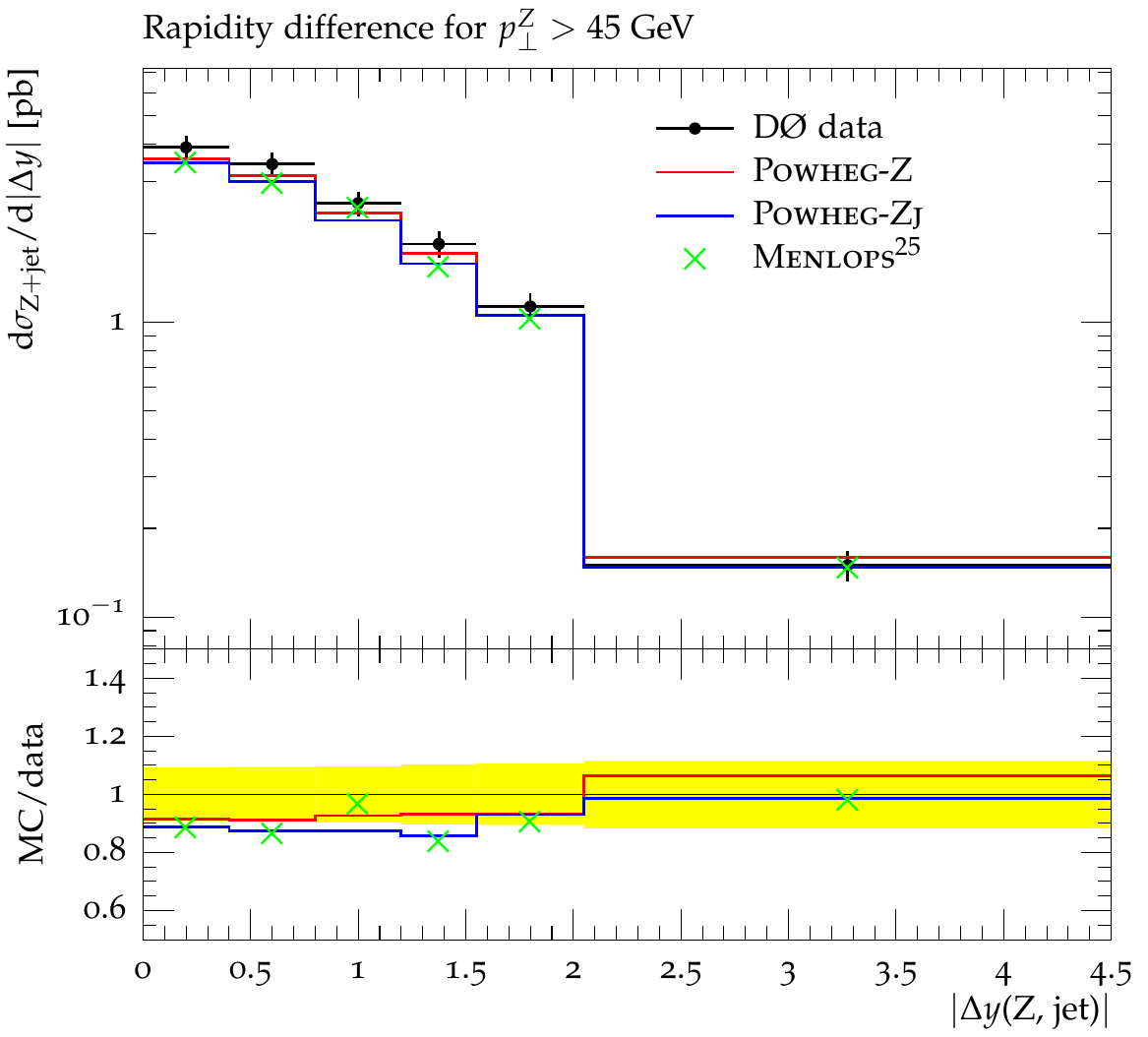}\hfill{}\includegraphics[width=0.4\textwidth]{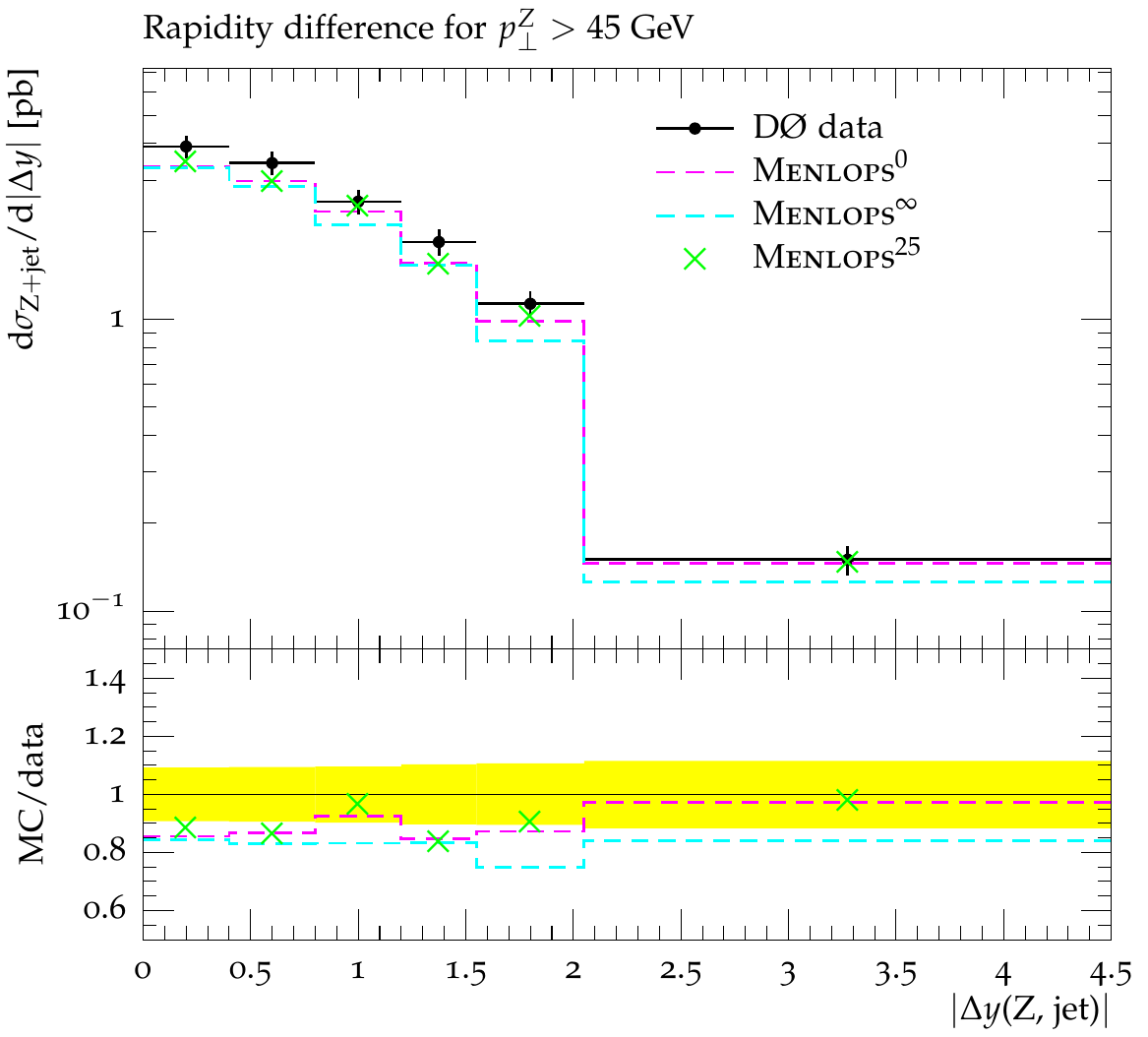}
\par\end{centering}

\caption{Here we compare \noun{Powheg-Z, Powheg-Zj }and \noun{Menlops$^{\text{25}}$}
predictions to Tevatron D\O~ collaboration measurements \cite{Abazov:2009pp}
of angular observables in the muonic decay channel. In the upper pair
of plots we show the azimuthal separation of the Z boson and leading
jet, while the lower plots show their rapidity separation. The colouring
of the various predictions is as in the previous figures.}

\label{fig:tevatron_Z_mm_angular_dists} 
\end{figure}
Displayed at the top of figure \ref{fig:tevatron_Z_mm_angular_dists}
are distributions of the azimuthal separation of the Z boson and leading
jet (above) in the muonic decay channel at the Tevatron \cite{Abazov:2009pp},
as well as the corresponding rapidity separation (below). Here the
description of the data is by-and-large fair in all of the approaches,
the most prominent source of discrepancy being in the region corresponding
to a large azimuthal separation of the leading jet and the Z boson
in the \noun{Powheg-Z }case. Again, as in the discussion surrounding
Fig.~\ref{fig:tevatron_Z_ee_jet_rates_and_pTs}, this is somewhat
expected due to the fact that this simulation resorts to a soft-collinear
parton shower approximation to generate events there, while the other
methods make use of the tree-level double emission matrix element.
In connection with this, one can again see a tendency for the \noun{Menlops
}predictions (right) to be in closer correspondence with one another
than the unimproved \noun{Powheg }ones (left), re-asserting that the
$p_{{\scriptscriptstyle \mathrm{T}}}^{{\scriptscriptstyle \mathrm{merge}}}$
dependence is minimized by building the merged sample from \noun{Menlops
}improved \noun{Powheg }simulations rather than unimproved ones.

\subsection{Comparison with LHC data\label{sub:LHC_comparison}}

\begin{figure}[H]
\begin{centering}
\includegraphics[width=0.4\textwidth]{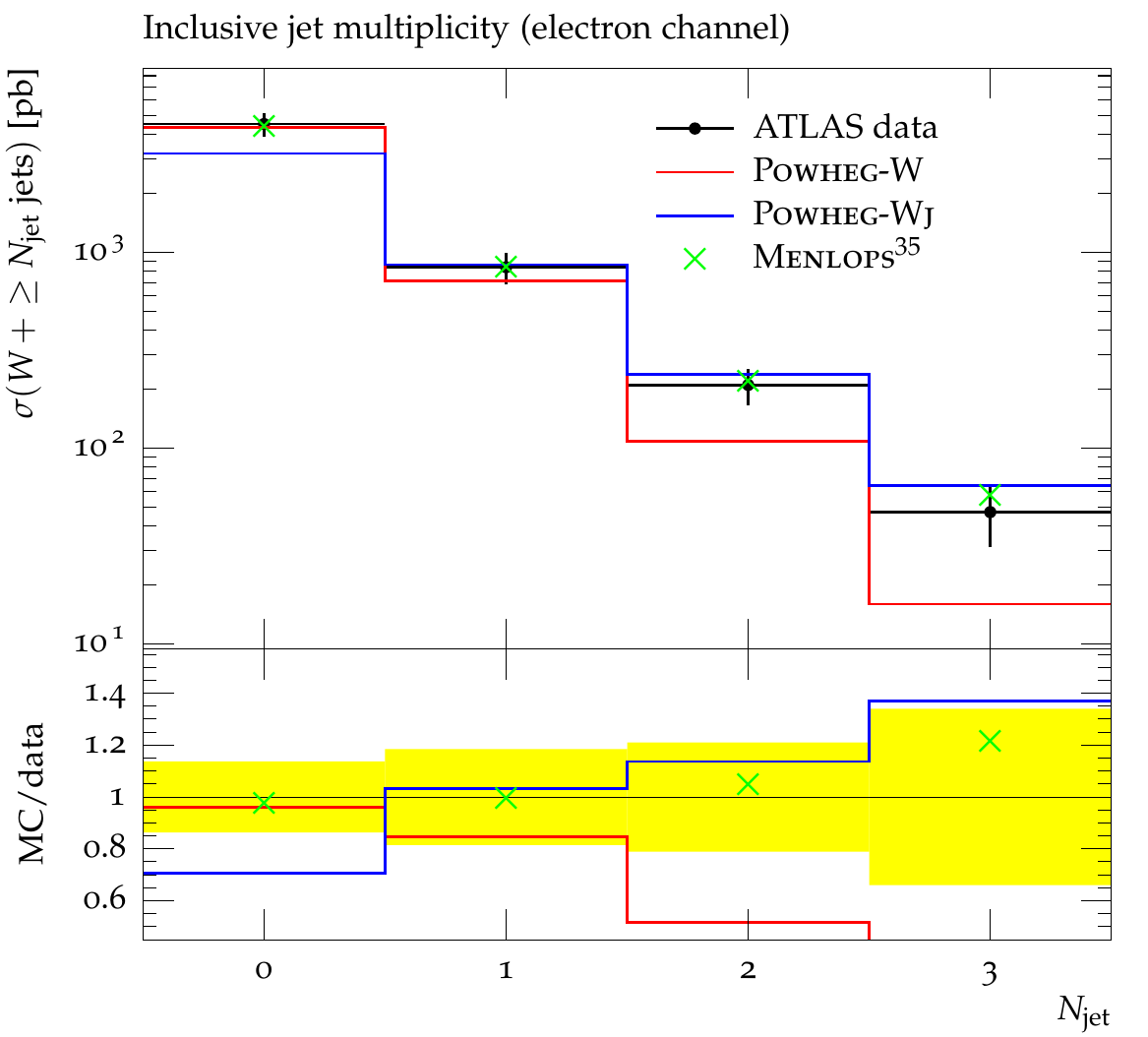}\hfill{}\includegraphics[width=0.4\textwidth]{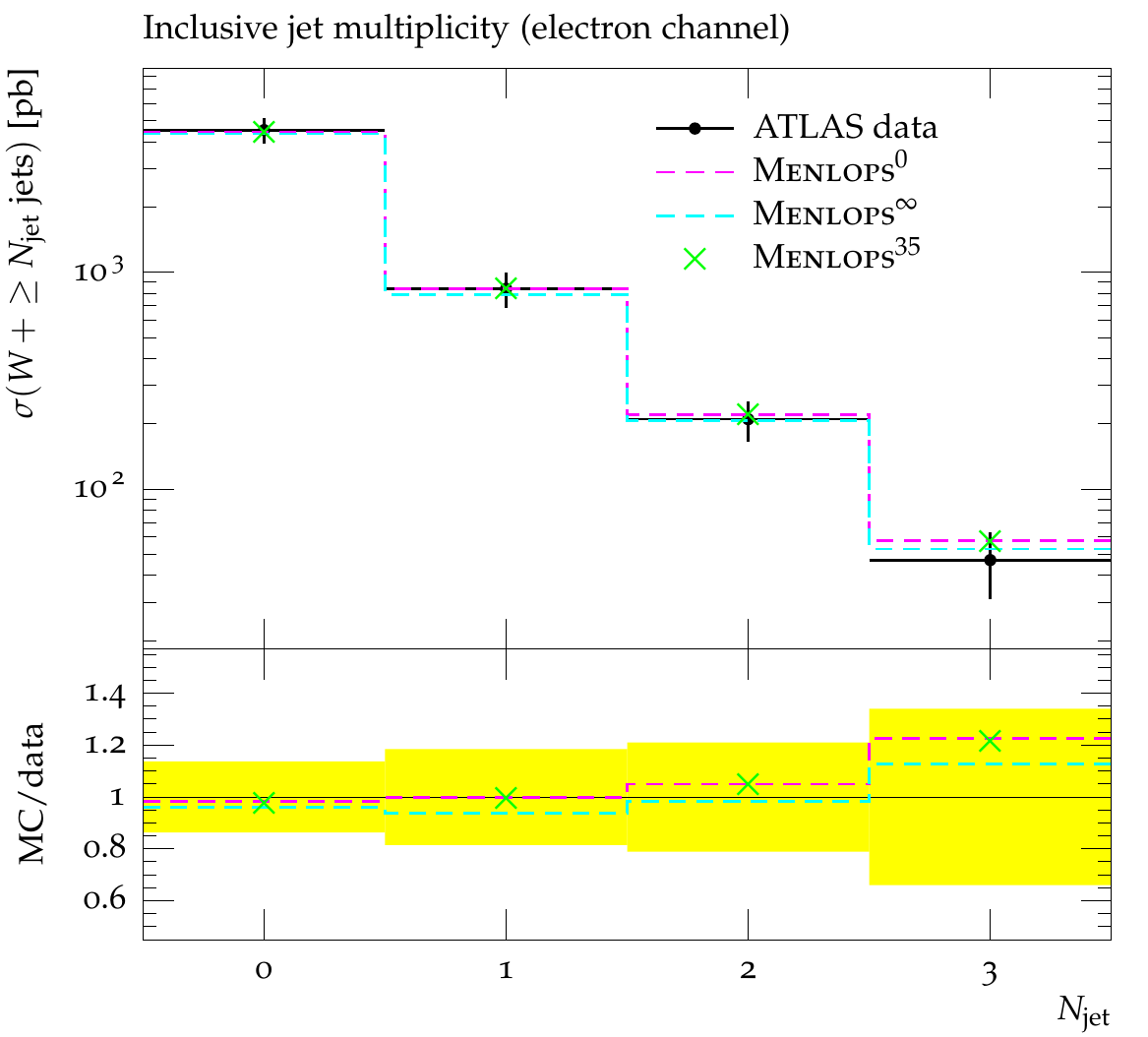}
\par\end{centering}

\begin{centering}
\includegraphics[width=0.4\textwidth]{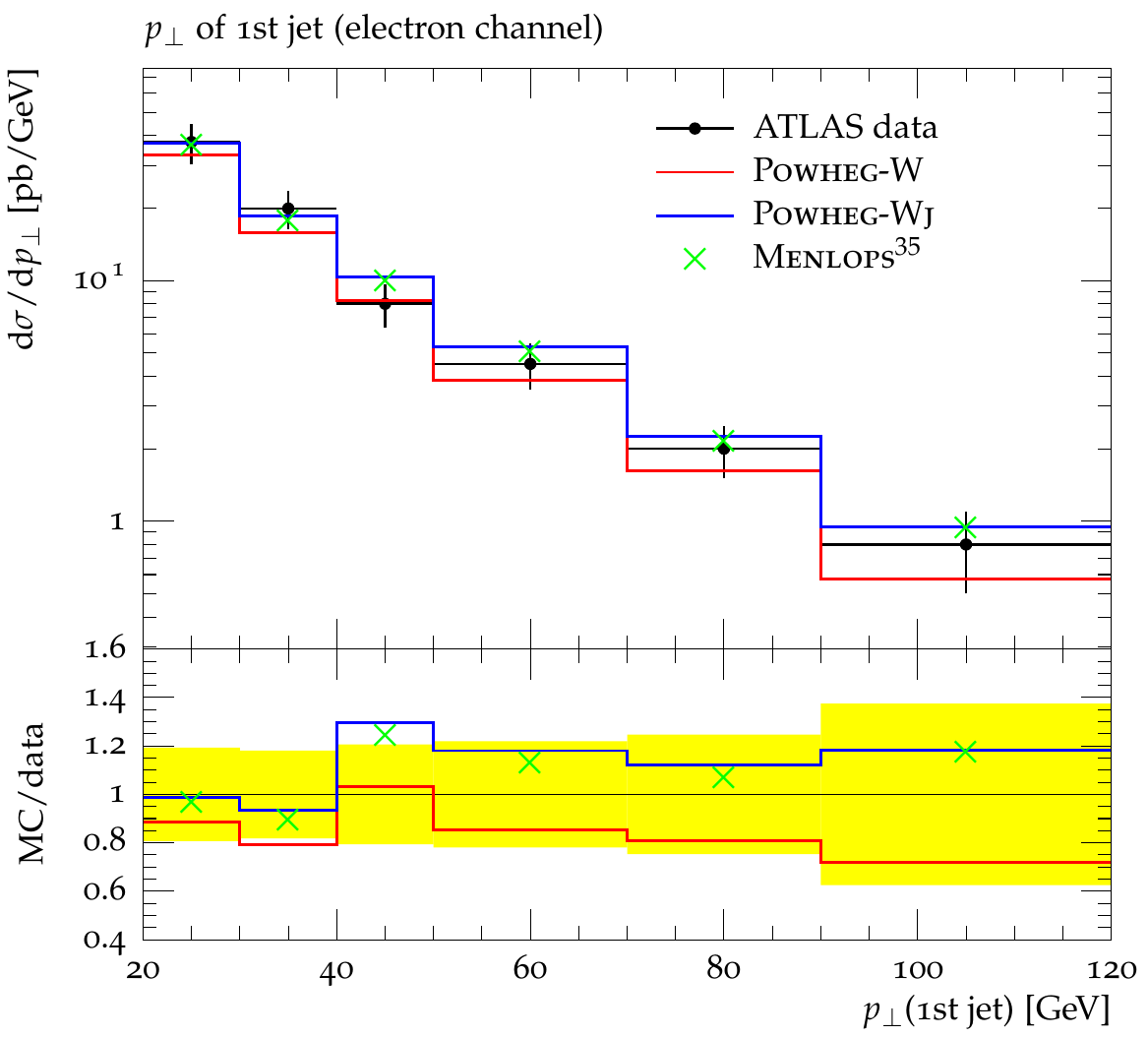}\hfill{}\includegraphics[width=0.4\textwidth]{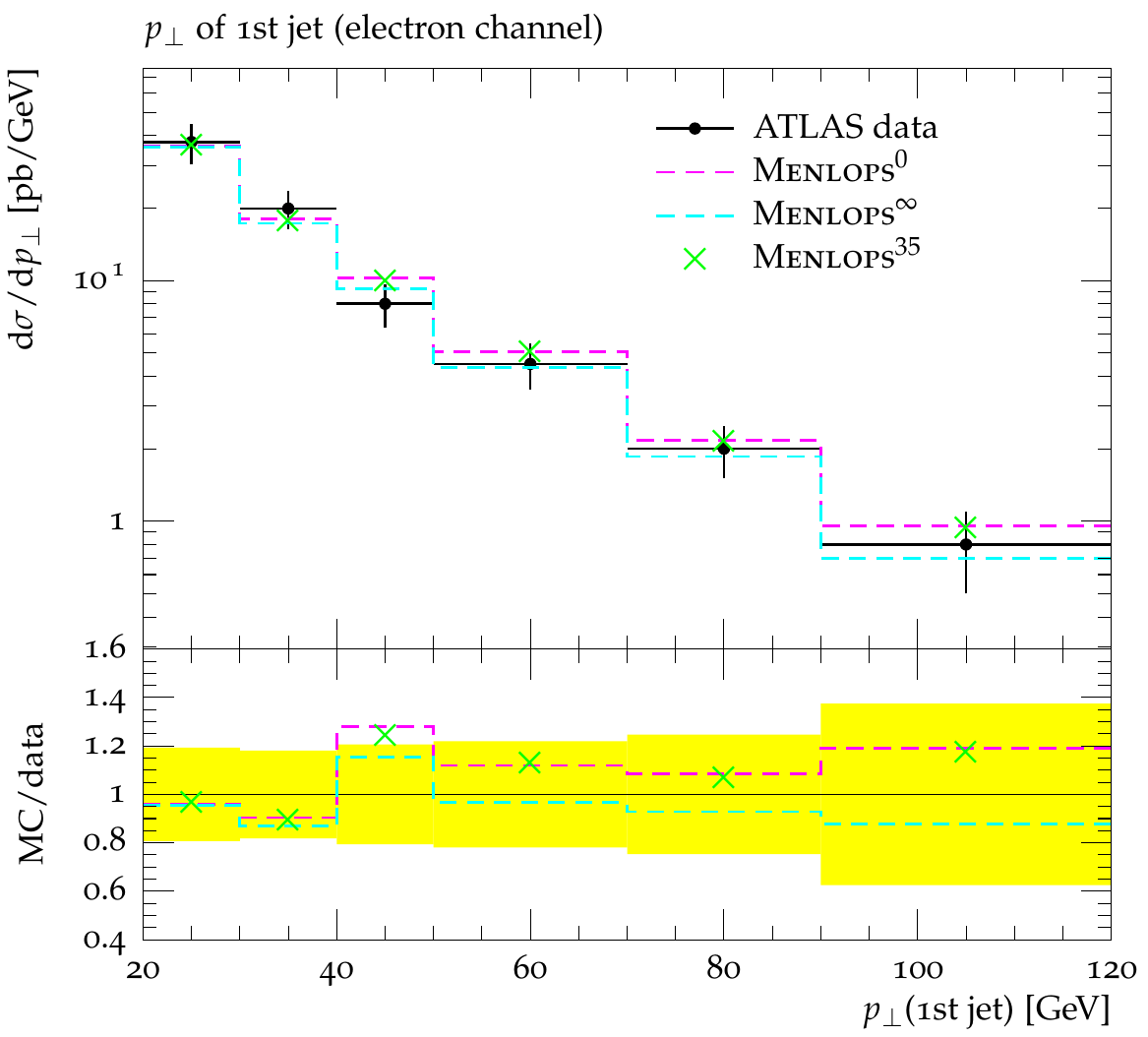}
\par\end{centering}

\begin{centering}
\includegraphics[width=0.4\textwidth]{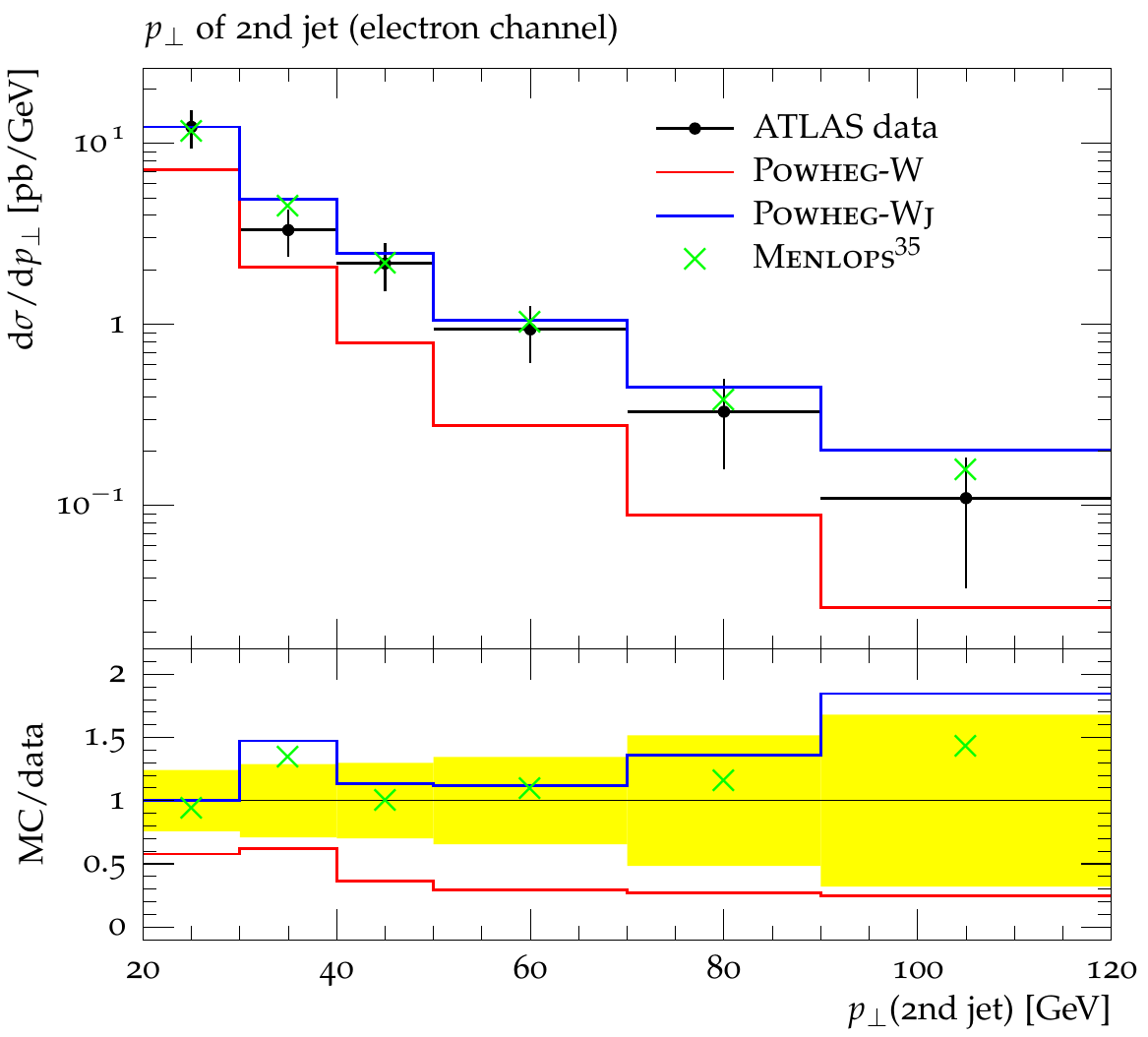}\hfill{}\includegraphics[width=0.4\textwidth]{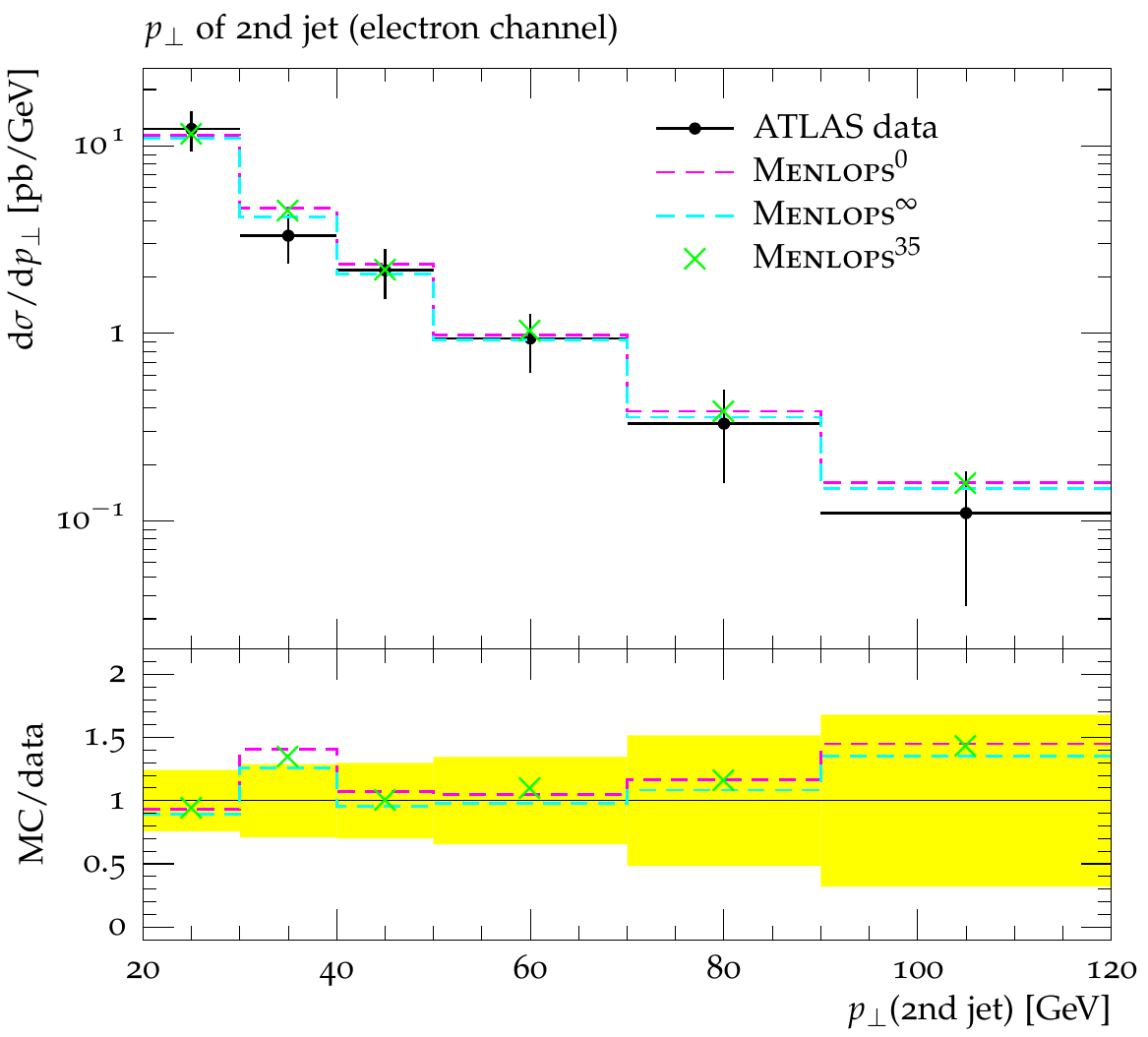}
\par\end{centering}

\caption{ \noun{Atlas} collaboration measurements \cite{atlas:2010pg} of jet
cross sections and jet $p_{{\scriptscriptstyle \mathrm{T}}}$ spectra
in the electron decay channel. On the left-hand side we compare \noun{Powheg-W,
Powheg-Wj }and merged \noun{Menlops$^{\text{35}}$} predictions. On
the right-hand side the merged \noun{Menlops$^{\text{35}}$} predictions
are shown alongside those obtained using the component samples it
derives from. The colouring of the different contributions is as in
previous figures in Sects.~\ref{sec:Method} and \ref{sec:Results}. }

\label{fig:lhc_W_jet_multiplicities_and_pTs_e_channel} 
\end{figure}

\begin{figure}[H]
\begin{centering}
\includegraphics[width=0.4\textwidth]{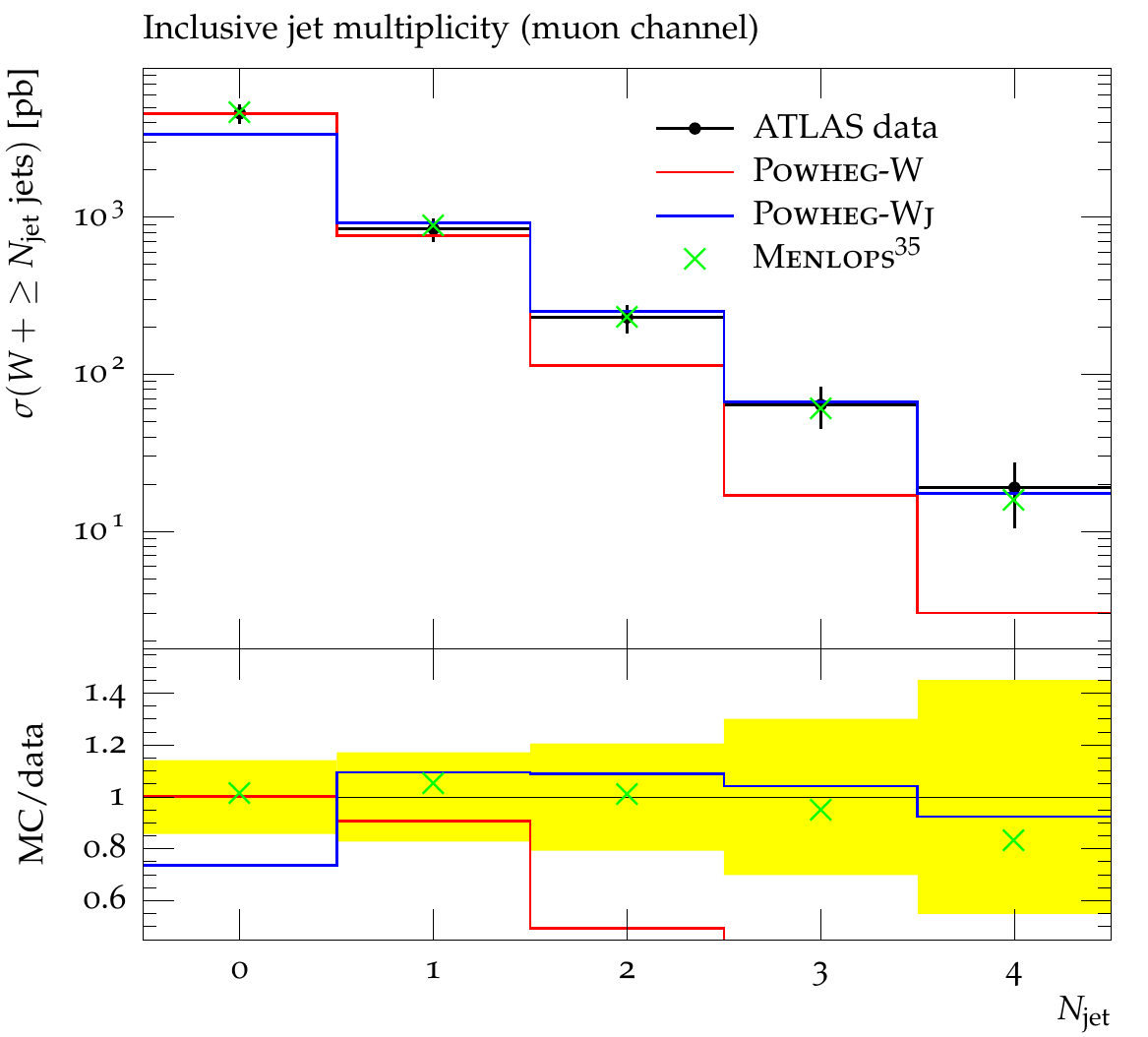}\hfill{}\includegraphics[width=0.4\textwidth]{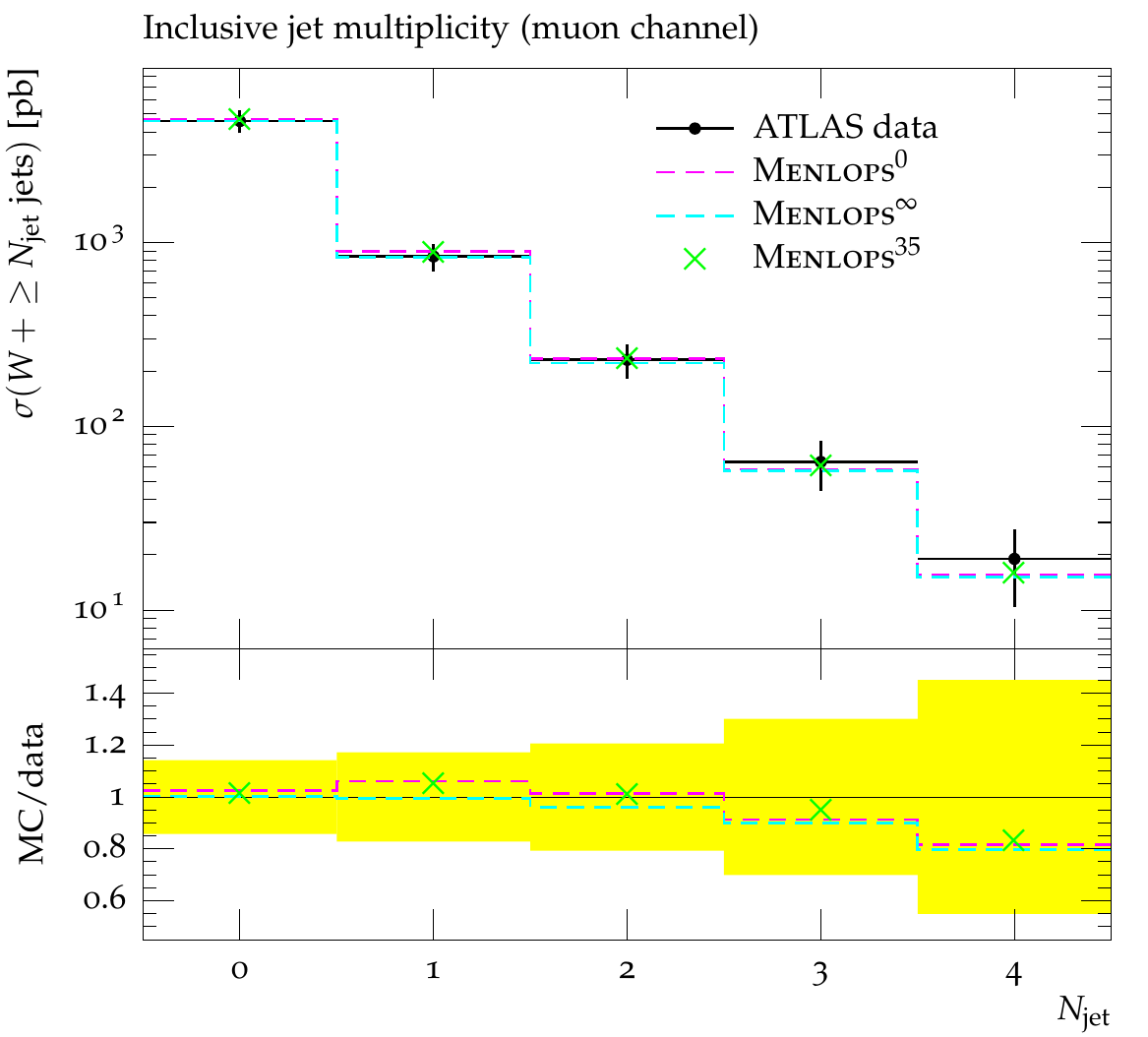}
\par\end{centering}

\begin{centering}
\includegraphics[width=0.4\textwidth]{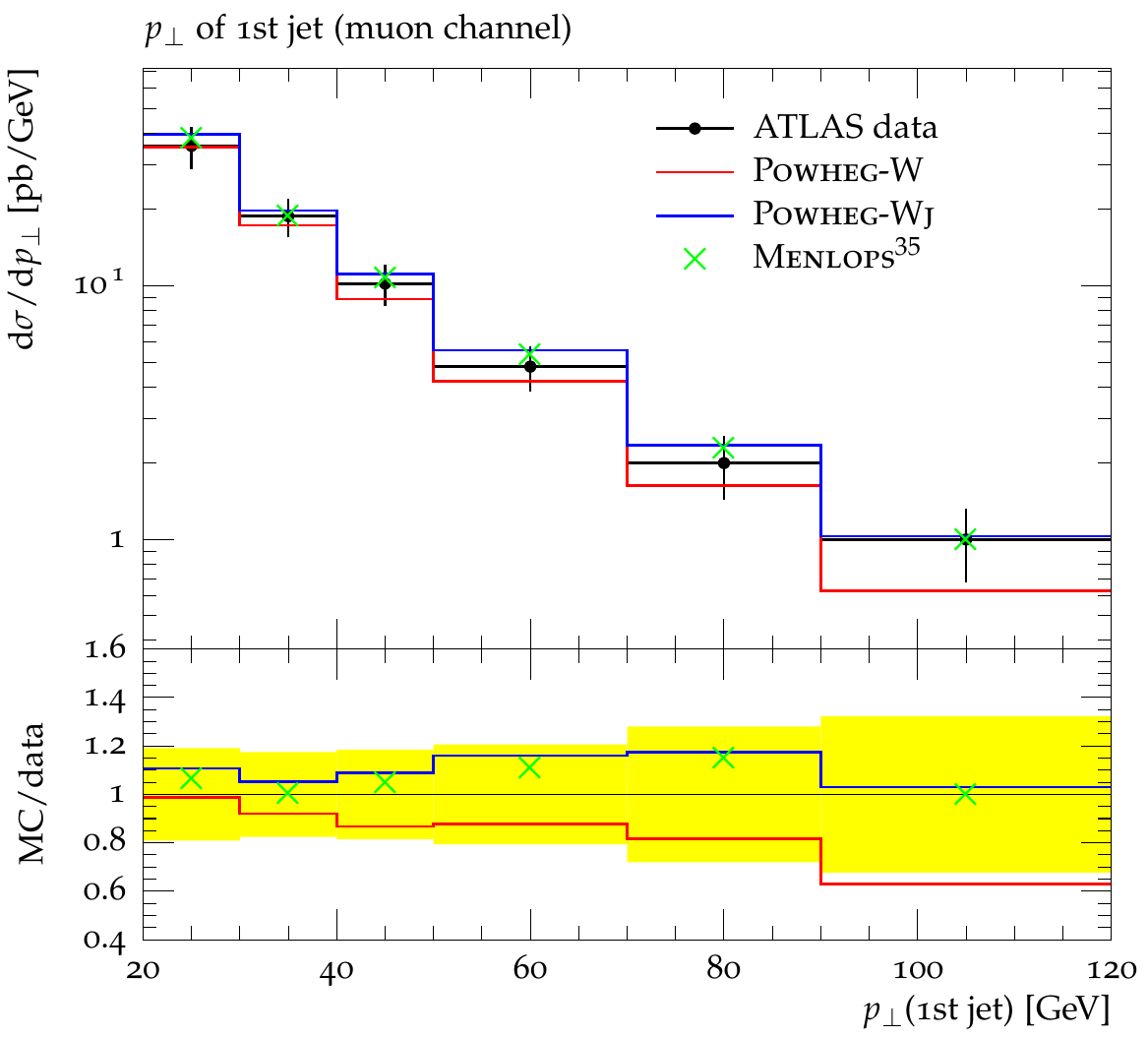}\hfill{}\includegraphics[width=0.4\textwidth]{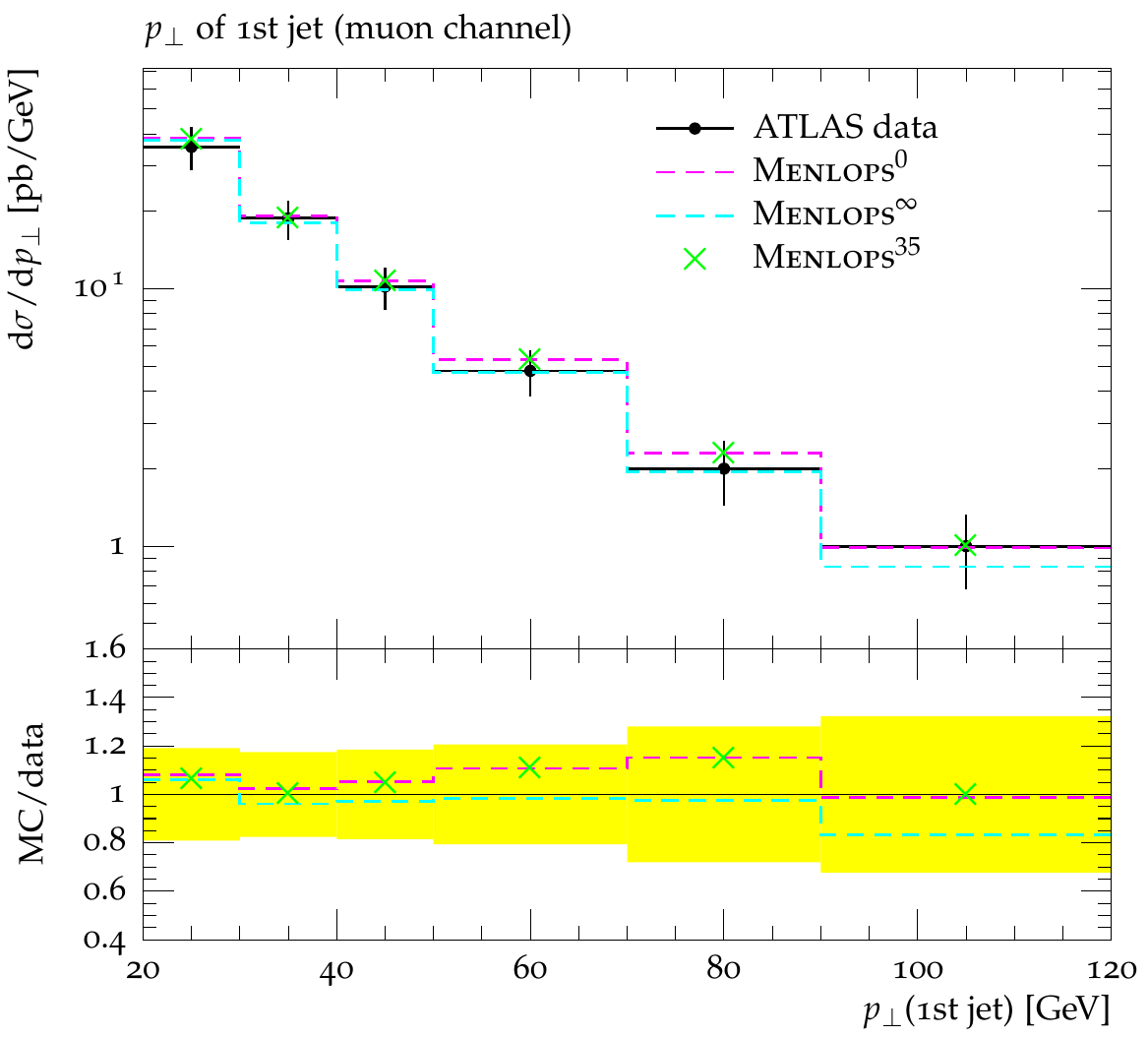}
\par\end{centering}

\begin{centering}
\includegraphics[width=0.4\textwidth]{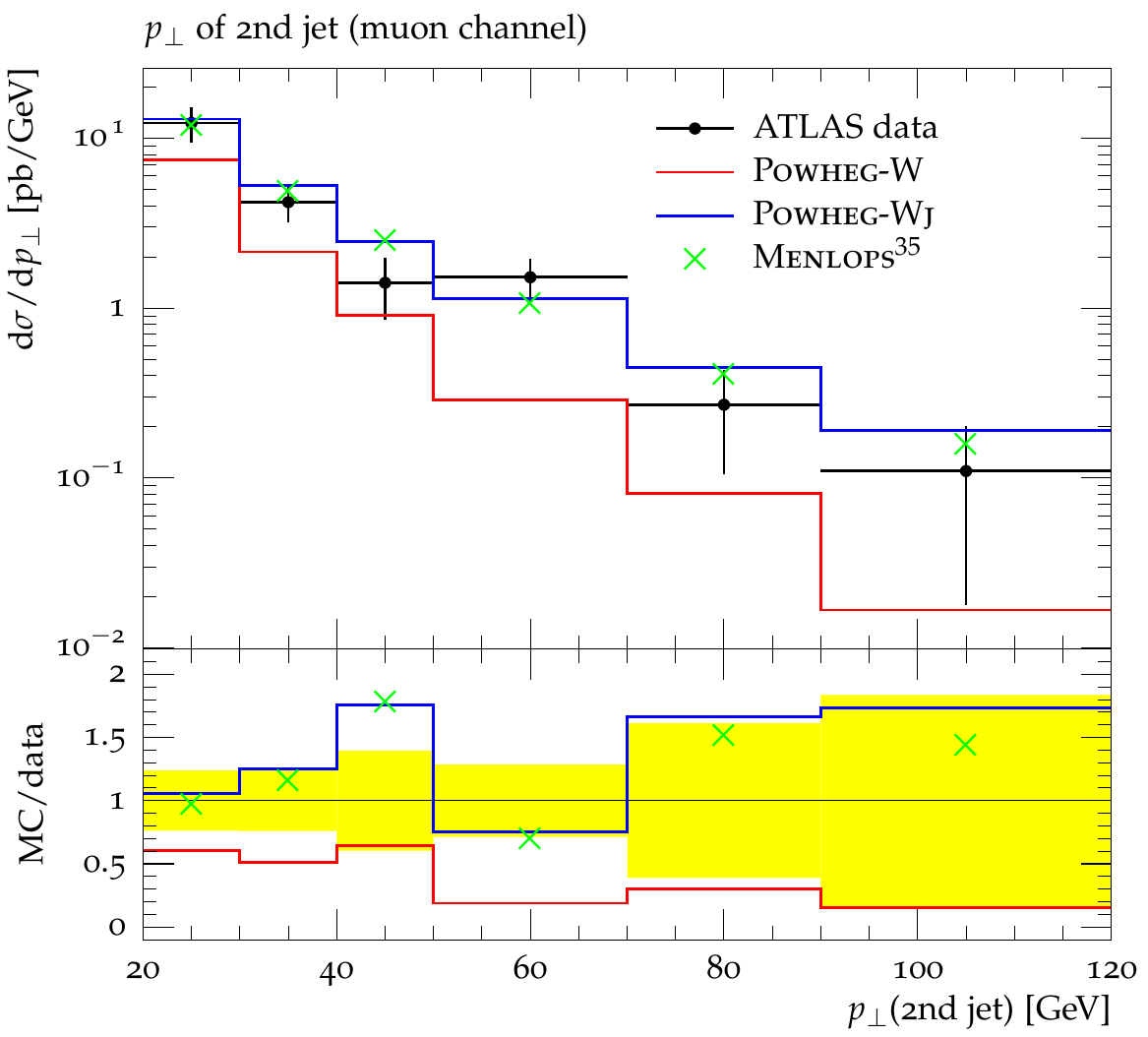}\hfill{}\includegraphics[width=0.4\textwidth]{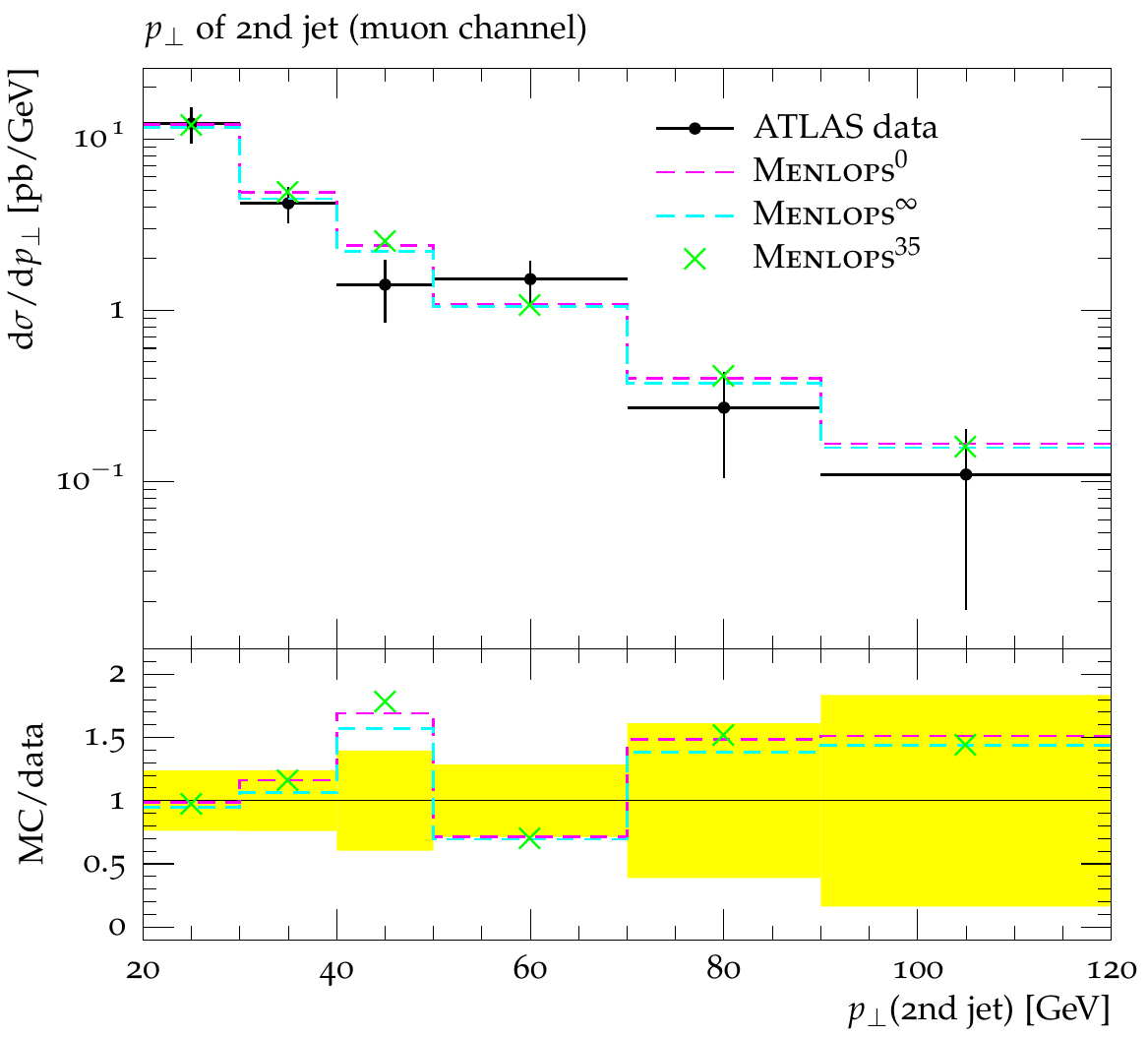}
\par\end{centering}

\caption{Here, on the left, we confront \noun{Powheg-W (}red\noun{), Powheg-Wj
}(blue) and \noun{Menlops$^{\text{35}}$} (green crosses) predictions
with \noun{Atlas} collaboration measurements \cite{atlas:2010pg}
of jet multiplicity and jet $p_{{\scriptscriptstyle \mathrm{T}}}$
spectra in the muon decay channel. On the right-hand side we show
the results from the corresponding \noun{Menlops$^{\text{0}}$ }(magenta)
and\noun{ Menlops$^{\text{\ensuremath{\infty}}}$ }(cyan dashes) improved
versions of these simulations, along with that obtained by merging
their output (\noun{\,Menlops$^{\text{35}}$\,).}}

\label{fig:lhc_W_jet_multiplicities_and_pTs_m_channel} 
\end{figure}

In figures \ref{fig:lhc_W_jet_multiplicities_and_pTs_e_channel} and
\ref{fig:lhc_W_jet_multiplicities_and_pTs_m_channel} we compare \noun{Powheg-W},
\noun{Powheg-Wj} and \noun{Menlops }predictions to \noun{Atlas }collaboration
measurements of the $\geqslant N$-jet cross sections ($0\leqslant N\leqslant4$),
as well as the leading and next-to-leading jet transverse momentum
spectra, in the electron and muon decay channels respectively \cite{atlas:2010pg}.

Regarding the leading and next-to-leading jet $p_{{\scriptscriptstyle \mathrm{T}}}$
spectra, all of the approaches, corrected and uncorrected, offer a
reasonable description of the data. The soft-collinear parton shower
approximation for the second jet in the \noun{Powheg-W }simulation
does not reveal itself to the same extent as in Fig.~\ref{fig:tevatron_Z_ee_jet_rates_and_pTs}
--- this should be exposed by greater statistics in future experimental
data sets --- nevertheless, one can see a systematic tendency for
the \noun{Powheg-W }program to undershoot the experimental measurements
in the higher $p_{{\scriptscriptstyle \mathrm{T}}}$ bins. On the
other hand, the relative softness of the radiation pattern in the
\noun{Powheg-W }program is much more in evidence in the $\geqslant N$-jet
cross sections where it can be seen to greatly underestimate that
of higher multiplicity events. As in Figs.~\ref{fig:tevatron_Z_ee_jet_rates_and_pTs},
\ref{fig:tevatron_Z_mm_angular_dists} this deficiency of the \noun{Powheg-W
}simulation is cured in the \noun{Menlops$^{\text{0}}$} enhanced
version (magenta lines).

From the point of view of the advantages to be had by the \noun{Menlops
}approaches here, considering just the jet $p_{{\scriptscriptstyle \mathrm{T}}}$
spectra, there is not much to be gained in practical terms with respect
to the \noun{Powheg-Wj }simulation; the discussion given earlier surrounding
figure \ref{fig:tevatron_Z_ee_jet_rates_and_pTs} largely applying
here too. On the other hand, looking also to the $\geqslant N$-jet
cross sections one can see, as expected, the \noun{Powheg-Wj }program
fails due to its unphysical description in the region where the vector
boson $p_{{\scriptscriptstyle \mathrm{T}}}$ is small, thus here the
\noun{Menlops }approach can start to pay dividends. The same advantage
will of course be present again in other measurements also covering
both the low- and high-$p_{{\scriptscriptstyle \mathrm{T}}}$ domains,
most obviously, the vector boson $p_{{\scriptscriptstyle \mathrm{T}}}$
spectrum, as seen already in Figs.~\ref{fig:tevatron_W_enue_pT}
and \ref{fig:tevatron_Z_mm_pT}. 

Having said that, let us re-emphasize the point made earlier in regards
to the jet $p_{{\scriptscriptstyle \mathrm{T}}}$ spectra in figure
\ref{fig:tevatron_Z_ee_jet_rates_and_pTs}, namely that we foresee
the main advantage in the \noun{Menlops }method being in the modeling
of W and Z production as background processes, where the significance
of the various jet multiplicity bins is not generally clear. Hence,
again, the real practical value of these plots is to demonstrate that
the \noun{Menlops }approach always embodies the best of the \noun{Nlops
}simulations and to show that the sensitivity of merged \noun{Menlops
}predictions to the merging scale is marginalized (as illustrated
by the plots on the right-hand side of the figures).

\pagebreak{}

\section{Conclusions\label{sec:Conclusions}}

Firstly in this article, we set out to investigate whether \noun{Powheg
}simulations of simple processes could be promoted to \noun{Menlops
}ones without approximation \cite{Hamilton:2010wh}. This was described
and demonstrated in Sect.~\ref{sub:v_menlops}, where only a fairly
modest effort was required; effectively recycling much of the radiation
generating machinery of \noun{Powheg} vector boson plus jet simulations
so as to generate next-to-next-to-leading order emissions according
to double emission matrix elements, instead of the parton shower approximation.
The method was shown to preserve the NLO accuracy of inclusive observables
and the logarithmic accuracy of those sensitive to Sudakov effects.
In all cases the \noun{Menlops} enhancement, labeled \noun{Menlops$^{\text{\ensuremath{\infty}}}$,
}provided a much improved description of Tevatron and LHC data with
respect to that of the underlying \noun{Powheg-V }program, specifically,
for observables sensitive to the emission of more than one jet. Given
the relative ease of the implementation and the findings of the validation
exercise, we believe that the same technique can be readily and successfully
applied to other processes, in particular, that it would be straightforward
to include the vector boson plus three partons matrix elements in
the same way. 

Our second goal was to seek improvement and similarly extend the reach
of vector boson plus jet simulations (the\noun{ Powheg-Vj} programs).
Here we have again worked in such a way as to minimize the interference
with existing programming, directly utilizing the radiation generating
apparatus of the \noun{Powheg-V }programs to induce Sudakov suppression
effects when the vector boson has low transverse momentum. The method
was shown to preserve the NLO accuracy of vector boson plus jet observables
and the all-orders resummation of Sudakov logarithms, while at the
same time giving predictions for fully inclusive observables in remarkably
good agreement with genuine NLO ones (better than one has right to
expect). The \noun{Menlops$^{\text{0}}$ }enhancements lead to a much
improved description of Tevatron and LHC data with respect to that
of the underlying \noun{Powheg-Vj }program, for the case where the
definition of the observables had some overlap with the Sudakov region,
\emph{e.g.} vector boson $p_{{\scriptscriptstyle \mathrm{T}}}$ and
rapidity distributions, as well as the 0-jet cross section (for which
the latter program exhibits unphysical behavior). We also remark that
the benefits here were not restricted to the program's output but
also extended to its run-time, which was found to be greatly reduced
due to it evaluating less complex NLO cross sections in the Sudakov
region.

Following these modifications we have considered the combination of
the resulting event samples in such a way as to yield a merged sample
improving on both of the components: inheriting both\emph{ }the NLO
accuracy of the enhanced vector boson production simulation \noun{Menlops$^{\text{\ensuremath{\infty}}}$
}and that of the corresponding vector boson plus jet simulation \noun{Menlops$^{\text{0}}$.
}We have discussed in Sect.~\ref{sub:v-vj_menlops} how the merging
scale must be bounded from below, to enforce that the fraction of
\noun{Menlops$^{\text{0}}$} events in the final sample not exceed
$\mathcal{O}\left(\alpha_{{\scriptscriptstyle \mathrm{S}}}\right)$
of the total, to maintain the NLO accuracy of inclusive observables.

As noted in Sect.~\ref{sub:v-vj_menlops}, if optimal renormalization
and factorization scale choices are used, the latter constraint should
effectively render the resummation in the vector boson plus jet simulations
null in the context of merging samples, with the \noun{Menlops$^{\text{0}}$
}sub-sample being\noun{ }open to replacement by one from the normal
\noun{Powheg-Vj }program. We therefore use the \noun{Menlops} improved
vector boson plus jet simulation in merging out of theoretical correctness.
On the contrary, the same is not true for the \noun{Menlops$^{\text{\ensuremath{\infty}}}$
}simulation, which if replaced by its unenhanced \noun{Powheg-V} program
would lead to two- and three-jet events being underestimated by the
parton-shower soft collinear approximation (exactly how much depends
on the value of the merging scale). In any case, by merging \noun{Menlops
}samples rather than \noun{Nlops }ones we minimize the merging scale
dependence: even for the most extreme variations we seldom find the
\noun{Menlops$^{\text{0}}$,} \noun{Menlops$^{\text{\ensuremath{\infty}}}$
}and merged \noun{Menlops }predictions out of agreement by more than
10\% (typically this occurs in regions were the predictions are all
no better than leading order anyway). The same cannot be said of the
unimproved \noun{Powheg} results. With this last point duly noted,
it seems likely that a true \emph{matching} of the \noun{Nlops} simulations,
without any merging scale dependence, would offer little in the way
of practical improvements beyond this pragmatic approach.\emph{ }

Good while this may sound, in regards to actually describing W and
Z (plus jet) production as well defined signal processes, none of
these enhancements seem to offer an improved description over what
could be done before, by taking care to use the right tool for the
right job (in the right way). Inclusive vector boson production measurements
are described as well by \noun{Powheg-V }programs as by any \noun{Menlops
}ones and the same is true in regards to the \noun{Powheg-Vj }programs
and vector boson plus jet measurements. In real terms, the main potential
improvement with regard to testing QCD is one of convenience, through
being able to perform inclusive and exclusive analyses with the one
simulation / event sample. On the other hand, when viewed as less
well defined background processes, one does not in general make any
clear distinction between vector boson and vector boson plus jet production.
It is in these applications that we believe the approach taken here
can be expected to offer real gains. Also, it is from this perspective
that the trouble taken to minimize the merging scale dependence, the
validation exercises and the favorable comparisons with real data,
take on their true significance. 

While the latter simple approach (Sect.~2.4) is not an exact theoretical
solution to the \noun{Nlops-Nlops }matching problem\noun{, }based
on the comparison with data in Sect.~\ref{sec:Results}, it would
appear to be an effective, convenient and extensible one. Comparisons
with a truly scale-free method will be interesting when one is realized
in the near future.

In conclusion we wish to recommend the \textsc{Menlops} procedures
documented here, in particular the merging of \noun{Menlops}-improved
\noun{Powheg }simulation output, as an efficient and clear means to
get the most from existing and future \noun{Nlops }simulations.

\section{Acknowledgments\label{sec:Acknowledgments}}

KH is very grateful to Paolo Nason and Torbjörn Sjöstrand for patiently
answering many questions regarding this work, and particularly to
Bryan Webber for reading and suggesting improvements to all versions
of the manuscript. KH thanks Italian Grid Infrastructure, TheoPhys
VO and INFN Milano-Bicocca computing support, without whose help and
resources this work would not have been possible. The work of SA is
supported by the Helmholtz Gemeinschaft under contract VH-HA-101 (Alliance
Physics at the Terascale) and by Deutsche Forschungsgemeinschaft in
Sonderforschungsbereich/Transregio 9. ER and SA also acknowledge financial
support from the LHCPhenoNet network under the Grant Agreement PITN-GA-2010-264564. 

\bibliographystyle{jhep}
\bibliography{meanlops}
 
\end{document}